\documentclass[12pt]{report}

\usepackage{cite,graphicx,amsmath,amssymb,bbm,color}
\usepackage{floatflt,epsf,pifont,amsfonts,mathrsfs}
\usepackage{caption}
\newcommand{\be}{\begin{equation}}
\newcommand{\ee}{\end{equation}}
\newcommand{\bea}{\begin{eqnarray}}
\newcommand{\eea}{\end{eqnarray}}

\newcommand{\dec}{{\rm \text{\tiny dec}}}
\newcommand{\IRc}{{\rm \text{\tiny IR,cr}}}
\newcommand{\IR}{{\rm \text{\tiny IR}}}
\newcommand{\UV}{{\rm \text{\tiny UV}}}
\newcommand{\RH}{{\rm \text{\tiny RH}}}
\newcommand{\pt}{{\rm \text{\tiny pt}}}

\newcommand{\eff}{{\rm \text{\tiny eff}}}
\newcommand{\maxx}{{\rm \text{\tiny max}}}
\newcommand{\minn}{{\rm \text{\tiny min}}}
\newcommand{\ym}{{\rm \text{\tiny YM}}}
\renewcommand{\Re}{\operatorname{Re}}
\renewcommand{\Im}{\operatorname{Im}}

\newcommand{\sgn}{\operatorname{sgn}}

\begin{document}
\thispagestyle{empty}
\begin{center}
\begin{normalsize}
\textbf{Dissertation\\
\vspace{.14cm}
submitted to the\\
\vspace{.14cm}
Combined Faculties for the Natural Sciences and for Mathematics\\
\vspace{.14cm}
of the Ruperto-Carola University of Heidelberg, Germany\\
\vspace{.14cm}
for the degree of\\
\vspace{.14cm}
Doctor of Natural Sciences\\}
\vspace{16.5cm}
\textbf{presented by \\
\vspace{.14cm}
Dipl.-Phys.~Benedict von Harling\\
\vspace{.14cm}
born in Celle\\
\vspace{.14cm}
Oral examination: 16th October 2008}
\end{normalsize}
\end{center}

\newpage
\thispagestyle{empty}
\textcolor{white}{.}

\newpage
\thispagestyle{empty}
\begin{center}
\vspace*{1cm}
\begin{Huge}
\textbf{Throat Cosmology}
\end{Huge}
\end{center}
\vspace{18cm}
\begin{normalsize}
\begin{tabbing}
\hspace*{2.5cm}\textbf{Referees:} \hspace*{2.5cm} \= \textbf{Prof.~Dr.~Arthur
Hebecker} \\
\> 
\vspace{.14cm}\textbf{Prof.~Dr.~Michael G.~Schmidt}
\end{tabbing}
\end{normalsize}

\newpage
\thispagestyle{empty}
\textcolor{white}{.}

\newpage
\thispagestyle{empty}
\paragraph{Throat-Kosmologie --- Zusammenfassung:}
In dieser Arbeit untersuchen wir\linebreak,\hspace{-.03cm},Throats" im
fr\"uhen, hei\ss en Universum. Throats sind ein h\"aufiges Merkmal in
der\linebreak
,\hspace{-.03cm},Land\-scape" der Typ-IIB-Stringtheorie. Wenn ein Throat
w\"ahrend der kosmologischen Ent\-wicklung aufgeheizt ist, wird nach und nach
Energie zu anderen Throats oder dem Standardmodell transferiert. Wir
berechnen die Transferrate von W\"armeenergie und die Zerfallsrate von im
Throat lokalisierten Kaluza-Klein-Moden in einem zehndimen\-sionalen Model.
Dazu benutzen wir die duale Beschreibung der Throats durch Eichtheo\-rien. Wir
diskutieren Modifikationen der Zerfallsrate, die in Fluss\-kompaktifizierungen
und f\"ur Klebanov-Strassler-Throats auftreten, und betonen die Rolle von
tachyo\-ni\-schen Skalaren in solchen Throats f\"ur die Vermittlung von
Zerf\"allen von Kaluza-Klein-Moden. Unsere Resultate sind auch anwendbar auf
den Ener\-gietransfer vom aufgeheizten Standardmodell zu Throats. Wir
bestimmen die daraus resultierende derzei\-tige Ener\-giedichte in Throats in
Abh\"angigkeit von den Infrarotskalen der Throats und der \linebreak
,\hspace{-.03cm},Reheating-Temperatur". Die Kaluza-Klein-Moden in den Throats
zerfallen in andere Sektoren mit einer stark unterdr\"uckten Rate. Falls ihre
Lebensdauer l\"anger als das Alter des Universums ist, sind sie ein
interessanter Kandidat f\"ur die Dunkle Materie. Wir zeigen, da\ss~Throats
mit
Infrarotskalen im Bereich von $10^5\hspace{+.12cm}\text{GeV}$ bis
$10^{10}\hspace{+.12cm}\text{GeV}$ die Dunk\-le Materie erkl\"aren k\"onnen,
wenn die Reheating-Temperatur $10^{10} - 10^{11}\hspace{+.12cm}\text{GeV}$
war. Wir finden zahlreiche Szenarien, in denen diese Form der Dunk\-len
Materie ausreichend langlebig ist, aber in denen Zerf\"alle zum Standardmodell
trotzdem durch Beobachtung von Gammastrahlung entdeckt werden k\"onnen.

\paragraph{Throat Cosmology --- Abstract:}
In this thesis, we study throats in the early, hot universe.
Throats are a common feature of the landscape of type IIB string theory. If a
throat is heated during cosmological evolution, energy is subsequently
transferred to other throats and to the standard model. We calculate the
heat transfer rate and the decay rate of throat-localized
Kaluza-Klein states in a ten-dimensional model. For the calculation, we employ
the dual description of the throats in terms of gauge
theories. We discuss modifications of the decay rate which arise in flux
compactifications and for Klebanov-Strassler throats and emphasize the role of
tachyonic scalars in such throats in mediating decays of Kaluza-Klein
modes. Our results are also applicable to the energy transfer from the heated
standard model to throats. We determine the resulting energy density in
throats at our epoch in dependence of their infrared scales and of
the reheating temperature. The Kaluza-Klein modes in the throats decay
to other sectors with a highly suppressed rate. If their lifetime is longer
than the age of the universe, they are an interesting dark matter candidate.
We show that, if the reheating temperature was $10^{10} -
10^{11}\hspace{+.12cm}\text{GeV}$, throats with infrared scales in the range
of $10^5\hspace{+.12cm}\text{GeV}$ to $10^{10}\hspace{+.12cm}\text{GeV}$
can account for the observed dark matter. We identify several scenarios where
this type of dark matter is sufficiently stable but where decays to the
standard model can be discovered via gamma-ray observations.

\newpage
\setcounter{page}{2}
\tableofcontents

\chapter{Introduction}
\section{String theory, flux and throats}
\label{introduction}
The standard model of particle physics is remarkably successful in that it
correctly predicts the outcome of a large number of experiments. It is a
certain type of quantum field theory, a theoretical framework which results
from the unification of quantum mechanics and special relativity. Yet there
are several reasons to believe that this theory is only an approximation
which is valid at comparatively low energies and that it has to be replaced
by a more fundamental theory at energies larger than that. In particular,
there is a combination of the velocity of light, Newton's constant and
Planck's constant which has the dimension of energy. For processes with
this Planck energy, gravity can no longer be neglected in standard model
interactions and a theory of quantum gravity is required. Such a
unification of general relativity with quantum field theory is still a
major open issue in fundamental physics.

A candidate for this unification is string theory in which the pointlike
particles of quantum field theory are replaced by tiny one-dimensional
objects. When viewed from large distances or, equivalently, at
small energies, these strings behave like pointlike
particles. String theory reduces to quantum field theory at such
low energies. For
processes at energies close to the Planck scale, on the other hand, the
fact that strings have a finite extent becomes important and quantum field
theory is no longer a good approximation.

The quantization of a classical string theory leads to a spectrum of
particles with various properties which correspond to different excitations
of the string. In particular, the spectrum contains a particle
which behaves like the graviton, the quantum of the gravitational field.
String
theory therefore provides a quantum theory of
gravity. In addition, gauge fields and particles which are charged under the
corresponding gauge groups appear naturally in the spectra of quantized
strings. For the appropriate gauge group and particle spectrum, the
standard model could thus follow as the low-energy limit of a
particular string theory.
Since the graviton is contained in the same spectrum, such a string
realization of the standard model would mean a unification of gravity with
the other known interactions. However, a completely satisfactory
realization has not been constructed so far.

The consistent quantization of string theory requires additional space
dimensions. To explain why these extra dimensions have escaped detection
so far, one assumes that they are curled up into a tiny space. The extra
dimensions are said to be compactified. At large distances or, equivalently,
low energies, our world then still seems to have only three space
dimensions. Although such extra dimensions are in principle an interesting
model building tool, the different ways in which they can be curled up lead
to a large variety of low-energy theories which follow from a
given string theory. 

This variety is even larger since additional choices can be made which
influence the low-energy theory. In certain string
theories, open strings are confined to hyperplanes in the higher-dimensional
space. These hyperplanes are called
Dp-branes, where p refers to the number of space dimensions in which
the plane is extended. After quantization
of the open strings, one finds a supersymmetric gauge theory which lives on
the world-volume of the D-brane. There are various possibilities
to embed these objects into a given compactification. The particular embedding
of the D-branes determines the gauge theory which lives on their world-volume.
In particular, it may be possible to realize the standard model
on D-branes. On the other hand, there is a large number of other gauge
theories which can be obtained in that way.

The spectrum of string theories contains differential form fields which are
generalizations of the Maxwell field of electrodynamics. The D-branes act as
sources for these form fields and can therefore be viewed as 
higher-dimensional generalizations of the pointlike sources of
electrodynamics. The latter sources lead to electric flux through a sphere 
which surrounds them. Similarly, in the aforementioned embeddings into a
compactification, the D-branes source form field flux which threads certain
compact submanifolds or cycles in the compact space. This flux
can, however, be switched on even in absence of
any D-brane sources. Due to the nontrivial topology of the compact
space, such a configuration remains stable. This is similar to the Dirac
monopole that can be viewed as a 
configuration of magnetic flux that is topologically stable because a
point has been removed from space. Furthermore, as is the case for
the Dirac monopole, the form field flux is quantized.

Typical compact spaces that one considers have a large
number of cycles. Through each of theses cycles, one can have a certain number
of
flux. Combined with the different possibilities for the compact
space and the number and embeddings of D-branes, this leads to a huge number
of compactified solutions of string theory. There are probably by far
more vacua in this so-called landscape than there are particles in
the visible universe. In principle, there are only
five consistent versions of string theory without any tunable parameters but
the landscape unfortunately introduces a large indeterminacy. In particular,
there may be
many solutions which look like the standard model at low energies but
which are very different from each other at high energies.

Flux is, on the other hand, interesting for model-building purposes.
The size of cycles in a
given compact space is initially unfixed. From a four-dimensional viewpoint,
these unfixed cycles lead to massless scalar fields which are called moduli.
The existence of such fields, however, would violate the equivalence principle
in a measurable way and is therefore excluded. The flux through a given
cycle, on the other hand, has a certain energy density. From a
four-dimensional
viewpoint, the potential energy will therefore
depend on the sizes of the cycles. Thus, flux can lead to their
stabilization. The moduli then become massive and are no longer a
phenomenological problem.

Since it has an energy density, the flux backreacts on the geometry. This
backreaction can lead to so-called warped regions in the compact space. In
these regions, there is a strong gravitational potential along certain
directions. As is well known from general relativity,
the deeper in such a gravitational potential a given physical process takes
place, the more does it appear redshifted for a fixed observer. Due to
this fact, large hierarchies can be generated in string compactifications
with flux.

More precisely, as Randall and Sundrum have
shown~\cite{Randall:1999ee,Randall:1999vf},
the four-dimensional graviton can be localized in a warped geometry. This
graviton is a Kaluza-Klein mode of the
higher-dimensional graviton and it therefore has a certain profile along the
compactified dimensions. Randall and Sundrum considered the five-dimensional
anti-de-Sitter
space. At fixed
positions along the fifth dimension of this spacetime, the geometry is
four-dimensional Minkowski space. The prefactor of the Minkowski metric,
however, depends exponentially on the position along the fifth dimension.
Accordingly, energy scales are exponentially redshifted or blueshifted if one
moves along this extra dimension. Randall and Sundrum chopped
this space off on two sides in such a way that they obtained a finite slice of
anti-de-Sitter space along the fifth dimension. The boundaries of this slice
are two copies of Minkowski space which are called the ultraviolet
brane and the infrared brane\footnote{These branes should not be confused with
D-branes from string theory. Both types of branes have in common that
they are hyperplanes in a higher-dimensional space.}, respectively.
Remarkably, the profile of the four-dimensional graviton in this geometry
turns out to be localized near the ultraviolet brane. Gravity is therefore
blueshifted compared to processes which are localized towards the infrared
brane. This fact allows for the generation of large hierarchies relative to
the
Planck scale. In particular, if the standard model is realized on the infrared
brane and
the redshift between the two branes corresponds to the hierarchy between the
Planck scale and the electroweak scale, this setup is a solution to the
hierarchy problem of the standard model.\footnote{In the setup that
we have described, this is actually not yet the case. A mechanism
is needed which stabilizes this geometry without too much fine-tuning.
Such mechanisms are known.}

It was shown by Giddings, Kachru and Polchinski~\cite{Giddings:2001yu} that
the Randall-Sundrum model can be realized in flux compactifications of type
IIB string theory. The backreaction of the flux on the geometry can lead to
the formation of a so-called Klebanov-Strassler throat~\cite{Klebanov:2000hb}
which plays the
role of the slice of anti-de-Sitter space in the Randall-Sundrum model. The
geometry of this warped region is smoothly terminated
in the infrared and thereby provides a realization of the infrared brane in
the Randall-Sundrum model. Furthermore, in the compactification proposed by
Giddings et al., the ultraviolet part of the Klebanov-Strassler throat is
smoothly embedded into the compact space. The unwarped part of the compact
space thus plays the role of the ultraviolet brane in the Randall-Sundrum
model. 

\section{String theory and cosmology}

It is expected that, in most vacua in the string theory
landscape, the strings are only slightly larger than the four-dimensional
Planck length.\footnote{Exceptions are compactifications with a large volume
and setups in which the standard model is localized in a strongly warped
region. String effects become important at much lower energies in these
setups.} The energies,
which are necessary to resolve this length scale and thus to test string
theory directly are many orders of magnitude larger than the energies
achieved in accelerators so far. It is therefore impossible to test string
theory directly in the near future and it may in fact never be possible. 

It is thus important to conceive other, indirect tests of string theory.
Since rather high energies have been reached in the early universe, it
is natural to consider cosmology for this purpose. String physics
may have led to observable signatures in several ways. As is well known, many
observations in cosmology imply a phase of inflation shortly after the big
bang. A possible string realization of such a phase is by means of two
D-branes which move slowly towards each other. Another realization employs
moduli with a sufficiently shallow potential. These inflationary mechanisms
lead to certain predictions for the anisotropies in the cosmic microwave
background and can thus be tested by experiments. However, tests of this kind
have to be interpreted carefully since in most cases other and in
particular field-theoretic realizations lead to similar signatures. 

The energy density of the universe becomes extremely diluted
during inflation and the universe has
to be reheated when inflation ends. Relics like topological defects may be
produced at this stage. For instance, during reheating after brane-antibrane
inflation in a Klebanov-Strassler throat, a network of cosmic strings is
formed. These cosmic strings influence the cosmic microwave background and
may in addition be detected by gravitational wave experiments. Note, however,
that the detection of cosmic strings would again be no clear signal of string
theory since they can also arise in field theories. 

Other relics like heavy particles may be produced by the reheating
mechanism and by thermal reactions in the hot standard model plasma. For
instance, Kaluza-Klein modes which are localized in a 
throat can be sufficiently light to be produced during reheating. As we have
discussed, energy scales of processes which are localized in warped
regions are redshifted. Similarly, Kaluza-Klein modes whose wave functions are
localized in a throat have redshifted and thus rather light masses. The decay
of relics which are produced in that way to standard model fields at later
stages of the cosmological
evolution may have led to observable signatures. On the other hand, if the
relics are sufficiently stable, they can provide a new explanation of the
observed dark matter. 

\section{Throats in the early universe}

A Klebanov-Strassler throat can be formed if flux threads
the cycles of a so-called conifold region (more precisely, of a deformed
conifold) in a compact Calabi-Yau space. It is expected that typical
Calabi-Yaus can have a large number of these conifold
regions. In the landscape of type IIB string theory vacua, flux is
distributed over the cycles of the Calabi-Yau in various ways. It is
then not surprising that in many cases the backreaction of the flux leads
to
a Klebanov-Strassler throat. These throats are indeed expected to be a common
feature of the type IIB landscape. It is therefore interesting to consider
possible observable consequences of throats in cosmology.

In the early, hot universe, these throats may have been heated to a certain
temperature. 
If the temperature of a Klebanov-Strassler throat is larger than a certain
critical energy scale, the backreaction of the thermal
plasma on the geometry can no longer be neglected and leads to the formation
of a black hole horizon which replaces the infrared end of the throat. Such a
black hole horizon emits Hawking radiation. Due to the warping of the
Klebanov-Strassler throat, this radiation has to tunnel through an effective
energy barrier before it can reach the unwarped part of the compact space
or other throats. Nevertheless, energy will be transferred to other throats
with a certain rate. This heat transfer rate is an important quantity
for the cosmology of throats.

In the first part of this thesis, based on our
publication~\cite{Harling:2007jy} (with Arthur Hebecker and Tatsuya
Noguchi), we calculate this rate in a simple setup: Two throats with
geometry
AdS$_5 \times$S$^5$ which are embedded into a six-dimensional torus. As
opposed to a Klebanov-Strassler throat, these throats are infinite in the
infrared direction. If a throat of this type is heated, the backreaction of
the thermal plasma leads to an AdS-Schwarzschild geometry. The heat
transfer rate from such a heated throat to another throat is determined by
the tunneling probability of the Hawking radiation. However, to determine
this probability, we have to solve a multi-dimensional tunneling
problem. Since this is quite difficult, we choose a different approach.

Consider a stack with a large number of D3-branes which is embedded into
flat
ten-dimensional space. The D3-branes have a certain energy density and
can therefore backreact on the geometry. Taking this backreaction into
account leads to a
background solution of type IIB supergravity which is known as a black
three-brane. Close to this object, the geometry is deformed to
AdS$_5\times$S$^5$, whereas far away it smoothly goes over to flat space.
The
crucial point is that, in this description, the D3-brane stack has
disappeared and is replaced by a curved geometry. The black three-brane is
therefore an \textit{alternative} description of the D3-brane
stack in flat space. This correspondence is the very basis of the AdS/CFT
duality: By taking the low-energy limit in both descriptions, one is led to
a
duality between string theory in the AdS$_5\times$S$^5$ region of the black
three-brane and the world-volume theory, a U$(N)$ gauge theory
with $\mathcal{N}=4$ supersymmetry, on the D3-brane stack.

In order to test the correspondence between a black three-brane and a
D3-brane stack, one can probe both objects with particles and compare
the corresponding absorption cross sections. Due to the warping, there is
again an effective energy barrier which separates the throat region from
the
asymptotically flat region of a black three-brane. The absorption cross
section of a particle, which is incident on the brane from the
asymptotically flat region, is determined by the corresponding tunneling
probability. The world-volume theory on the D3-brane stack, on the other
hand, couples to supergravity in the embedding flat space. The
absorption of a particle by a D3-brane stack is due to these couplings.
The corresponding calculations are particularly simple for the absorption of
a dilaton. Remarkably, the resulting absorption cross sections by a
D3-brane stack and by a black three-brane agree exactly.\footnote{It turns out
that, if
the geometry due to the backreaction is weakly curved and the description
in terms of an AdS$_5 \times$S$^5$ throat is applicable, the world-volume
theory on
the D3-brane stack is strongly coupled. For the aforementioned comparison,
however, the absorption process of a dilaton
by a D3-brane stack is calculated only at tree-level. The
fact that the resulting absorption cross section nevertheless agrees
with the absorption cross section by a black three-brane is explained by
a nonrenormalization theorem in the $\mathcal{N}=4$ supersymmetric U$(N)$
gauge theory.}

Heat transfer between two throats is due to tunneling and absorption of
Hawking radiation. Since the
absorption cross sections agree, we can replace
the two AdS$_5 \times$S$^5$ throats in our setup by two equivalent D3-brane
stacks. In particular, the heated throat is replaced by a stack with a
heated world-volume gauge theory. Both stacks are coupled to each other by
supergravity fields in the embedding torus. If we perform a
Kaluza-Klein expansion of these supergravity fields, we obtain a purely
four-dimensional description of our setup: A heated gauge theory which is
coupled to another
gauge theory by a tower of Kaluza-Klein modes. This four-dimensional
description is much easier to analyse than the initial ten-dimensional setup
with two throats in a torus. The heat transfer rate between the two throats
is the same as the corresponding rate between the two gauge theories. The
calculation of the latter is a straightforward exercise in quantum field
theory. 

\begin{figure}[t]
\begin{center}
\includegraphics[scale=2.2]{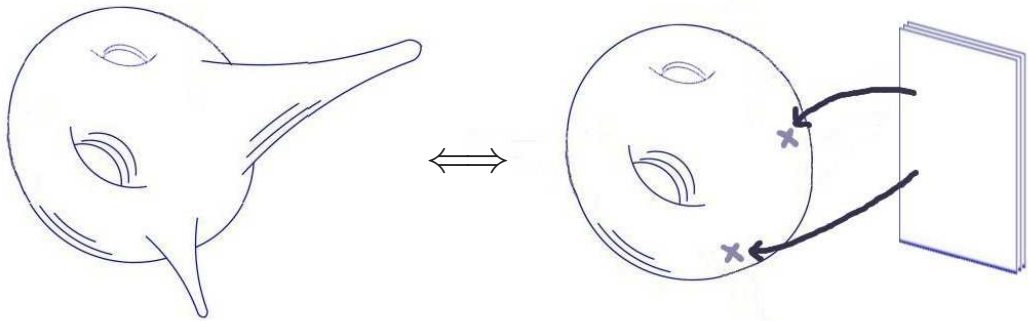} \put(-250,65){\Large$\Longleftrightarrow$}
\caption{Warped regions or throats in a compact space can 
equivalently be described by D-brane stacks. The four-dimensional
world-volume of these D-branes is aligned along the four
uncompactified dimensions. The D-branes are therefore pointlike in
the extra-dimensional compact space.}\label{TB}
\end{center}
\end{figure}

A Klebanov-Strassler throat can be understood as the result of the
backreaction of D3-branes and fractional D3-branes which are placed on a
conifold singularity. Such a throat can therefore equivalently be described by
a D-brane stack (cf.~Fig.~\ref{TB}). In this description, the supergravity
dynamics in the throat region is replaced by the dynamics of the world-volume
gauge theory. We will correspondingly refer to these two descriptions as the
gravity picture and the gauge theory picture, respectively.

The Klebanov-Strassler throat is finite in the infrared direction. In
the dual gauge theory, this is
reflected by the existence of a 
confinement scale. The gauge theory therefore has massive
glueball states which correspond to the Kaluza-Klein modes in the
throat. During cosmological evolution, the energy density in a throat
or, equivalently, in the dual gauge theory is diluted by the expansion of
the universe. If this energy density reaches the confinement scale, a phase
transition takes place and a gas of glueballs is formed. Similarly, the
black hole horizon in the throat is replaced by the infrared end of that
throat during the phase transition. The glueballs or
Kaluza-Klein modes which are formed during the phase transition can decay to
other throats or gauge sectors.
The corresponding decay rate is again important for the cosmology of throats.

As before, we calculate this rate in the gauge theory picture instead
of the gravity picture. In this description, the glueball decays to a
Kaluza-Klein mode of supergravity in the embedding space which in turn
decays to gauge fields on another brane stack. However, since the glueball is
a nonperturbative
object, the vertex between a glueball on a brane stack and supergravity
fields in the embedding space can not be read off from any Lagrangian. To
determine this vertex, we calculate the decay rate in a simple
example: The decay of a dilaton from the throat region of a black
three-brane to the asymptotically flat region. We can then determine the
decay vertex of a glueball on the equivalent brane stack by the
requirement
that the decay rate be reproduced from the gauge theory picture. Once we
have this vertex, it is again a straightforward exercise in quantum field
theory to calculate the decay rate of a glueball or a Kaluza-Klein mode to
another sector. We compare certain limiting cases of this decay rate with
results from calculations in the gravity picture.

The heat transfer rate that we derive is
applicable to more general geometries than our simplified setup with two
AdS$_5 \times $S$^5$ throats in a torus. In particular, we can consider a
different geometry for the embedding compact space. That the parametric
dependence of the heat transfer rate does not change is particularly easy to
see if the distance between the throats is of the order of the size of the
embedding space. We find that the heat transfer rate is then
dominated by
the mediation by zero-modes of supergravity fields in the embedding space.
The wave function of these lowest Kaluza-Klein modes, however, does not
depend on the geometry of the embedding space. Similarly, our result stays
correct for more general
throat geometries such as Klebanov-Strassler throats. Moreover, our results
can also be applied to small stacks of
D-branes. If the
standard model is realized on a small number of D-branes, we can determine
the rate of heat transfer to a throat.

For the decay rate in flux compactifications, on the other hand,
modifications of our result can arise. Certain moduli, which are massless
in our simplified setup and which mediate decays, become very massive due
to the flux. This fact can lead to a
suppression of the decay rate. In addition, the decay rate of Kaluza-Klein
modes in a Klebanov-Strassler throat is in general difficult to determine.
The equations of motion of these modes are involved because the 
flux in such a throat mixes field
fluctuations in a complicated way. In order to determine the glueball vertex
as
discussed above, we would have to solve these equations of motion. On the
other hand, we find that the glueballs in a given gauge sector (dual to a
throat) can decay
to a certain lightest glueball with the emission of a graviton. This process
is
similar to the decay of a hadron into a lighter hadron with the emission of
a photon. These decays typically happen on cosmologically short timescales.
From the point of view of cosmology, it is then sufficient to analyse 
decays to other sectors only for these lightest glueballs or the
corresponding lightest Kaluza-Klein modes. We show that there is a flat
direction for the supergravity fields in the Klebanov-Strassler throat.
The field which parameterizes this
flat direction has the same equation of motion as the dilaton in an AdS$_5
\times$S$^5$ throat and therefore couples to supergravity in the
embedding space with the previously derived vertex. We then argue that the
lightest Kaluza-Klein mode mixes
with this flat direction and thus also couples with the previously derived
vertex. Taking this fact and the suppression due to the
flux-stabilization of moduli
into account, we can determine the decay rate of this Kaluza-Klein mode.

A stronger decay vertex arises if the Kaluza-Klein mode mixes with a
tachyon in the Klebanov-Strassler throat. The reason is that the wave
function of a tachyon, i.e.~a scalar with a negative mass squared, is
less suppressed than the wave function of a dilaton if one moves in the
ultraviolet direction in the throat. A tachyon therefore couples with a
stronger vertex to supergravity fields in the embedding space. On the other 
hand, there is a compensating effect. In anti-de-Sitter space, scalars
with negative mass squared down
to the so-called Breitenlohner-Freedman bound~\cite{Breitenlohner:1982bm} do
not lead to
instabilities. In a Randall-Sundrum model, or a string realization
thereof, however, a tachyon must have a large mass on the ultraviolet brane
in order to avoid tachyonic Kaluza-Klein modes. This mass is similar to the
aforementioned mass of mediating fields in flux compactifications and again
leads to a suppression of the decay rate. In order to determine the
relative importance of the enhancing effect and the
suppressing effect, we calculate the decay rate of Kaluza-Klein
modes of a tachyon between two throats in a five-dimensional 
model. To this end, we approximate one throat by a Randall-Sundrum model.
The other throat is then dual to a gauge theory which lives on the
ultraviolet brane of the Randall-Sundrum model. The decay rate of Kaluza-Klein
modes between the two throats can be calculated as the decay rate of
Kaluza-Klein modes in the Randall-Sundrum model to gauge fields on the
ultraviolet brane. The corresponding part of the thesis is based
on~\cite{TachyonPaper} (with Sebastian Halter and Arthur Hebecker). 

\section{Dark matter in throats}

The glueball decay rates that we find are highly suppressed. If the
mean lifetime of the glueballs is larger than the age of the universe, these
particles are an interesting dark matter
candidate. Furthermore, a small fraction of these glueballs decays
to the standard model already at our epoch and can lead to
interesting observable signals. In the second part of this thesis, based
on our publication~\cite{Harling:2008px} (with Arthur Hebecker), we study
this new dark matter candidate. 

For definiteness, we assume that the standard model is realized
on some D-branes in the unwarped part of a compact space which also
has throat regions. The standard model and possibly the throats are heated by
the reheating mechanism after inflations ends. If the throats were heated to
the same temperature as the standard model, however, they would lead to too
much dark radiation during big bang nucleosynthesis and/or would overclose the
universe. Accordingly, the reheating mechanism has to interact more weakly
with the throats than with the standard model. The resulting abundance of
Kaluza-Klein dark matter depends on the model at hand. In order to provide a
model-independent lower bound on this abundance, we assume that the throats
receive no energy from the reheating process. Even under this minimal
assumption, energy is
deposited in a given throat due to heat transfer from the standard
model. 

The resulting energy density in the throat is diluted by the expansion of
the universe. In order to calculate the abundance of glueballs at our epoch,
we have to
determine the scaling behaviour of this energy density with the expansion
of the universe. This analysis is easiest in the gauge theory picture. As we
will show, if the energy density in a given gauge sector was above
the confinement scale after reheating, the gauge theory thermalizes. The
energy density then scales like radiation until a confinement phase
transition takes place. Afterwards, it scales like matter. 

The situation is different if the energy density was never above the
critical energy density for a confinement phase transition. The energy
density is produced by the annihilation of standard model particles into
some gauge theory states. This is similar to the annihilation of an
electron
with a positron into a quark and an antiquark. In QCD, this process leads
to two jets which subsequently hadronize. The particles, which are produced
in that way, are ultrarelativistic and would scale like radiation with the
expansion of the universe. As we will discuss in more detail, there are no
jets in the
strongly coupled gauge theories which are dual to throats. Instead, after
hadronization, the energy is completely in the form of slow glueballs.
These particles then immediately scale like matter. This fact partially
balances a large suppression of the heat transfer rate by four powers of
the four-dimensional Planck scale that we find. In particular, the energy
density in the corresponding throats can
have the right magnitude to account for the observed dark matter already
for moderately high reheating temperatures.

More precisely, we find that throats with infrared scales between
$10^5\hspace{+.09cm}\text{GeV}$ and $10^{11}\hspace{+.09cm}\text{GeV}$ can
account for the observed dark matter if
the reheating temperature was approximately $10^{10} -
10^{11}\hspace{+.09cm}\text{GeV}$. The
lifetime of Kaluza-Klein modes in these throats or, equivalently, of
glueballs can be made considerably longer than the age of the universe.
Nevertheless, for certain choices of parameters, glueball decays to the
standard model can lead to interesting observable signatures. In
particular, these decays produce hadrons which in turn decay to relatively
soft photons. These photons contribute to the diffuse $\gamma$-ray
background at low energies and may be detected by experiments like GLAST.
In addition,
certain glueballs species can decay directly into a pair of photons.
Decays of this type 
in the halo of our galaxy lead to a sharp line in the $\gamma$-ray
spectrum at high energies. This line may be observed by experiments like
HESS.

Since throats are expected to be a common feature in the landscape of type
IIB vacua, it is interesting to analyse scenarios with a large number of
throats. To this end, we use the estimate of the expected number of
throats in dependence of their infrared scales from
Ref.~\cite{Hebecker:2006bn}.
The decay rates that we find depend strongly on the mass of the
glueballs. In scenarios
with a large number of throats, it is therefore possible that glueballs
from certain sectors decay already at early epochs while other glueballs
are sufficiently stable to account for the observed dark matter. Decays to the
standard model at early epochs can influence primordial nucleosynthesis and
are therefore severely constrained. We check in a specific example
that the corresponding bounds can be fulfilled. 

\section{Organization of the thesis}
This thesis is organized as follows. In Chapters~\ref{throats}
and~\ref{StringRealizations},
we review several aspects of throats and their dual gauge theories that are
relevant for this thesis. We begin in Chapter~\ref{throats} by
recapitulating the Randall-Sundrum models. We then introduce black-three
branes and D3-branes and review a calculation showing that these branes
are just two descriptions of the same object. We discuss the AdS/CFT
conjecture and how it can be motivated by taking the low-energy limit of the
black 3-brane and the equivalent D3-brane stack. We also introduce the
Klebanov-Strassler throat and review the high-temperature phase of throats
and their dual gauge theories. In Chapter~\ref{StringRealizations}, we first
recapitulate how
the Randall-Sundrum II model can be obtain from string theory. We then
revisit the string realizations of the Randall-Sundrum I model which can
arise in flux compactifications. Finally, we conclude
Chapter~\ref{StringRealizations} with a
discussion of the statistical probability of throats in the landscape of flux
vacua.

In Chapter~\ref{et5d}, we first present a motivation for the central topic of
Chapters~\ref{et5d} to~\ref{modifications}: The transfer of energy in
different forms between throats in the early universe. We then review a
calculation of the decay rate of Kaluza-Klein modes between throats in a
simple five-dimensional geometry. We also present two other ways to derive
this decay rate, one way being based on our paper~\cite{Harling:2007jy}. In
Chapter~\ref{energytransfer}, following again~\cite{Harling:2007jy}, we derive
the energy loss rate of a heated throat to another throat in a simple
ten-dimensional model. This calculation is performed by modelling both throats
by equivalent stacks of D-branes. It is then straightforward to derive the
heat transfer rate by summing over the contributions of bulk Kaluza-Klein
modes coupling to both D-brane stacks. We also present a cross-check in which
we compare the heat loss of a throat and an equivalent D-brane stack to flat
space.

In Chapter~\ref{KKdecay}, which is still based on~\cite{Harling:2007jy}, we
describe an analogous calculation for the decay rate of Kaluza-Klein modes
localized in one throat to fields in a distant throat. In the gauge theory
picture, the decaying Kaluza-Klein modes are represented by glueballs. Thus,
we first derive the effective vertex for the coupling of these glueballs to
bulk fields. After that, the calculation proceeds analogously to that in the
previous chapter. Finally, we compare certain limiting cases of our result
with calculations in the gravity picture and with formulae from the
literature. 

Modifications to our results from Chapters~\ref{energytransfer}
and~\ref{KKdecay}, that arise in setups which are more realistic than our
simple ten-dimensional model, are discussed in Chapter~\ref{modifications}.
This chapter contains results from our papers~\cite{Harling:2007jy},
\cite{TachyonPaper} and~\cite{Harling:2008px}. We first argue that our heat
transfer rate is applicable to other geometries of the throats and the
embedding compact space. The decay rate of Kaluza-Klein modes from a
Klebanov-Strassler throat, on the other hand, is in general difficult to
determine. We discuss several aspects of the Kaluza-Klein spectrum of such a
throat and then show that all Kaluza-Klein modes decay quickly to a lightest
scalar state and its superpartner. As we argue, these Kaluza-Klein modes
decay again with the vertex from Chapter~\ref{KKdecay}.
Taking a suppression due to flux-induced masses into account, we derive the
decay rate of the lightest scalar Kaluza-Klein mode and its superpartner. On
the other hand, we find that Kaluza-Klein modes mixing with tachyons in a
Klebanov-Strassler throat decay with a stronger vertex than that from
Chapter~\ref{KKdecay}. Finally, we discuss the energy transfer between throats
and the standard model. 

Glueballs as a dark matter candidate are studied in
Chapters~\ref{thermalproduction} and~\ref{scenarios}. We describe the
thermal production of this form of dark matter in
Chapter~\ref{thermalproduction}. Using the heat transfer rate
from Chapter~\ref{energytransfer}, we determine the energy density
deposited in a throat by the heated standard model. We discuss how the
initially created gauge theory states hadronize and in which situations the
gauge theory thermalizes. Taking the resulting scaling of the energy density
with the expansion of the universe into account, we calculate the late-time
abundance of glueballs as a function of the reheating temperature and the
confinement scale of the gauge theory. 

Cosmological scenarios are discussed
in Chapter~\ref{scenarios}. First, we analyse scenarios with a single throat.
We find that a moderately long throat gives a promising dark matter candidate
which may allow for a discovery by new $\gamma$-ray experiments.
Then, we consider scenarios with a large number of throats, using results on
the distribution of multi-throat configurations reviewed in
Chapter~\ref{StringRealizations}.
We find that a throat in the required range of lengths is in many cases
present. We also discuss some issues in scenarios with low-scale
supersymmetry breaking and the relation of our dark matter scenario to
earlier work in the literature.

Our conclusions are given in Chapter~\ref{conclusions}. Some calculations
are relegated to appendices: The Kaluza-Klein decomposition of the graviton
and a tachyon in a Randall-Sundrum model is determined in
Appendices~\ref{KKmodes} and~\ref{TachyonAppendix}, respectively. In
Appendix~\ref{integral}, an integral is evaluated which is needed in
Chapters~\ref{energytransfer} and~\ref{KKdecay}. Finally,
Appendix~\ref{additional} discusses a process which takes place after the
confinement phase transition of gauge theories dual to finite throats.

Finally, let us fix some notations and conventions. Throughout this thesis,
we use the `mostly-plus' signature for the metrics. Our notation
for the form fields of type IIB supergravity is
e.g.~as in~\cite{Giddings:2001yu}. We
denote the (reduced) Planck scale in $d$ dimensions by $M_d$, the string scale
by $M_s$ and the string coupling by $g_s$. The symbol $\sim$ that we will
use frequently signifies equality up to $\mathcal{O}(1)$ prefactors, whereas
the symbols $\approx$ and $\simeq$ are reserved for results which hold to a
higher precision.

\chapter{Warped geometries and dual gauge theories}
\label{throats}

In this chapter, we will review several aspects of warped geometries. This
will fix our notation and will also provide results that will be needed in
later chapters. There are in particular two reasons for the enormous
interest in these geometries in recent years: As Randall and Sundrum have
shown, AdS-spaces allow for the generation of large hierarchies of scales.
Moreover, Maldacena has argued that string theory on AdS$_5$ times some
compact manifold is dual to a gauge theory. More exhaustive reviews
of these subjects can be found
e.g.~in~\cite{Aharony:1999ti,Rattazzi:2003ea,Gherghetta:2006ha}.

\section{The Randall-Sundrum models}
\label{RSmodels}

We will now review the Randall-Sundrum (RS) models
\cite{Randall:1999ee,Randall:1999vf} which will play an important role in
this thesis. In particular, the RSI model \cite{Randall:1999ee} offers a
possible solution to the hierarchy problem. However, we should clarify
at this point that we will not be concerned with the hierarchy
problem later in this thesis. Instead, we will more generally be interested
in the generation of large hierarchies in geometries of the RS type.

We consider a 5-dimensional theory and take the 5th dimension to be
compactified on an $S^1/\mathbb{Z}_2$ orbifold of length $\ell$. Thus, the
following equivalence relations hold for the 5th coordinate $y$:
\be
y \, \sim \, y + 2 \, \ell \qquad  y \, \sim \, - y  \,.
\ee
We denote the coordinates of the 4 uncompactified dimensions by $x$.
Points in spacetime
which remain fixed under the orbifold $\mathbb{Z}_2$-action $y \rightarrow -y$
form
$(3+1)$-dimensional hypersurfaces which are located at $y=0$ and $y = \ell$.
These hypersurfaces, which form the boundaries of the 5th dimension, are
called the ultraviolet (UV) brane and the infrared (IR) brane,
respectively. The branes can
support (3+1)-dimensional field theories. We denote the corresponding
Lagrangian on the UV brane by $\mathcal{L}_\UV$ and the Langrangian on the
IR brane by $\mathcal{L}_\IR$, respectively. Furthermore, the branes can
have
certain tensions $V_\UV$ and $V_\IR$. Taking also a 5-dimensional
cosmological constant $\Lambda$ into account, the action of the theory is
given by
\begin{equation}
\label{5daction}
S \, = \, \int d^4 x \int_{-\ell}^{\ell} dy \,  \sqrt{g} \,  \Bigl( 2 M^3_5
\, \mathcal{R} -
\Lambda  + \delta(y) \,  (\mathcal{L}_\UV - V_\UV) +
\delta(y- \ell)  \, (\mathcal{L}_\IR - V_\IR)\Bigr) \,.
\end{equation}
Here, $M_5$ is the 5-dimensional Planck mass.
We look for solutions of the Einstein equations that satisfy an ansatz of
the form
\be
\label{ansatz}
ds^2 \, = \, e^{-2 A(y)} \, \eta_{\mu \nu} dx^\mu dx^\nu + dy^2 \,.
\ee
Using this ansatz, the Einstein equations
which follow from the action Eq.~\eqref{5daction} reduce to
\begin{align}
A'^2 \, & = \, \frac{- \Lambda}{24 \, M_5^3} \label{EE1} \\
A'' \, & = \, \frac{1}{12 M_5^3} \Bigl( V_\UV \delta(y) + V_\IR
\delta(y -  \ell) \Bigr) \,.\label{EE2} 
\end{align}
A solution to Eq.~\eqref{EE1} which respects the orbifold
symmetry $y \rightarrow -y$ is 
\be
\label{solution}
A(y) \, = \, k \, |y|\, , \quad \text{where} \quad k \, \equiv \,
\sqrt{\frac{-
\Lambda}{24 M_5^3}} \,.
\ee
This solution is meaningful only if the 5d cosmological constant
is negative, $\Lambda < 0$. The geometry in between the two
branes accordingly is a slice of 5-dimensional anti-de-Sitter space
(AdS$_5$). The
curvature scale of this geometry is given by $k$. We can therefore trust
our solution only if $k \ll M_5$. Keeping in mind
that $y \sim y +2 \ell$, Eq.~\eqref{solution} also solves Eq.~\eqref{EE2} if
\be
V_\UV \, = \, - V_\IR \, = \, \sqrt{- \Lambda \, 24 M_5^3} \,.
\ee
This is a tuning of two parameters and is required in order to get a static
solution with 4d Poincar\'e symmetry. A fine-tuning of parameters
in a proposed solution to the hierarchy problem may seem strange. A
detailed
analysis shows (see e.g.~\cite{Rattazzi:2003ea} for a pedagogical
discussion) that one tuning is necessary to get a flat potential for the
scalar field which parameterizes the length of the 5th dimension (the radion).
Otherwise, in absence of a stabilization mechanism for the radion, the 5th
dimension would either collapse or decompactify. By stabilizing the radion
(which is anyway required in a phenomenologically viable theory), this tuning
is no longer necessary. The remaining tuning corresponds to a vanishing 4d
cosmological constant and is common to other solutions to the hierarchy
problem.

To discuss the effective 4d gravity theory from the point of view of brane
observers, we promote the background metric $\eta_{\mu \nu}$ in
Eq.~\eqref{ansatz} to
a dynamical field $\smash{g^{(4)}_{\mu \nu}}$. The 4d Ricci scalar
$\smash{\mathcal{R}^{(4)}}$ constructed from $\smash{g^{(4)}_{\mu \nu}}$ is
contained
in
the 5d Ricci scalar $\mathcal{R}$:
\begin{gather}
\label{gravityaction}
2 M^3_5 \, \int d^5 x  \, \sqrt{g} \, \mathcal{R} \quad \subset \quad 2
M^2_4 \, \int
d^4 x \, \sqrt{g^{(4)}} \, \mathcal{R}^{(4)} \,, \\ \label{5dPlanckMass}
\text{where} \quad
M^2_4 \, = \, M_5^3
\int^\ell_{-\ell} dy \, e^{- 2 k |y|} \, = \, \frac{M_5^3}{k} \left( 1 - e^{-2
k \ell} \right) \,.
\end{gather}
From the last equation, we see that the 4d Planck mass $M_4$ hardly
depends on the size $\ell$ of the extra dimension. In particular, it stays
finite in the limit $\ell \rightarrow \infty$. This suggests that we can
recover 4d gravity (with small corrections) on the UV brane even with an
infinite 5th dimension. This was analysed in great detail in
\cite{Randall:1999vf,Garriga:1999yh,Giddings:2000mu}. The corresponding
setup, in which the IR brane is sent to
infinity, is known as the RSII model. The setup in which the IR brane is
kept, on the other hand, is called the RSI model.

The Kaluza-Klein (KK) expansion of gravity in a RS model is discussed in
Appendix \ref{KKmodes}. In particular, one finds that the wavefunction of
the 4d
graviton is localized near the UV brane. The massive KK modes, on the other
hand, are localized in the IR and their couplings to the UV brane are
strongly suppressed. These KK modes therefore give
only small corrections to 4d gravity on the UV brane even if the IR brane
is sent to infinity. If the IR brane is kept, on the other hand, the masses
$m_n$ of the KK modes are quantized in units of
the warped AdS scale:
\begin{equation}
\label{massesKK}
 m_n \, \sim \, n \, m_\IR \,, \qquad  \text{where} \;\,  m_\IR  \equiv 
e^{-k \ell}  k \;\, \text{and} \;\, n \in \mathbb{N} \,.
\end{equation}

Next, we consider the field theory which is confined to
the IR brane in a RSI model in more detail. From Eq.~\eqref{5daction}, the
relevant part
of the action is
\be
\label{actionIR}
\int d^4 x  \sqrt{g^\IR} \, \mathcal{L}_\IR\left( g_{\mu \nu}^\IR, \phi, m
\right) \,.
\ee
Here, $\phi$ and $m$ collectively denote any fields and mass scales which
may appear in the action and $g^\IR$ is the induced metric on the IR brane.
Using Eqs.~\eqref{ansatz} and
\eqref{solution}, we have
\be
\label{rescaling}
\smash{g^\IR_{\mu \nu }= g_{\mu \nu}(x,y= \ell) = e^{- 2 k \ell} g^{(4)}_{\mu
\nu}}
\,.
\ee
The crucial point is that the action on the IR brane,
Eq.~\eqref{actionIR}, is given in terms of the metric $g^\IR$ which is
rescaled by a factor $e^{- 2 k \ell}$. The gravity part of the action,
Eq.~\eqref{gravityaction}, on the other hand, depends on the unrescaled metric
$g^{(4)}$. To compare the scales $M_4$ and $m$ in Eqs.~\eqref{gravityaction}
and \eqref{actionIR}, we have to agree on one metric to be used in both
actions. For the moment, we choose $g^{(4)}$ to be this metric. Rewriting
Eq.~\eqref{actionIR} in terms of $g^{(4)}$ by using Eq.~\eqref{rescaling}
brings factors of $e^{- 2 k \ell}$ into the action. These factors can be
absorbed into field redefinitions and the mass parameters. Thus, the
action is invariant except for the mass parameters which transform as
\be
\label{rescaling2}
m \; \longrightarrow \; e^{- k \ell} \, m \,.
\ee
The mass scales which appear in the Lagrangian using $g^{(4)}$ as 
reference metric are rescaled by a factor of $e^{- k \ell}$! In
particular, for $k \ell \simeq 35$, a mass of the order of the Planck scale
is scaled down to the TeV scale. Note however that, to solve the
hierarchy problem, a mechanism is required which stabilizes the
branes at the right distance (i.e. $k \ell \simeq 35$) without too much
fine-tuning. An example of such a mechanism is due to Goldberger and Wise
\cite{Goldberger:1999uk}.

Alternatively, we can use $g^\IR$ as the reference metric in
Eqs.~\eqref{gravityaction} and \eqref{actionIR}. In this case, we get a factor
of $e^{2 k \ell}$ into the action in Eq.~\eqref{gravityaction}. Absorbing
this
factor into the 4d Planck scale, we have
\be
\label{PlanckMass}
M^2_4 \,  = \, \frac{M_5^3}{k} \left( e^{2 k
\ell} -1\right) \,.
\ee
The mass scales $m$ in the action Eq.~\eqref{actionIR}, on the other
hand,
are not rescaled. From this viewpoint, we can have the 5d Planck scale and the
AdS scale in the TeV range and still get a sufficiently large 4d Planck scale.
According to Eq.~\eqref{PlanckMass}, the resulting hierarchy is again given by
$\smash{e^{k \ell}}$. In the following, however, we use the definition of
scales corresponding to Eqs.~\eqref{5dPlanckMass} and~\eqref{rescaling2}.

\section{D3-branes and black 3-branes}
\label{branes}
String realizations of the RS models exist which we will discuss in
Chapter~\ref{StringRealizations}. In these constructions, D-branes and
supergravity branes play an important role. Furthermore, these objects are
also the starting point from which one can motivate the AdS/CFT
conjecture. 

We consider type IIB string theory. A D3-brane is a $(3+1)$-dimensional
hyperplane in 10-dimensional space on which open strings end. At low
energies, the open strings give rise to a field theory which is
confined to the world-volume of the brane. For a D3-brane which is
embedded into 10d
Minkowski space, it is straightforward to guess this field theory: A
D-brane
preserves $1/2$ of the $\mathcal{N}=2$ supersymmetry of type IIB string
theory in the bulk~\cite{Polchinski:1995mt}. The field theory on
the 4d world-volume of the brane thus has $\mathcal{N}=4$ supersymmetry.
Furthermore, oscillations of a D3-brane in the 6
directions transverse to its world-volume lead to 6 scalar
fields.\footnote{These scalar fields are the Goldstone bosons due to broken
Lorentz invariance in the transverse directions
\cite{Sundrum:1998sj,Low:2001bw}.} Now, 6 scalars are
contained in an $\mathcal{N}=4$ vector multiplet.
Accordingly, the world-volume theory on a D3-brane is the $\mathcal{N}=4$
supersymmetric U$(1)$ gauge theory.

More precisely, a D3-brane and its interactions with supergravity fields
in the embedding space is governed by the so-called DBI
action. It reads
\begin{multline}
\label{DBI}
S_{\text{DBI}}\, = \, - \, T_3 \, \int d^4x \sqrt{- \det \left(
G_{\alpha \beta} +e^{- \phi/2}\left( B_{\alpha \beta} +  F_{\alpha
\beta}\right)  \right)
} \;, \\
\text{where} \quad \, 
G_{\alpha \beta} \, = \, \frac{\partial X^M}{\partial x^\alpha} \,
\frac{\partial X^N }{x^\beta} \, g_{M N}
\end{multline}
is the pullback of the 10d metric $g_{M N}$ to the 4d brane world-volume
parametrized by $x^\alpha$ ($\alpha = 0 \dots 3$) and the $X^M(x^\alpha)$
($M= 0 \dots 9$) describe the embedding of the brane into 10d space.
Similarly, $B_{\alpha \beta}$ is the pullback of the NS 2-form $B_2$,
whereas $\phi$ is the dilaton. The field strength of the U$(1)$ gauge
boson is denoted by $F_{\alpha \beta}$. We
have not written out
fermionic contributions to the action as well as interactions with RR-form
fields. Furthermore, $\smash{T_3=\sqrt{\pi}
M_{10}^4}$ is the tension of a D3-brane, where $M_{10}$ is the 10d Planck
scale. Note that, here and below, we work in the 10d Einstein frame.  Using
the static gauge $X^\alpha
= x^\alpha$ ($\alpha = 0 \dots 3$), an expansion of Eq.~\eqref{DBI} in a
10d
Minkowski background gives
\be
\label{DBIexp}
S_{\text{DBI}}\, = \, T_3 \, \int d^4x \left( 1 
-\frac{1}{4} e^{-\phi} F_{\alpha \beta}^2 - \frac{1}{2} \partial^\alpha X^m
\partial_\alpha X_m  + \text{interactions} \right) \,,
\ee
where the $X^m$ ($m= 4 \dots 9$) describe oscillations of the brane. We
have written out the interaction of the dilaton with the gauge field
strength because it will be needed later on.

A D3-brane is charged under the RR 4-form $C_4$ and thus sources a unit of
the corresponding flux. The resulting repulsion of two D3-branes due to
this charge is precisely cancelled by the mutual attraction due to gravity.
Thus, one can place $N$ D3-branes on top of each other. The world-volume
theory on this D3-brane stack is a $\mathcal{N}=4$ U$(N)$ gauge theory. 

As can be seen from the first term in Eq.~\eqref{DBIexp}, the tension
$T_3$ of a D3-brane acts as a localized cosmological constant. Therefore,
D3-branes backreact on the geometry. Assuming for the moment that the
resulting curvature is weak, the backreaction can be described by the low
energy limit of type IIB string theory, type IIB supergravity. The
corresponding solution is called a black 3-brane. It has the metric
\cite{Horowitz:1991cd}
\begin{multline}
\label{threebrane}
ds^2 \, = \, H^{-1/2}(r) \; \eta_{\mu \nu} dx^\mu dx^\nu \, + \,
H^{1/2}(r) \, \left[ dr^2 + r^2 d\Omega_5^2 \right] \,, \\
\text{where} \qquad H(r) \, = \, 1 + \frac{R^4}{r^4} \,, 
\end{multline}
$R$ is a curvature scale and $\smash{d\Omega_5^2}$ is the line element of
a 5-sphere. Furthermore, the 5-form field strength $F_5$ of the RR
4-form $C_4$ has an $r$-dependent vacuum expectation value whereas the
axion $C$ and the dilaton $\phi$ are constant. 

The geometry in Eq.~\eqref{threebrane} can be understood as the result
from the backreaction of a
D3-brane stack at $r = 0$. A natural question then is: What has happened
to the D3-brane stack? In the limit $r\ll R$, Eq.~\eqref{threebrane} takes
the form
\be
ds^2 \, = \, \frac{r^2}{R^2} \, \eta_{\mu \nu} dx^\mu dx^\nu \, + \,
\frac{R^2}{r^2} \, dr^2 \, + \, R^2 d\Omega_5^2 \,.
\ee
After a coordinate transformation to $\smash{y \equiv - k^{-1} \ln[k
r]}$, where $\smash{k \equiv R^{-1}}$, the metric reads
\be
ds^2 \, = \, e^{-2 k y}\, \eta_{\mu \nu} dx^\mu dx^\nu \, + \, dy^2 \, + \, 
R^2 d\Omega_5^2 \,.
\ee
As we have discussed in Section~\ref{RSmodels}, the first two terms in this
metric determine the AdS$_5$ geometry (cf.~Eqs.~\eqref{ansatz} and
\eqref{solution}) with curvature scale $k=R^{-1}$. The third term
describes an S$^5$ of radius $R$. The geometry of the brane in the region
$r \ll R$ thus is AdS$_5 \times$S$^5$. Using the above relation between
the coordinates $r$ and $y$, we see that $r \rightarrow 0$ corresponds to
$y \rightarrow \infty$. Thus,
instead of the initial D3-brane stack, one encounters an infinitely deep
throat region if one moves towards smaller $r$. The black 3-brane is therefore
believed to give an \textit{alternative} description of D3-branes: The
dynamics of open strings on a D3-brane is replaced by closed string dynamics
in the black 3-brane background. 

The curvature scale $R$ due to the
backreaction of $N$ D3-branes (or, equivalently, due to $N$ units of 5-form
flux through S$^5$) is
\begin{equation}
\label{RN}
R^4 \, \sim \,  N M_{10}^{-4} \, \sim \, N \, g_s \, M_s^{-4} \,,
\end{equation}
where $M_s$ is the string scale and $\smash{g_s=e^{\langle \phi \rangle}}$
is the string coupling. A
description in terms of classical supergravity is valid if $\smash{R \gg 
M_{10}^{-1}}$ and $\smash{R \gg M_s^{-1}}$. These conditions are
fulfilled if $N \gg 1$ and $N \, g_s \gg 1$. The Yang-Mills coupling
$g_\ym$ is related to the string coupling $g_s$ by $\smash{g_\ym^2
=g_s}$. Therefore, when the description in terms
of a black 3-brane is applicable, the gauge theory on the corresponding
stack of D3-branes is at large 't Hooft coupling $\lambda \equiv g_\ym^2 N$
and perturbation theory breaks down. When the gauge theory is in the
perturbative regime $\lambda < 1$, on the other hand, the supergravity
description is no longer valid. 

In the limit $r \gg R$, the function $H$ in the
metric Eq.~\eqref{threebrane} is approximately 1 and the geometry is 10d
Minkowski space. This corresponds to the fact that the geometry is due to
the backreaction of a D3-brane stack in 10d Minkowski
space. In particular, similar to the D3-brane, we can think of the black
3-brane as a localized object embedded into flat space.

\section{Absorption of a dilaton by a brane}
\label{3ba}
To check whether D3-branes and black 3-branes are really just two
descriptions of the same object is in general difficult, as we have seen in
the last section: If one description is weakly coupled, the other one is
strongly coupled and vice versa. We will now describe a
calculation~\cite{Klebanov:1997kc} which nevertheless allows such a check.

\begin{figure}[t]
\begin{center}
\includegraphics[scale=0.8]{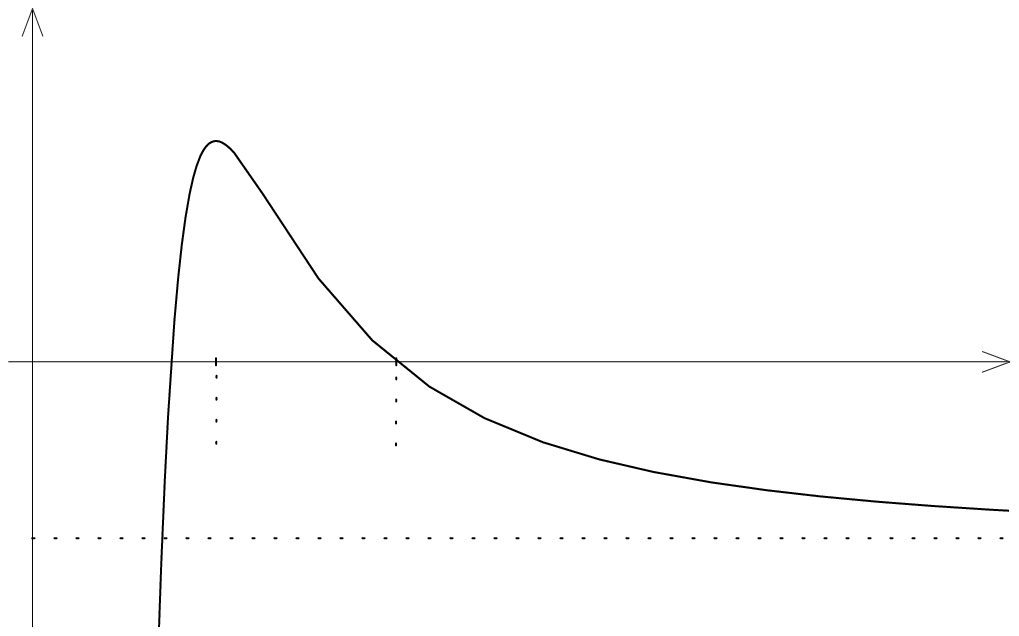}  \put(-240,160){$V/ \omega^2$}  
\put(-245,28){$-1$}  \put(-195,38){\footnotesize$\sim \omega R$}  
\put(-155,38){\footnotesize$\sim (\omega R)^{-1}$} 
\put(-50,55){$\omega R/\rho =z/R$}
\caption{Potential in the effective Schr\"o\-din\-ger equation for the
dilaton in a throat\label{pot}}
\end{center}
\end{figure}

Consider a dilaton $\phi$ which is incident on a black 3-brane. Its
equation of motion is the Laplace equation in the background
geometry given by Eq.~\eqref{threebrane}:
\be
\label{LE}
\partial_M \left( \sqrt{g} g^{MN} \partial_N \phi \right) = 0 \,.
\ee
Performing an expansion into eigenwaves of the angular Laplacian on S$^5$,
the equation of motion of an incident dilaton with energy $\omega$ reads
\be
\label{dilatonEOM}
\left(\rho^{-5} \frac{d}{d \rho} \rho^5 \frac{d}{d \rho} + 1 +
\frac{(\omega R)^4}{\rho^4} - \frac{l (l + 4)}{\rho^2} \right) \phi^{(l)}
\, = \, 0 \,.
\ee
Here, $\rho \equiv \omega r$
and $l$ determines the eigenvalues of the angular Laplacian. Introducing a
new coordinate
$\smash{z \equiv R^2/ r}$ and substituting $\smash{\psi \equiv
z^{-3/2} \phi}$, this can be written as
\be
\label{se1}
- \frac{d^2}{d z^2} \, \psi^{(l)} \, + \,  \left(\frac{15/4 + l (l +
4)}{z^2} -
\frac{\omega^2 R^4}{z^4} - \omega^2  \right) \psi^{(l)} \,
= \, 0 \,.
\ee
This has the form of a Schr\"odinger equation with a potential which is
given by the term in brackets. A schematic plot of this potential is
shown in Fig.~\ref{pot}. As one can see, a wave coming from the
asymptotically flat region ($z \rightarrow 0$ corresponding to $\rho
\rightarrow \infty$) has to tunnel through an effective barrier to reach
the inner region ($z \rightarrow \infty $ corresponding to $\rho
\rightarrow 0$).\footnote{As we will see in Section~\ref{comparison}, by
using cartesian coordinates for the directions transverse to the brane,
one again gets a Schr\"odinger-like equation. However, in this case
there is no barrier an incoming wave would have to tunnel
through. Instead, the reflection of a large part of the incoming wave
is due to the steepness of the potential well.} 

We will now solve Eq.~\eqref{se1} by applying the so-called matching
technique: For $z \gg \omega R^2$ corresponding to $\rho \ll 1$, the term
$\smash{\omega^2 R^4/z^4}$ in Eq.~\eqref{se1} can be ignored. An
approximate solution in this region is given by
\be
\label{solution1}
\phi^{(l)} \, = \,  \rho^{-2} \left[ J_{2+l}\Bigl(\frac{(\omega
R)^2}{\rho}\Bigr)+ i Y_{2+l}\Bigl(\frac{(\omega
R)^2}{\rho}\Bigr) \right] \,.
\ee
Here, $J$ and $Y$ are Bessel functions and the linear combination
corresponds to an incoming wave in the direction of small
$\rho$. This can be seen from the asymptotic forms of the Bessel functions
for large arguments: Eq.~\eqref{solution1} 
approximately is $\smash{\psi^{(l)} \propto \rho^{-3/2} e^{i (\omega R)^2/
\rho}}$ for $\rho \ll (\omega R)^2$. Another way of
writing Eq.~\eqref{dilatonEOM} is by substituting $\smash{\tilde{\psi}
\equiv \rho^{5/2} \phi}$:
\be
\left( \frac{d^2}{d \rho^2}  -\frac{15/4 + l (l + 4)}{ \rho^2} + 1 +
\frac{(\omega R)^4}{\rho^4} \right) \tilde{\psi}^{(l)} \, = \, 0 \,.
\ee
For $\smash{\rho \gg ( \omega R )^2}$, the term $\smash{(\omega
R)^4/\rho^4}$ can be ignored and an approximate solution in this region is
\be
\label{solution2}
\phi^{(l)} \, = \,\rho^{-2} \left[ A \, J_{2+l}(\rho) + B \,
Y_{2+l}(\rho)\right] \,.
\ee
The regions of validity of Eqs.~\eqref{solution1} and \eqref{solution2}
overlap for 
\be
\label{expansionparameter}
\omega R \ll 1 \,.
\ee
We restrict ourselves to this case from now on. Using the asymptotic forms
of the Bessel functions for small arguments, we can then match both
solutions in the overlapping region $( \omega R)^2 \ll \rho \ll 1$. This
fixes the constants $A$ and $B$ in Eq.~\eqref{solution2} and we find, to
lowest order in $\omega R$ and up to numerical prefactors:
\begin{gather}
\label{A}
A \;  \sim \; \left(\omega R \right)^{-4-2 l} \\
\label{B}
B \; = \; 0 \,.
\end{gather}
The absorption probability $\mathcal{P}$ of a dilaton by a black 3-brane
follows from the ratio of the flux at $\rho = 0$ and $\rho = \infty$.
Using
Eqs.~\eqref{solution1} and \eqref{solution2}, we have
\be
\label{tp}
\mathcal{P} \, \simeq \, A^{-2} \, \sim \, \left(\omega R
\right)^{8+4 l} \,.
\ee
In $(6+1)$ dimensions (where we mean the 6 dimensions transverse to the
world-volume of the 3-brane), the relation between the absorption
probability $\mathcal{P}$ and the absorption cross section $\sigma$ is
$\sigma \sim \mathcal{P} / \omega^5$, again neglecting numerical
prefactors. The absorption cross section per 4d world-volume of the 3-brane
thus is
\be
\label{cs1}
\sigma \, \sim \, \omega^{3 + 4 l} \, R^{8 + 4 l} \,.
\ee

Alternatively, we can consider the decay of a dilaton onto an 
equivalent stack of D3-branes. The kinetic term of the dilaton from the type
IIB 
supergravity action is
\be
S_{\text{IIB}}\, \supset \, -\frac{M_{10}^8}{4}\int d^{10}x\sqrt{g}
(\partial\phi)^2 \,.
\ee
The dilaton couples to the operator $\smash{F_{\alpha \beta}^2}$ on a
single D3-brane according to Eq.~\eqref{DBIexp}. Canonically normalizing
the kinetic terms of the dilaton and the fields $X^i$ and allowing
for brane fluctuations, we
get~\cite{Klebanov:1997kc}
\begin{equation}
\label{coupling2}
S \supset \frac{M^{-4}_{10}}{2^{3/2}}\left[ \int\! d^{4}x \, \phi(x,
\langle \vec{X} \rangle) \,  F_{\alpha \beta}^2 + \sum_{l=1}^{\infty} \int\!
d^{4}x \frac{M_{10}^{-2 l} }{ l!\pi^{l/4} } \left(\partial_{i_1}
\cdots \partial_{i_l} \phi \right)  X^{i_1} \cdots X^{i_l}
\, F_{\alpha \beta}^2 \right]\!,
\end{equation}
where $\smash{\langle \vec{X} \rangle}$ is the position of the brane. 
The coupling of a dilaton s-wave to the world-volume theory is due to
the first term in Eq.~\eqref{coupling2}, whereas higher partial waves
couple via the other terms. In additon, there are couplings of the dilaton
to fermionic terms which we can neglect for an order of magnitude estimate.
The absorption
cross section of the $l$-th partial wave by a single D3-brane follows
from the vertices in Eq.~\eqref{coupling2} as
\be
\sigma \,  \sim \, M_{10}^{-8-4 l } \omega^{3 + 4l} \,,
\ee
up to numerical prefactors. For a stack of $N$ D3-branes, the absorption cross
section has an extra factor of $ \sim N^{2+ l}$ since the gauge fields and
the fields $X^i$ are in the adjoint representation of
the U$(N)$ world-volume gauge theory.\footnote{For $N$ D3-branes, one has
traces over U$(N)$ indices in the DBI action and accordingly in
Eq.~\eqref{coupling2}. For example, the $(l = 1)$-vertex has the structure
$X^I_J F^J_K F^K_I$, where $I,J,K$ are U$(N)$ indices. The three summations
lead to a factor $\sim N^3$.} Using Eq.~\eqref{RN}, we then have
\be
\label{cs2}
\sigma \, \sim \, \omega^{3 + 4 l} \, R^{8 + 4 l} \,.
\ee
This is the same result as in Eq.~\eqref{cs1}! Thus, we have found 
the same parametric dependence of the dilaton absorption cross section for a 
black 3-brane and a D3-brane stack. We have ignored numerical prefactors in
our calculation. Remarkably, even these prefactors turn out to agree exactly 
\cite{Klebanov:1997kc,Gubser:1997yh,Klebanov:1999xv}. This is evidence that
black 3-branes and D3-branes are just two descriptions of the same object.

Taking a closer look, the exact agreement of the absorption cross sections
from both calculations comes as a surprise. The absorption cross section for
the black 3-brane actually is a perturbative expansion in $\omega R$
(cf. Eq.~\eqref{expansionparameter}). Corrections to Eq.~\eqref{cs1} in higher
orders of $\omega R$ are the only possible corrections in the limit
of classical supergravity (i.e.~for large $N$ and large
$\lambda$, cf.~Section~\ref{branes}).\footnote{For an s-wave process, these
corrections were calculated to higher orders in $\omega R$
in \cite{Gubser:1998kv,GubserHashimoto}.} The gauge theory on
the corresponding D3-brane stack, on the other hand, has a large 't Hooft
coupling $\lambda$. The absorption cross section therefore has the general
form
\be
\label{cs3}
\sigma \, \sim \, \omega^3  R^8 \, \left( 1 + b_1 \, \lambda + b_2
\lambda^2  + \dots \right) \, + \, \text{higher orders in $\omega R$} \,,
\ee
where we have taken $l=0$ for simplicity. If the coefficients $b_i$ in
Eq.~\eqref{cs3} were nonvanishing, the results from the calculations for
the black 3-brane and the D3-brane stack would disagree. It was shown
\cite{Gubser:1997se}, though, that all the coefficients $b_i$ are zero due
to a nonrenormalization theorem in 4d $\mathcal{N}=4$
super-Yang-Mills theory.\footnote{Actually, this was shown for the absorption
cross section
of a graviton polarized parallel to the D3-brane stack. A similar
nonrenormalization theorem is expected to hold also for the dilaton
absorption process.}

\section{The Maldacena or AdS/CFT conjecture}
\label{AdS/CFT}
As before, let us consider a black 3-brane and an equivalent stack of
D3-branes. We want to ana\-lyse the low-energy limit while keeping the
parameters $M_s$, $g_s$ and $N$ fixed.\footnote{In this section, we
follow closely the analysis in~\cite{Aharony:1999ti}.} In the black 3-brane
background, two
kinds of excitations can have arbitrarily low energies (when measured by a
fixed observer): There are massless closed string states in the
asymptotically flat region. These states give rise to type IIB supergravity
in 10d Minkowski space. In addition, there are closed string states (not
necessarily massless) in the throat region at smaller and smaller $r$ (cf.
Section~\ref{RSmodels}). These states form type IIB string theory
on AdS$_5 \times$S$^5$. Interactions between the two types of excitations
vanish in the low-energy limit. For example, the absorption cross section
of a dilaton which is incident on the brane from the asymptotically flat
region, Eq.~\eqref{cs1}, goes to zero for $\omega \rightarrow 0$. In turn,
states which are localized at smaller and smaller $r$ find it more and more
difficult to climb the gravitational potential to escape to the
asymptotically flat region. Thus, we end up with two decoupled theories in
the low-energy limit: Type IIB supergravity in flat space plus type IIB
string theory on AdS$_5 \times$S$^5$.

For the D3-brane stack, it is more convenient to keep the energy fixed and
to take the limit $M_s \rightarrow \infty$ instead. Since we keep $g_s$
fixed, this means that also $\smash{M_{10} \sim g_s^{-1/4} M_s
\rightarrow \infty}$. Interactions between the gauge theory on the stack
and the theory in the surrounding flat space vanish in this limit. This can
be seen, for example, from Eq.~\eqref{coupling2} which gives the leading
interactions between the dilaton and the gauge field strength. Since string
corrections vanish for $M_s \rightarrow \infty$, the theory in the
surrounding 10d Minkowski space is type IIB supergravity. Moreover, higher
derivative terms from the DBI action also vanish in this limit as can be
seen from
an expansion of Eq.~\eqref{DBI} (see e.g.~\cite{Gubser:1998kv}). The
remaining theory on
the D3-brane stack is the pure $\mathcal{N}=4$ U$(N)$ gauge theory which is
a conformal field theory (CFT). Thus, we again have two decoupled theories
in the low-energy limit (or, equivalently, for $M_s \rightarrow \infty$):
Type IIB supergravity in flat space and $\mathcal{N}=4$ U$(N)$
super-Yang-Mills.

In summary, we find two decoupled theories in the low-energy limit of the
black 3-brane as well as the D3-brane stack. The fact that one
theory is the same for both types of branes (namely type IIB supergravity
in flat space) led Maldacena to the conjecture \cite{Maldacena:1997re} that
also the other two theories are identical. The (Maldacena or AdS/CFT)
conjecture thus is that type IIB string theory on AdS$_5 \times$S$^5$ is
the same as (or dual to) $\mathcal{N}=4$ SU$(N)$ super-Yang-Mills in 4d
Minkowski space.\footnote{Up to some $\mathbb{Z}_N$ identifications, U$(N)$
is SU$(N) \times$U$(1)$. The U$(1)$ factor is related to the center of mass
motion of the branes. On the black 3-brane side, it corresponds to certain
low-energy modes which live in the transition region between the throat and
flat space. These modes and the U$(1)$ can be omitted from the
correspondence (see e.g.~\cite{Aharony:1999ti}).}

As an immediate test of the correspondence, one can compare the symmetries
on both sides. The isometry group of AdS$_5$ and the conformal group in 4d
Minkowski space are both SO$(4,2)$. In particular, a
translation along the radial coordinate of AdS$_5$ maps to a conformal
transformation
on the gauge theory side. The radial coordinate of AdS$_5$ can therefore be
viewed as the dual of the renormalization scale of the gauge
theory.\footnote{Of course, since the gauge theory is conformal and
AdS$_5$ is homogenous in the radial direction, both sides of the duality
are invariant under such a transformation.}
The gauge theory has an SU$(4)$ $R$-symmetry which can be identified with
the
isometry group SO$(6)$ of S$^5$. Furthermore, one can show that even the
supergroups which contain the aforementioned bosonic symmetries agree.

More generally, it is believed that other theories on AdS$_5$ which
include 
gravity are as well dual to a CFT in 4d Minkowski space. In particular,
one can consider the RS models in the light of the AdS/CFT duality. In the
RSII~model, AdS$_5$ is
cut off in the UV by a brane (cf.~Section~\ref{RSmodels}). Since the radial
coordinate of AdS$_5$ is dual to the renormalization scale, the RSII model
is dual to a CFT which has a cutoff at a certain UV scale.
Furthermore, due to the cutoff, the RSII model has a normalizable 4d
graviton. One therefore expects that the dual CFT is coupled to 4d
gravity as well. Let us describe a test of this correspondence. To
this end, one calculates corrections to the 4d Newton potential on both
sides of the duality. On the RSII side, these corrections are due to the
exchange of KK modes between two test particles and result in a leading
correction which goes like distance$^{-3}$. On the CFT side, on the other
hand, one considers the coupling of the energy-momentum tensor $T$ of the
CFT to the 4d graviton. The graviton propagator is corrected by insertions
of the 2-point function $\langle T T\rangle$. Even though the CFT is
strongly coupled, this 2-point function is fully determined by conformal
invariance. The leading insertion again results in a
distance$^{-3}$-correction to the Newton potential
\cite{Gubser:1999vj}.

In the RSI model, the AdS$_5$ space is additionally cut off in the IR by
another brane. The dual theory correspondingly has a
certain IR scale at which the conformal symmetry is broken (see
e.g.~\cite{ArkaniHamed:2000ds}). Below this scale, we expect a discrete
spectrum of particle-like states to which we refer as glueballs. 
This spectrum simply corresponds to the tower of KK modes in the RSI
model. The latter fact can be used to determine the masses of glueballs
from a KK expansion.

\section{The Klebanov-Strassler throat}
\label{KS}
Up to now, we have considered D3-branes (and their backreaction) in 10d
Minkowski space. Another interesting option for the embedding space is a
product of 4d Minkowski space and the conifold. The latter is a
6-dimensional submanifold
of $\mathbb{C}^4$ defined by the equation
\be
\label{conifold}
z_1^2 + z_2^2 + z_3^2 + z_4^2 \, = \, 0 \,.
\ee
One can show that this space is a cone\footnote{This can be
seen from the fact that Eq.~\eqref{conifold} is invariant under
$z_i \rightarrow t z_i$ for real $t$.} over the 5-dimensional
manifold $T^{1,1} \equiv$
(SU$(2)\times$SU$(2))/$U$(1)$. It has a singularity at
$(z_1,z_2,z_3,z_4)=0$ \cite{Candelas:1989js}. The metric can be written as 
\be
ds^2 \, = \, dr^2 \, + \, r^2 d\Omega_{T^{1,1}}^2 \,,
\ee
where $\smash{d\Omega_{T^{1,1}}^2}$ is the line element on $T^{1,1}$.
The space is Ricci-flat and an explicit Calabi-Yau metric is known 
\cite{Candelas:1989js} but will not be needed in the following. 
We place a large number $N$ of parallel D3-branes, which are aligned along
4d Minkowski space, on the singularity of the conifold. The metric due to
their backreaction is
\begin{multline}
\label{threebraneC}
ds^2 \, = \, H^{-1/2}(r) \; \eta_{\mu \nu } dx^\mu dx^\nu \, + \,
H^{1/2}(r) \, \left[dr^2 + r^2 d\Omega_{T^{1,1}}^2\right] \,, \\
\text{where} \qquad H(r) \, = \, 1 + \frac{R^4}{r^4}
\end{multline}
and $R$ is given in Eq.~\eqref{RN}. The gauge theory on the world-volume of
the D3-brane stack
was determined in \cite{Klebanov:1998hh}: It is a conformal
$\mathcal{N}=1$
SU$(N) \times$SU$(N)$ super-Yang-Mills with some chiral superfields and a
certain superpotential. Studying the
low-energy limit, we are led to a duality between this gauge theory and
string theory on the small-$r$ part of Eq.~\eqref{threebraneC}, which is
AdS$_5 \times$T$^{1,1}$. 

It is interesting to look for nonconformal theories. To this
end, one can place a certain number of fractional D3-branes on the conifold
singularity in addition to D3-branes~\cite{Klebanov:1999rd}. Topologically,
$T^{1,1}$ is S$^3 \times$S$^2$.
Fractional D3-branes are D5-branes which are wrapped over the
2-cycle of $T^{1,1}$. Each fractional D3-brane sources a unit of
$F_3$-flux through the 3-cycle of $T^{1,1}$, where $F_3$ is the field
strength of the RR 2-form $C_2$.\footnote{Our notation for the form-fields of
type IIB supergravity is e.g. as in~\cite{Giddings:2001yu}.} This flux causes
the 5-form flux through $T^{1,1}$ to depend on the radial coordinate $r$
\cite{Klebanov:1999rd,Klebanov:2000nc}:
\be
\label{N_eff}
N_{\text{eff}}(r) \, \sim \, N \, + \,  g_s \, M^2 \ln(r/r_0) \,.
\ee
Here, $N$ and $M$ are the numbers of D3-branes and fractional D3-branes,
respectively, $r_0$ determines a UV scale and we have suppressed an
$\mathcal{O}(1)$ factor in the second term. 

The fluxes backreact on the geometry and this backreaction produces a
so-called Klebanov-Strassler (KS) throat~\cite{Klebanov:2000hb}:
Sufficiently far away from the
conifold singularity, the metric again has the form given in
Eq.~\eqref{threebraneC}. Furthermore, the curvature scale $R$ is fully
specified
by the number of 5-form flux $N_{\text{eff}}$ as in Eq.~\eqref{RN} and the
warp factor thus is
\be
\label{lw}
H(r) \; = \; 1 + \frac{ R_\UV^4 + R_\IR^4 \,  \ln(r/r_0)}{r^4} \; = \;  1
+ \frac{R_\IR^4 \, \ln(r/r_s)}{r^4} \,. 
\ee 
Here, $R_\UV$ and $R_\IR$ are, respectively, the AdS radius at the UV scale
defined above and at an IR scale whose meaning will become clear in a
moment. The AdS radii are determined by the corresponding numbers of 5-form
flux $N_\UV = N_{\text{eff}}(r_0) = N$ and $N_\IR \sim g_s
M^2$ (cf.~Eq.~\eqref{RN}). 

Comparing Eqs.~\eqref{threebrane} and \eqref{lw} for small $r$,
we see that the warping of the KS throat deviates logarithmically from AdS
warping. Due to the relation between the radial coordinate $r$ and the
renormalization scale of the dual gauge theory, we conclude that the latter
is no longer conformal. Indeed, the dual gauge theory, which has again
$\mathcal{N}=1$ supersymmetry, has running gauge couplings and performs a
so-called duality cascade: The
rank of the gauge group, which is SU$(N+M)\times$SU$(N)$ in the UV, is
repeatedly reduced by a series of Seiberg duality transformations along
the renormalization group flow towards the IR.

Close to $r=r_s$,
the warp factor Eq.~\eqref{lw} vanishes and the metric becomes singular.
This singularity is unphysical and can be removed if one
replaces the conifold by the so-called deformed
conifold~\cite{Klebanov:2000hb}. The latter is a 
submanifold in $\mathbb{C}^4$ defined by the modified equation
\be
\label{deformedconifold}
z_1^2 + z_2^2 + z_3^2 + z_4^2 \, = \, \epsilon \,.
\ee
Due to the small constant $\epsilon$, the singularity is replaced by a
finite S$^3$ at the bottom of the deformed conifold. The KS throat is thus
finite. This fact corresponds to the existence of a confinement scale in
the dual gauge theory. Indeed, the duality cascade is stopped at a certain
IR scale and the remaining gauge group confines.

\section{Heated branes, throats and gauge theories}
\label{HeatedBranes}
The black 3-branes that we have considered in Section~\ref{branes} are also
known as extremal 3-branes because they fulfill a BPS condition. A
generalization are the non-extremal 3-branes with background metric
\begin{multline}
\label{metricne3brane}
ds^2 \, = \, H^{-1/2}(r) \, \left[-f(r) \, dt^2 + dx^i dx_i \right] \, + \,
H^{1/2}(r) \, \left[f^{-1}(r) \, dr^2 + r^2 d\Omega_5^2 \right] \,, \\
\text{where} \qquad 
f(r) \, = \, 1 + \frac{r_0^4}{r^4}
\end{multline}
and the warp factor $H(r)$ is as before. This brane has a black hole
horizon at $r=r_0$ which in turn has a certain Hawking temperature.
This brane is therefore dual to a 
stack of D3-branes on which the world-volume gauge theory is heated to the
same temperature (see \cite{GKP}). Absorption calculations in order to test
this correspondence where performed in \cite{Satoh:1998ss}. The results
for the D-brane stack (at zeroth order in the 't Hooft coupling $\lambda$ but
taking finite-temperature effects into account) and the non-extremal
3-brane have the form
\be
\label{csNE}
\sigma_T \, \sim \, \sigma_0 \; f\Bigl(\frac{\omega}{T}\Bigr) \,.
\ee
Here, $\omega$ is the energy of the incident dilaton and $T$ is the Hawking
temperature. Furthermore, $\sigma_0\sim \omega^3 R^8$ is the absorption
cross section by an extremal 3-brane which was determined
in Section~\ref{3ba}. It was found in~\cite{Satoh:1998ss} that the function
$f$ differs for the 3-brane and the D-brane stack. This is not
surprising: The fact that the zeroth order (in $\lambda$) calculation gave
the correct result in Section~\ref{3ba} was due to a nonrenormalization
theorem in 4d $\mathcal{N}=4$ super-Yang-Mills theory. Such a
nonrenormalization theorem
is usually related to supersymmetry. At nonzero temperature, however,
supersymmetry is broken and we cannot expect the nonrenormalization theorem
to hold any more.

As in Section~\ref{AdS/CFT}, we can study the low-energy limits of the
non-extremal 3-brane and the heated D3-brane stack. In that way, we are
led to a duality between string theory in the small $r$ part of
Eq.~\eqref{metricne3brane}, the so-called AdS-Schwarzschild space, and the
$\mathcal{N}=4$ SU$(N)$ gauge theory at nonzero temperature \cite{W}. 

It is also interesting to analyze theories with IR cutoff or confinement
scale at nonzero temperature. A simple example is provided by the RSI model
and its dual gauge theory. The low-temperature phase can be
described by a gas of KK modes or glueballs. From the gauge theory point
of view it is clear that, if we raise the temperature above the confinement
scale, the gauge theory undergoes a deconfinement phase transition. As
we have discussed above, this deconfined phase is dual to an
AdS-Schwarzschild geometry. In this phase, the IR brane is thus replaced by
a Schwarzschild horizon. In turn, if the theory is cooled below the
confinement scale, a confinement phase transition takes place. The
corresponding transition between AdS-Schwarzschild and AdS$_5$ with IR
cutoff was analyzed in \cite{Creminelli:2001th}: It is first order and
involves the nucleation of bubbles of the IR brane phase from the
Schwarzschild horizon. Similarly, if the KS throat is heated above
its critical temperature, it develops a horizon. The corresponding
supergravity solution was found in \cite{BHinThroat}. The phase
transition between this phase and the KS throat was studied in \cite{KSpt}
using the full 10d geometry and in \cite{Hassanain:2007js} using a 5d model
of the KS throat developed in \cite{Brummer:2005sh}.

\chapter{String realizations of the Randall-Sundrum model}
In this chapter, we will continue our review of material that will be
relevant for this thesis. We will modify the geometries from
the last chapter in order to obtain approximate RS models. The resulting
solutions are examples of flux compactifications of type IIB string
theory. In particular, we will see that approximate RS models are quite
common in the landscape of type IIB flux vacua. More exhaustive reviews can be
found e.g.~in~\cite{Frey:2003tf,Grana:2005jc,Douglas:2006es}.

\label{StringRealizations}
\section{The Verlinde compactification}
\label{verlinde}
In the small-$r$ region, the geometry of black 3-branes is AdS$_5
\times$S$^5$. These
objects are therefore an interesting building block for a string
realization of the RS model. Black 3-branes are not terminated at large
$r$, however, but go over to 10d Minkowski space. In order to obtain an
(approximate) RS model, we have to add an UV cutoff to this geometry. To
this end, we consider an orientifold of type IIB string theory on a
6-torus $T^6$ \cite{Verlinde:1999fy}. This compactification has $64$
O3-planes which are located at all the half-way points of the $T^6$. The
charge of an O3-plane is $-1/4$ times that of a D3-brane. To fulfill the
tadpole cancellation (or vanishing charge) condition, we place 16
D3-branes into the orientifold.\footnote{More precisely, there are 32
D3-branes in the $T^6$. These are pairwise identified under the orientifold
$\mathbb{Z}_2$-action.} If these D3-branes are on top of each other, their
backreaction on the geometry creates an AdS$_5 \times $S$^5$ throat which
is glued into the torus \cite{Verlinde:1999fy}. To see this, recall that
a black 3-brane, which is dual to a D3-brane stack, can approximately be
viewed as a localized object which is embedded into
flat space. The diameter of this object is given by the AdS scale. As long
as the size $L$ of the torus
is larger than this diameter $R$, it should be possible to glue the black
3-brane into the torus. Moreover, even though the number of D3-branes is
not very large, there is still a parameter range in which we can trust the
supergravity approximation. Using Eq.~\eqref{RN} with $N=16$ and inserting
the omitted numerical prefactor, we have to require that 
\be
  g_s \, \gg \, \frac{1}{64 \, \pi} \,.
\ee
In principle, we also have to take the backreaction of the O3-planes into
account (see \cite{Verlinde:1999fy}). For simplicity, we will neglect it in
the following. The metric is then given by\footnote{The backreaction of
O3-planes leads to additional terms in the function $H(\vec{x}_\perp)$.}
\begin{multline}
\label{3branetorus}
ds^2 = H(\vec{x}_\perp)^{-1/2} \,  \eta_{\mu \nu} \, dx_\shortparallel^\mu
dx_\shortparallel^\nu  + H(\vec{x}_\perp)^{1/2}
\,  dx_\perp^{\textcolor{white}{.} i} dx_{\perp
 \textcolor{white}{.} i}^{\textcolor{white}{0}}\,,\\
\text{where} \qquad H(\vec{x}_\perp) \, = \, 1 + \sum_{\vec{n} \in
\mathbb{Z}^6} \frac{R^4}{|\vec{x}_\perp - \vec{A} + \vec{n} L |^4} \,.
\end{multline}
The coordinates along the 4 uncompactified
dimensions are denoted by $x_\shortparallel$, the
$\smash{x_\perp}$ refer to coordinates in the torus and
$\vec{A}$ is the position of the D3-brane stack. The sum in the warp factor
$H$ is due to mirror effects in the torus. In particular, we see that close
to
the D3-brane stack, i.e.~for small $\smash{r=|\vec{x}_\perp - \vec{A}|}$,
the space is indeed AdS$_5 \times$S$^5$. At large $r$, on the other hand,
the space is cut off by the compactness of the torus.

We thus have obtained an approximate RSII model from string theory. The
torus
plays the role of the UV brane, whereas the AdS$^5 \times $S$^5$ throat
replaces the slice of AdS$_5$. As in the RSII model, we expect a
normalizable 4d graviton. The 4d Einstein-Hilbert action is contained in
the 10d action:
\begin{multline}
\label{PlanckScale}
M_{10}^8 \int d^{10}x \,  \sqrt{g_{10}} \, \mathcal{R}_{10} \, \supset
\,
M_4^2 \int d^4x_\shortparallel \sqrt{g_4} \, \mathcal{R}_4 \,, \\ \text{where}
\qquad M_4^2 \, = \,M_{10}^8 \int_{T^6} d^6x_\perp \,
H(\vec{x}_\perp) \, \sim \, M_{10}^8 \, L^6 \,.
\end{multline}
We have used Eq.~\eqref{3branetorus} and the fact that $L>R$ as discussed
above. The integral thus yields a finite result and the 4d graviton is
indeed normalizable.

\section{Flux compactifications \`a la GKP}
\label{GKP}
In the UV, the geometry of a KS throat (i.e.~the small-$r$ part of the
geometry in Eqs.~\eqref{threebraneC} and \eqref{lw}) is approximately
AdS$_5 \times$T$^{1,1}$, whereas it is smoothly terminated in the IR. The
KS
throat is therefore an interesting building block for a string realization
of the RSI model. A way to add a smooth UV cutoff to this geometry was
found in a seminal paper by Giddings, Kachru and Polchinski
(GKP)~\cite{Giddings:2001yu}:

Many Calabi-Yau manifolds develop conifold
singularities at certain points in their moduli space. Close to this
singularity, the manifold is then described by Eq.~\eqref{conifold}. More
precisely, we consider an orientifold of such a Calabi-Yau and place $N$
D3-branes and $M$ fractional D3-branes on the conifold singularity. The
tadpole cancellation (or vanishing charge) condition requires
negative-charge objects somewhere in the rest of the compact space. These
can be provided e.g.~by some O3-planes. The backreaction
of the D-branes on the geometry creates a KS throat and this yields
a string realization of the RSI model: The IR end of the throat plays the
role of the IR brane in the RSI model. Since the throat is embedded into
the Calabi-Yau orientifold, it is also terminated in the UV. The (rest of
the) Calabi-Yau orientifold thus corresponds to the UV brane of the RSI
model. 

The KS throat in this setup can equivalently be understood as a result of
the backreaction of 3-form fluxes on a deformed conifold. This can be seen as
follows: Each fractional D3-brane
sources a units of $F_3$-flux through the 3-cycle of the $T^{1,1}$. This
flux together with the $F_5$-flux from the D3-branes acts as a source
for $B_2$. Moreover, the $F_5$-flux can be absorbed into $B_2$ (see
\cite{Klebanov:2000nc}). Altogether, this leads to a certain number of
$H_3$-flux (recall that $H_3=dB_2$) through another 3-cycle which
consists of the 2-cycle of the $T^{1,1}$ and the radial
direction.\footnote{This cycle is the Poincar\'e dual of the 3-cycle of
the $T^{1,1}$.} Since we consider a deformed conifold which is embedded into a
compact space, this submanifold is indeed compact. Denoting the two
3-cycles by $A$ and $B$, we have
\begin{equation}
\begin{split}
\label{flux}
\left(\frac{M_s}{2 \pi}\right)^2 \int_A \, F_3 \, & = \,  M \\
\left(\frac{M_s}{2 \pi}\right)^2 \int_B \, H_3 \, & = \, -  K \,.
\end{split}
\end{equation}

In order to determine the resulting hierarchy, we analyse the setup from 
a 4d viewpoint. A compactification of type IIB supergravity on a
Calabi-Yau orientifold leads to an $\mathcal{N}=1$ supersymmetric
low-energy theory. Certain flux choices may also break this
amount of supersymmetry but we restrict ourselves to a supersymmetric setup
for simplicity. From the 4d perspective, the blow-up parameter $\epsilon$ in
the defining equation of the deformed conifold, Eq.~\eqref{deformedconifold},
is promoted to a 4d scalar field $z(x)$. Without fluxes, this scalar is
massless and as such is an example of a modulus in string compactifications.
The fluxes enter into a superpotential of the Gukov-Vafa-Witten type
\cite{Gukov:1999ya,Giddings:2001yu}:
\be
W \, = \, \int G_3 \wedge \Omega \, = \, \left(\frac{2 \pi}{M_s}\right)^2
\left( M
\int_B \Omega - K \, \tau \int_A \Omega \right) \,.
\ee
Here, $\smash{\tau= C + i e^{-\phi}}$ is the axio-dilaton, $G_3 = F_3 -
\tau H_3$ and $\Omega$ is the holomorphic 3-form. It is a well-known result
\cite{Candelas:1990pi} that\footnote{The first equation is usually
taken as a definition for the modulus $z$. We have already defined $z$ in
a different way, namely via Eq.~\eqref{deformedconifold} (recall that
$\epsilon \rightarrow z$).} 
\be
\int_A \Omega \, = \, z \quad \text{and} \quad \int_B \Omega \, = \,
\mathcal{G}(z) \, \equiv \, \frac{z}{2 \pi i}  \ln z \, +  \,
\text{holomorphic} \,.
\ee
As we will see in a moment, the holomorphic terms are not important
for a leading-order analysis. For a supersymmetric vacuum, we have to
require
that $D_z W = 0 $, where $D_z = \partial_z + (\partial_z \mathcal{K})$ is the
K\"ahler covariant derivative and $\mathcal{K}$ is the K\"ahler potential:
\be
0 \; = \; D_z W \; \propto M \partial_z \mathcal{G} - K \tau + \partial_z
\mathcal{K} \left(M \mathcal{G} - K \tau z \right) \; \sim \; \frac{M}{2 \pi
i} \ln z - i \frac{K}{g_s} \,.
\ee 
For the last result, we have assumed that $\smash{K \gg g_s}$ and that
$z$ is exponentially small. Solving for $z$, we see that the latter
assumption can be justified if $M$ is not too large:
\be
\label{stabilization}
z \, \sim \, e^{- 2 \pi K / M g_s} \,.
\ee
In order to estimate the generated hierarchy between the UV end, where
the throat goes over to the embedding Calabi-Yau orientifold, and the IR
end, we can neglect
the logarithmic running of the AdS scale $R$ in a KS throat. Deep in the
throat, the warp factor
Eq.~\eqref{lw} is $\smash{H(r) \simeq (R/r)^4}$. As we have seen in
Section~\ref{RSmodels}, 4d energy scales of processes which are located
at a
certain distance $r$ in the throat are rescaled by a factor of $h^{-1
}\equiv H(r)^{-1/4} \simeq r/R$ (cf.~Eq.~\eqref{rescaling2}). Due to the
deformation of the conifold, the S$^3$ inside the T$^{1,1}$ does not shrink
to zero
size at the tip (as for the singular conifold) but it only shrinks to a
certain radius $r_c$. The throat is then terminated in the IR at this value
of the radial coordinate $r$. From the explicit supergravity solution from
\cite{Klebanov:2000hb}, we know that the size of the S$^3$ at the bottom of
the KS throat is $\smash{r_c \sim z^{1/3} R}$. The hierarchy between the
unwarped
part of the compact space and the bottom of the KS throat thus is 
\be
\label{hierarchy}
h \, \sim \, e^{2 \pi K / 3 M  g_s} \,.
\ee
To give an example: For $K=9$, $M=5$ and $g_s=0.1$, we find $\smash{h \sim
10^{16}}$, the hierarchy between the Planck scale and the electroweak
scale. 

We see from the result in Eq.~\eqref{stabilization} that the
(complex-structure) modulus $z(x)$ is stabilized by the 3-form flux. More
generally, it was shown in \cite{Giddings:2001yu} that the other
complex-structure moduli of the Calabi-Yau orientifold can be
stabilized as well by using 3-form flux. In addition, it is possible to fix
the dilaton at a small value (as we have assumed in deriving
Eq.~\eqref{stabilization}). 

For completeness, let us mention that it is not possible to stabilize the
K\"ahler moduli along the lines of \cite{Giddings:2001yu}: They remain flat
directions of the scalar potential. This flatness is due to a cancellation in
the scalar potential which in turn results from the form of the tree-level
K\"ahler potential (which is of the so-called no-scale type). Perturbative
corrections to the K\"ahler potential may lead to the stabilization of the
K\"ahler moduli (see e.g.~\cite{vonGersdorff:2005bf} and references
therein).\footnote{Recall that the K\"ahler potential, as opposed to the
superpotential, is not protected from perturbative corrections.}
Alternatively, non-perturbative corrections to
the superpotential may depend on the K\"ahler moduli. This fact was
employed in the seminal work by Kachru, Kallosh, Linde and Trivedi (KKLT) to
construct stabilized de Sitter vacua \cite{Kachru:2003aw}: Using
D3-instantons or gaugino condensation on D7-branes, the universal K\"ahler
modulus can indeed be stabilized but the resulting vacuum has a negative
cosmological constant. This vacuum can be uplifted by adding an
anti-D3-brane in a warped region. By varying the length of this throat, the
contribution of the brane to the vacuum energy can be fine-tuned in order
to yield a vacuum with a small, positive cosmological constant. Although
this vacuum is only metastable, its lifetime can be shown to be
considerably larger than the age of the universe.

\section{Statistics}
\label{statistics}
Calabi-Yau manifolds can have a large number of 3-cycles. Through each of
these 3-cycles, one can have some 3-form flux as in Eq.~\eqref{flux}. The
resulting variety of different flux configurations on different Calabi-Yau
orientifolds is known as the landscape of type IIB flux vacua. The large
number of vacua ($\sim\hspace{-.07cm}10^{500}$ in some estimates) in this type
of
constructions makes them amenable to statistical analysis
\cite{Douglas:2003um}. In particular, given a Calabi-Yau orientifold, one
can ask for the number of vacua in which the complex-structure moduli (and the
dilaton) are stabilized at a particular value.
In \cite{Ashok:2003gk,Denef:2004ze}, it was shown how to calculate this
number from the
metric on the corresponding moduli space.

We will particularly be interested in the number of vacua which have
strongly warped regions. Such throats are formed if the modulus $z$, which
controls the size of the 3-cycle at the tip of a conifold, is stabilized by
flux at an exponentially small value. Here, $z=0$ corresponds to a situation
with a vanishing 3-cycle and a singularity at the tip of the conifold. For a
Calabi-Yau orientifold with a single conifold singularity, the number of
vacua with $|z| < |z_*|$ was found to be \cite{Denef:2004ze}
\be
\label{number2}
\mathcal{N} \, \propto \, \frac{1}{\log (1/|z_*|)} \,.
\ee
Let us consider a more general Calabi-Yau orientifold. We denote the number
of 3-cycles by $K$. The moduli space will contain regions in which certain
3-cycles shrink to zero size. In these regions, a singularity develops in the
Calabi-Yau orientifold. In \cite{Hebecker:2006bn}, it was argued that often,
presumably even for an $\mathcal{O}(1)$ fraction, these singularities 
are conifold singularities. Let us choose coordinates on moduli space such
that these conifold singularities arise for $z_i=0$ with $i=1 \dots K$. It
was furthermore argued in \cite{Hebecker:2006bn} that the probability that
a randomly chosen vacuum is near $z_j=0$ for some $j$ will have the same
parametric dependence as Eq.~\eqref{number2}. If we assume that the
probabilities are uncorrelated, we can calculate the expected number of
3-cycles which are smaller than some value $|z_*|$. Using also the relation
between moduli coordinates $z$ and generated hierarchy $h$ ($h \sim
z^{-1/3}$ as we have found at the end of Section~\ref{GKP}), the
expected number of throats with a hierarchy larger than some $h_*$ follows
as \cite{Hebecker:2006bn}
\be
\label{expectation}
\bar{n}(h > h_*) \, = \, \frac{K}{3 c \log h_*} \,.
\ee
Here, $c$ is an unknown constant of order 1 which is related to the
normalization of the aforementioned probabilities.

\chapter{Energy transfer between throats}
\label{et5d}
\section{A motivation: Reheating after brane-antibrane inflation}
\label{motivation}
We will now discuss a motivation for the topics which will be analysed in
subsequent chapters. A different application of the results from these
chapters will be presented in Chapters~\ref{thermalproduction}
and~\ref{scenarios}.

Branes allow for an interesting realization of the inflationary scenario
\cite{Dvali:1998pa}. Two branes, which are initially separated in the extra
dimensions and which exert an attractive force on each other, will move
towards each other. From a 4d viewpoint, the relative distance
between these branes is a scalar field which rolls down a potential.
If this potential is sufficiently flat, the brane motion can lead to
slow-roll inflation. 

A string theory realization of this idea is provided by a D3-brane and an
anti-D3-brane which are aligned along the noncompact dimensions
\cite{Alexander:2001ks,braneantibraneinflation}.\footnote{Recall that two
D3-branes exert no force on each other, hence the anti-D3-brane.}
Unfortunately, for branes in flat extra dimensions, the potential is usually
too steep to achieve slow-roll inflation with enough e-foldings. The
potential is
much flatter if the branes instead move in a warped region
\cite{Kachru:2003sx}. More precisely, we consider a KS throat in a
flux compactification as in Section~\ref{GKP}. We place the anti-D3-brane at
the tip of the KS throat. It will approximately remain fixed at this
position due to the background fluxes.
As a result of the warping, the potential between the brane and the antibrane
is very flat. If the D3-brane starts far away from the tip, slow-roll
inflation with a sufficient number of e-foldings is thus
possible.

Inflation ends when the brane comes close to the antibrane. The subsequent
annihilation of the branes produces a large amount of KK modes which are
localized in the throat.\footnote{More precisely, an open string tachyon
develops at substringy distances \cite{Sen} whose condensation
produces a large amount of massive, closed string states
\cite{Lambert:2003zr}. The final phase of brane-antibrane inflation is thus
similar to the waterfall regime of hybrid inflation. The massive string states
in turn decay quickly to massless string states, respectively to KK modes of
the corresponding supergravity fields.} If the standard model is realized
on some D-branes in that same throat, the KK modes decay to standard 
model fields and thereby lead to the reheating of the visible sector.
There are reasons,
though, to prefer that the standard model and inflation are realized in
different throats. Most importantly, in order to reproduce the observed
amplitude of density fluctuations, the throat in which inflation takes
place should have an IR scale of the order of
$10^{14}\hspace{+.09cm}\text{GeV}$
\cite{Kachru:2003sx}. If the standard model lives in such a throat, a
solution of the hierarchy problem \`a la RS is not possible.

The KK modes in the inflation throat, however, have only very weak couplings
to the rest of the compact manifold. Namely, due to the warping, they
have to climb a potential well before they can reach the unwarped part or
other throats. It is therefore not immediately clear whether reheating will
be successful if the standard model is not realized in the inflation
throat. A detailed analysis in a series of papers
\cite{Barnaby:2004gg,Chialva:2005zy,Kofman:2005yz,Frey:2005jk,CT} has
shown that,
at least for a certain range of parameters, viable reheating is possible.

Obviously, a crucial quantity in this respect is the decay rate of KK modes
from one throat to another throat. The energy density of KK modes is
diluted by the expansion of the universe. The reheating temperature of the
standard model therefore depends on this decay rate. It was calculated in a
5d model in \cite{Dimopoulos:2001ui} and we will
review this calculation in the next section. In Chapter~\ref{energytransfer},
we will present a calculation of the decay rate in a 10d model.

\section{The tunneling calculation using a 5d model}
\label{decayrate5d}
\begin{figure}[t]
\begin{center}
\includegraphics[scale=0.35]{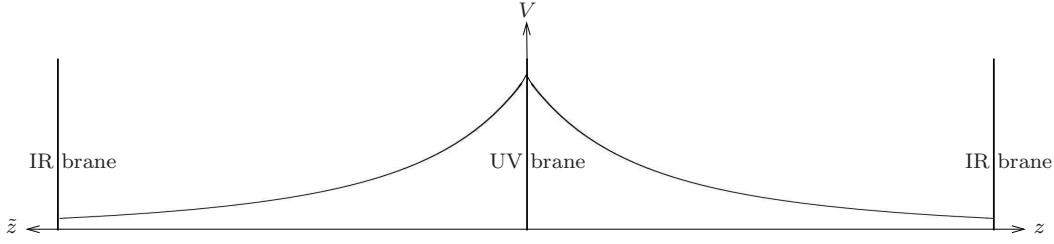}  
\put(-192,82){\scriptsize $V$} 
\put(-203,25){\scriptsize UV brane} 
\put(-377,25){\scriptsize IR brane}
\put(-23,25){\scriptsize IR brane}
\put(-386,0){\scriptsize $\tilde{z}$}
\put(3,0){\scriptsize $z$}
\caption{Two throats in a compact space approximated by two RSI
models which are glued together at a common UV brane. A KK mode
in one throat has to tunnel through an effective energy barrier
before it can decay to the other throat.\label{pot5d}}
\end{center}
\end{figure}

We consider a Calabi-Yau orientifold with two strongly warped regions. We do
not need to specify the precise form of these throats, but
we assume that they are finite and reasonably well approximated by a
slice of AdS$_5$ times some compact 5d manifold $\mathcal{M}$. The prime
example certainly is a KS throat. For simplicity, we also assume that
both throats have the same AdS scale $R$. The size $L$ of the embedding
manifold is larger than this AdS scale, $L \gtrsim R$, since otherwise
the throats could not be glued into the manifold. If the embedding manifold
is of minimal size, $L \sim R$, KK modes with masses $\smash{m_n \ll
R^{-1}}$ cannot resolve its precise geometry.  We can then describe the
embedding manifold by the UV brane in a RS model. A setup with two such
throats
can correspondingly be approximated by two RSI models which are glued
together at a common UV brane (cf.~Fig.~\ref{pot5d}) times the compact
manifold $\mathcal{M}$. The metric for this setup (ignoring
$\mathcal{M}$) is given by Eqs.~\eqref{ansatz} and \eqref{solution} but the
5th coordinate $y$ now runs from $y_1 < 0$ to $y_2 > 0$ and negative and
positive coordinate values are no longer identified. We want to determine
the transition rate of KK modes from one throat to the other throat. For
simplicity, we consider only KK modes of the 4d graviton and restrict
ourselves to s-waves with respect to the compact manifold $\mathcal{M}$. A
convenient parameterization for these spin-2 fluctuations is 
\be
\label{fluctuation2}
ds^2 \, = \, e^{-2 k|y|} \, \left( \eta_{\mu \nu} + h_{\mu \nu}(x,y)
\right) \, dx^\mu dx^\nu + dy^2 \,,
\ee
where $\smash{k=R^{-1}}$ is the inverse AdS radius.

The linearized equation of motion for the fluctuation $h_{\mu \nu}$ is a
Laplace equation \cite{Csaki:2000fc,Firouzjahi:2005qs}
\be
\label{LaplaceEq}
\partial_M \left( \sqrt{g} g^{MN} \partial_N h_{\mu \nu} \right) = 0 \,,
\ee 
where $g^{MN}$ is the background metric. We consider eigenstates of the 4d
d'Alembertian, 
\be
\label{sep}
h_{\mu \nu}(x,y) \, = \, \epsilon_{\mu \nu} \, e^{ipx} \, \psi(y) \,,
\ee
where $\epsilon_{\mu \nu}$ is a polarization tensor, $p^2=-m^2$ and $m$ is the
4d mass of the mode. We focus on the first throat (i.e.~$y<0$) for the moment
and introduce a new (dimensionless) coordinate $\smash{z \equiv k^{-1} e^{k
|y|}}$ and a rescaled field $\smash{\chi \equiv z^{-3/2} \psi}$ in this
throat. Using Eq.~\eqref{sep} in Eq.~\eqref{LaplaceEq}, the equation of
motion can be written as
\begin{equation}
\label{se2}
- \frac{d^2}{dz^2} \, \chi(z) +  \frac{15} {4 z ^2} \, \chi(z)
 = m^2 \, \chi(z) \,.
\end{equation}
This is a Schr\"odinger equation with a potential which is given by the
second term on the l.h.-side and which we have
plotted, together with the corresponding potential for the second throat,
in Fig.~\ref{pot5d}.\footnote{As expected, Eq.~\eqref{se2} is the same as
Eq.~\eqref{se1}
for $l=0$, up to the term proportional to $\smash{z^{-4}}$.} As one can see, a
light mode has to tunnel through an effective
energy barrier before it can reach the other throat. The mode has to fulfill
certain boundary conditions at the IR branes and the UV brane. Note, however,
that we do not take the boundary conditions on the IR
branes into account. It was shown in \cite{Chialva:2005zy} that, at least
for orbifold boundary conditions, the results from the following calculation
are not changed up to $\mathcal{O}(1)$ prefactors by the IR boundary
conditions. 

The
equation of motion can be solved in terms of Bessel functions. Going back
to the unrescaled field, we have
\be
\label{solution3}
\psi(z) \, = \, A \, (m z)^2 \, H^+_2 \hspace{-1mm} \left(m z
\right) + B \, (m z)^2 \, H^-_2 \hspace{-1mm} \left(m z \right) \,.
\ee
For later convenience, we have written the solution in terms of Hankel
functions
$H_2^\pm\hspace{+.05cm}=\hspace{+.05cm}J_2\hspace{+.05cm}\pm\hspace{+.05cm}
i\hspace{+.05cm}Y_2$, where $J_2$ and $Y_2$ are Bessel
functions. The constants $A$ and $B$ will be determined in a moment.

Next, we introduce a new coordinate $\tilde{z}$
for the second throat (i.e.~for $y>0$) and denote the unrescaled field and
the rescaled field in that throat by $\tilde{\psi}$ and $\tilde{\chi}$,
respectively. We have
\be
\label{solution4}
\tilde{\psi}(\tilde{z}) \, = \, \tilde{A} \, (m \tilde{ z})^2 \,
H^+_2 \hspace{-1mm} \left(m \tilde{z}\right) + \tilde{B} \, (m \tilde{z})^2 \,
H^-_2 \hspace{-1mm} \left(m \tilde{z}\right) \,.
\ee
We want to determine the transition (or tunneling) probability of a mode
from the first throat to the second throat. Accordingly, we
consider a wave which is incoming in the second throat. This corresponds to
$\tilde{A}=1$ and $\tilde{B}=0$. To see this, note first that the mass of
the tunneling KK
mode is quantized in units of $\smash{m_\IR = e^{-k \ell_1} k }$ according to
Eq.~\eqref{massesKK}, where $\ell_1$ is the length of the first throat.
We assume that the second throat is longer than that,
i.e. $\ell_2 > \ell_1$. Near the
IR brane of the second throat we then have $m \tilde{z} \gg 1$ and
can use the asymptotic forms of the Bessel functions for large arguments.
We find $\smash{\tilde{\psi}\propto \tilde{z}^{3/2} e^{i m
\tilde{z}}}$ which is an incoming wave as one can see from the
time-dependence in Eq.~\eqref{sep}. 

We
determine the constants $A$ and $B$ from the requirement that both solutions,
Eqs.~\eqref{solution3} and \eqref{solution4}, as well as their first
derivatives match smoothly at $z=\tilde{z} = R$, corresponding
to $y=0$.\footnote{Note that there is no jump in the first derivative at
$y=0$ if the fluctuation is parametrized as in Eq.~\eqref{fluctuation2}. This
follows from Eq.~\eqref{LaplaceEq}.} That is, we have to require that
$\smash{\psi(R)= \tilde{\psi}(R)}$ and $\smash{\psi'(R)= -
\tilde{\psi}'(R)}$.\footnote{The minus sign is due to the fact that both
coordinates, $z$ as well as $\tilde{z}$, grow in the directions away from the
UV brane.} Since $m R \ll 1$ (cf.~Eq.~\eqref{massesKK}), we can use
the asymptotic forms of the Bessel functions for small arguments:
\begin{gather}
\psi(z) \,  \sim \, ( A + B) \, (m z)^4 + i \, ( B - A ) \left(1 +
(m z)^2 \right) \,  +  \dots \\
\tilde{\psi}(\tilde{z}) \,  \sim \,  
(m \tilde{z})^4 - i \, \left(1 +
(m \tilde{z})^2 \right) \,  +  \dots
\end{gather}
The ellipsis represent higher order corrections and we have neglected
numerical prefactors. The tunneling probability $\mathcal{P}$ is given by
the ratio of amplitudes of the outgoing wave in the first throat and the
incoming wave in the second throat. Thus, we have to calculate $\mathcal{P}
\sim |\tilde{A}/B|^2 = |1/B|^2$. Matching and solving for $B$, we find
\cite{Dimopoulos:2001ui}
\be
\label{tp5d}
\mathcal{P} \, \sim \, \left( m  R \right)^4 \,.
\ee

A mode that is initially localized in the first throat is described by a
wave packet in that throat. This wave packet can be decomposed into two
sets of modes which move in the IR direction and in the UV direction,
respectively. If the barrier on the UV side were impenetrable, the modes
would be reflected
entirely. However, since a small fraction of
the incoming flux is able to penetrate the barrier, the wave leaks out of
the throat. A wave packet initially localized in the throat will thus
decohere. The incoming flux at the barrier $j_{\text{in}}$ and the
tunneling probability $\mathcal{P}$ determine the decay rate $\Gamma$:
\begin{equation}
\label{dr}
\Gamma \, = \, j_{\text{in}} \mathcal{P} \,.
\end{equation}

To determine $\Gamma$, we need solutions to Eq.~\eqref{se2} describing
waves which are reflected back and forth between the UV barrier and
the IR end of the throat. From these we can calculate the incoming
flux $j_{\text{in}}$. For $\smash{z \gg m^{-1}}$, we can neglect the last term
in the potential in Eq.~\eqref{se2}, keeping only the constant term. In this
limit, the solution
is simply given by plane waves:
\begin{equation}
\label{planewave}
\chi(z) \simeq A \, \cos m z + B \, \sin m z \,.
\end{equation}
The approximation is valid for $\smash{z_\IR \geq z  \gg m^{-1} \sim
z_\IR/n}$, where $\smash{z_\IR= k^{-1} e^{k \ell_1}}= m_\IR^{-1}$ is the
position of the IR brane and we have used Eq.~\eqref{massesKK}. If $n$ is
not too small, the mode is well approximated by a plane wave in a large
portion of the throat. Deviations from this form for $z \lesssim m^{-1} $
are due to reflection at and tunneling through the effective barrier.

To calculate $j_{\text{in}}$ from Eq.~\eqref{planewave}, we have to
determine the normalization of the solution in physical terms. As a
simplification, we consider a complex scalar and a plane wave moving
around an $S^1$ parametrized by $z \in [0,z_\IR)$. Going to the rest
frame with respect to momenta parallel to the brane and reinstating
time dependence, we have
\begin{equation}
\label{testwave}
\chi(z) = \mathcal{N} \, e^{i m (z + t)}
\end{equation}
for the plane wave moving towards the UV barrier. To determine the
normalization constant $\mathcal{N}$, we use the standard charge
density for a Klein-Gordon particle, $\smash{j^0=\Im(\chi^*
\partial_t \, \chi)}$. It has to be normalized according to
\begin{equation}
\label{normalization}
1 = \int_0^{z_\IR} dz \, j^0 \; \; \Rightarrow \; \; \mathcal{N} =
\frac{1} {\sqrt{m z_\IR}} \,.
\end{equation}
The flux is then given by $j_{\text{in}}=j^z = \Im(\chi^*
\partial_z \, \chi)$. Using the solution of
Eq.~\eqref{testwave} with the normalization of
Eq.~\eqref{normalization}, we find
\begin{equation}
\label{TunnelFlux}
j_{\text{in}} = \frac{1}{z_\IR}=m_\IR \,.
\end{equation}
Using this result and Eq.~\eqref{tp5d}, the decay rate of a graviton KK mode
follows from Eq.~\eqref{dr} as \cite{Dimopoulos:2001ui}
\begin{equation}
\label{Gamma5d}
\Gamma \sim m_\IR (m R)^4 \,.
\end{equation}

\section{Two other ways to derive the decay rate}
\label{OtherWays}
We will now derive the decay rate Eq.~\eqref{Gamma5d} of KK modes between
two throats in a different way. Similar to the last section, we assume that
the AdS scale of the first throat is of the same order as the size of the
embedding manifold, $R_1 \sim L$, and that this throat can well be
approximated by a RSI model times a compact manifold $\mathcal{M}$. We
restrict ourselves to KK modes of the graviton
and to an s-wave with respect to the compact manifold $\mathcal{M}$. 
Let us furthermore assume that the second throat has the geometry AdS$_5
\times$S$^5$. In this case, it can equally well be described by a stack
of D3-branes. Its AdS scale $R_2$ cannot be larger than the size $L$ of the
embedding manifold, and since we have assumed that $L \sim R_1$, we have
$R_1 \gtrsim R_2$. The corresponding number $N_2$ of D3-branes follows from
Eq.~\eqref{RN} as $N_2 \sim M_{10}^4 R_2^4$. Now, when viewed from the
first throat, the U$(N_2)$ gauge theory on the stack of $N_2$ D3-branes
resides on the UV brane. Therefore, the graviton KK modes in this throat
couple directly
to the energy-momentum tensor of the gauge theory.

The KK
expansion of the graviton in a RS model is reviewed in
Appendix \ref{KKmodes}. In particular, the coupling of the KK modes
$\smash{h_{\mu \nu}^{(n)}}$ to the energy-momentum tensor $\smash{T^{\mu
\nu}}$ on the UV brane is given in Eq.~\eqref{5daction2}:\footnote{The
usual orbifold boundary conditions were taken for the derivation of coupling
strengths and masses of graviton KK modes. It is not immediately clear
whether the same boundary conditions follow from a reduction to 5d of
a 10d geometry since the effective theory is defined on an interval
instead of an $\smash{S^1 / \mathbb{Z}_2}$ orbifold. However, one can
rederive
the RS model on an interval if one takes Gibbons-Hawking
terms~\cite{Gibbons:1976ue} at the IR and the UV brane into
account. Varying with respect to the metric yields a condition similar
to the Israel junction condition, to be evaluated only at one side of
the brane. Inserting the background metric, one finds the relation
between the cosmological constants on the brane and in the bulk as
well as the usual boundary conditions for the fluctuations (see
e.g.~\cite{Chamblin:1999ya} for a derivation of the Israel junction
condition using Gibbons-Hawking terms).}
\begin{equation}
\label{KKcoupling}
S \, \supset \, \frac{g_n}{M_{10}^4 R_1^3}  \int d^4 x  
\, h^{(n)}_{\mu \nu} \, T^{\mu \nu} \,.
\end{equation}
We have used Eq.~\eqref{PlanckScale} and the fact that $L \sim R_1$.
The coupling constants $g_n$ are given in Eq.~\eqref{gn2},
\begin{equation}
\label{gn}
g_n \sim \sqrt{m_n m_\IR} \, R_1 \,,
\end{equation}
where we have used $\smash{k= R_1^{-1}}$ as well as Eq.~\eqref{massesKK} for
the masses $m_n$ of the KK modes and the IR scale $m_\IR$.

Using
Eqs.~\eqref{KKcoupling} and \eqref{gn}, the decay of graviton KK modes into
the second
throat can be calculated as a decay into gauge fields.\footnote{There
are also decays into the fermions and scalars in the gauge
theory. However, the corresponding decay rates have the same order of
magnitude as the decay rate into gauge fields.} By the standard
formula, the decay rate of a graviton KK mode into one species of gauge fields
is
\begin{equation}
\Gamma \sim \frac{m_n^4 m_\IR}{M_{10}^8 R_1^4} \,.
\end{equation}
There are $N_2^2$ gauge fields in the adjoint representation of
U($N_2$). Summing and using Eq.~\eqref{RN}, the total
decay rate follows:
\begin{equation}
\label{Gamma5d2}
\Gamma \sim \frac{R_2^8}{R_1^4} \, m_n^4 \, m_\IR \,.
\end{equation}
For $R_1 \sim R_2$, Eq.~\eqref{Gamma5d2} gives the same result as
Eq.~\eqref{Gamma5d}, including the factor of $m_\IR$! Note that this decay
process is just the reverse of the energy loss by the heated UV brane
considered e.g. in~\cite{HMR,Langlois:2002ke}. 

There is yet another way to derive Eq.~\eqref{Gamma5d}, which is due
to Ref.~\cite{Langfelder}. To this end, one considers a KK expansion in the
geometry with two throats. As in the last section, one models this
geometry by two RS models with AdS scale $R$ which are glued together at a
common UV brane. As before, it is assumed that the second throat is much
longer than the first throat. It turns out \cite{Langfelder} that the
amplitudes of KK modes are sizeable in the first throat only around certain
resonance peaks. These peaks occur for the discrete set of
masses $m_n$ which would follow from a KK expansion in the first throat if
that throat would be taken as an isolated system. The width of the peaks is
\be
\label{spread}
\delta m_n \, \sim \, m_n^5 R^4 \,.
\ee

Now, consider a mode which is created in the first throat. This mode
can be viewed as a wave packet which is fully localized in that throat,
i.e.~whose
amplitude vanishes in the second throat. The spread in frequencies of this
wave packet is roughly given by the width $\delta m_n$ of the amplitude peaks.
Due to the different time evolution of the constituent modes, the wave packet
decoheres after a time $t_\dec$ and the amplitude no longer vanishes in the
second throat. This is the analogue of the KK mode decay that we have
discussed before. The decoherence time can be estimated as $\smash{t_\dec
\sim \delta m_n^{-1}}$. Thus, in this approach the decay rate follows as
\be
\label{Gamma5d3}
\Gamma \, \sim \, \delta m_n \, \sim \, m_n^5 R^4 \,.
\ee
For the light modes, $m_n \sim m_\IR$, this once again is the same result
as Eq.~\eqref{Gamma5d}.

\chapter{Heat transfer between throats from a 10d perspective}
\label{energytransfer}
\section{Preliminaries}
\label{INTROet}
\begin{figure}[t]
\begin{center}
\includegraphics[scale=0.5]{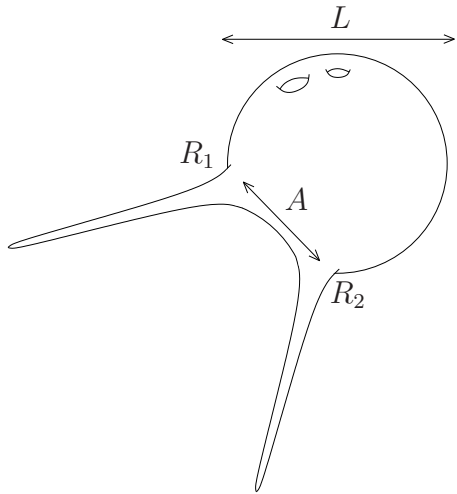} \put(-105,125){$R_1$}
\put(-48,72){$R_2$} \put(-65,107){$A$} \put(-48,176){$L$}
\caption{Two throats with AdS radii $R_1$ and $R_2$ separated at
distance $A$ inside a Calabi-Yau orientifold of total size
$L$.}\label{TwoThroats}
\end{center}
\end{figure}

For the calculations in the last chapter, we have assumed that the size $L$
of
the embedding manifold is of the same order of magnitude as the AdS scales $R$
of the throats. Of course, this is not always the case. In the following, we
will determine the rates of heat transfer for larger embedding manifolds.
In this case, a 5d model is no longer sufficient and we have to take the full
10d geometry into account. To this end, we consider two throats which
are separated at distance $A$ and which are embedded into a compact manifold
of size $L$ (cf.~Fig.~\ref{TwoThroats}).

We will perform the calculation for a simple example -- two semi-infinite
AdS$_5 \times$S$^5$ throats embedded in a 6-dimensional torus. This is
similar to the Verlinde compactification discussed in
Section~\ref{verlinde}. As it stands, though, our model is not a consistent
compactification since negative-charge objects are needed to fulfill the
tadpole cancellation condition. However, in the course of our calculation we
will argue that including these and other objects (e.g. further D-branes) as
well as using a different embedding manifold and a different throat geometry
only leads to $\mathcal{O}(1)$ corrections.

The calculation of the heat transfer rate between these two throats turns
out to be a multi-dimensional tunneling problem. Since such a problem is
difficult to solve, we choose a different approach: As we have discussed in
Section~\ref{branes}, AdS$_5 \times$S$^5$ throats are the near-horizon
geometries of black 3-branes which in turn correspond to stacks of
D3-branes.\footnote{Recall that the number $N$ of branes in each stack is
related to the S$^5$ radius $R$ of the corresponding throat by
Eq.~\eqref{RN}.} Due to this fact, instead of the aforementioned geometry with
throats, we can equally well consider a torus with two D3-brane stacks.

\begin{figure}[ht]
\begin{center}
\includegraphics[scale=0.5]{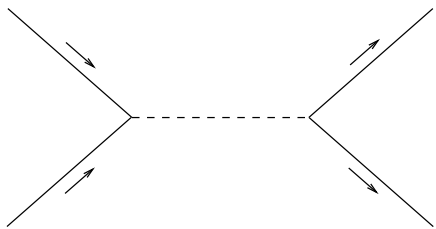}
\caption{Feynman diagram for the scattering of fields on one brane
stack into fields on another brane stack.}\label{scat}
\end{center}
\end{figure}

In this picture, the heat transfer from throat to throat is rephrased
as heat transfer between the two world-volume gauge theories. This reduces
the problem to the calculation of simple processes in quantum field
theory. To see this, consider a throat which is heated to a certain
temperature $T$. As we have discussed in Section~\ref{HeatedBranes}, such a
throat is the near-horizon geometry of a non-extremal black 3-brane which is
dual to a D3-brane stack with a heated world-volume gauge theory. The latter
is coupled to the world-volume theory on the second brane stack (corresponding
to the second throat) by the supergravity fields in the embedding space.
Heat transfer between the two throats then is, in this picture, due to
processes of the type shown in Fig.~\ref{scat}, where fields in the thermal
plasma on one brane stack scatter into fields on the other brane
stack.
 
We will now calculate this heat transfer rate from a heated throat to
another throat. An application of the result will be
presented in Section~\ref{et}. The decay rate of KK modes between two
throats, which
generalizes the results from the last chapter, will be determined in
Chapter~\ref{KKdecay}, using similar methods. We will restrict our
calculation to the mediation by the dilaton, the RR scalar and the graviton
polarized parallel to the branes. In the gravity picture these three fields
satisfy the same wave equation~\cite{Gubser:1997yh}. Correspondingly, in the
gauge theory picture their effect in mediating heat transfer is
parametrically the same.\footnote{This can also be inferred from the relevant
part of the DBI action, which couples them to the world-volume theories on the
D3-branes.  } Hence, we can further restrict our calculation to one of
the three fields, which we take to be the dilaton. In particular, we
will not consider the effect of fermions living in the embedding
manifold. In fact, in~\cite{Hosomichi:1998ei} the absorption cross
section of dilatinos by 3-branes was calculated and found to agree
with the result for the dilaton. Therefore, we expect the fermions to
give parametrically the same contribution as the fields that we
consider.

Recall from Section~\ref{branes} that the (low-energy) world-volume
theory on $N$ parallel D3-branes is $\smash{\mathcal{N}=4}$ U($N$) super
Yang-Mills. Its field content is given by the field strength $F_{\alpha
\beta}$ in the adjoint representation, six adjoint scalars $X^i$ corresponding
to the positions of the branes, and fermionic superpartners. The
relevant couplings between the dilaton $\phi$ and the world-volume theory 
are given in Eq.~\eqref{coupling2}:
\begin{equation}
\label{coupling3}
S \supset \frac{M^{-4}_{10}}{2^{3/2}}\left[ \int\! d^{4}x \, \phi(x,
\langle \vec{X} \rangle) \,  F_{\alpha \beta}^2 + \sum_{l=1}^{\infty} \int\!
d^{4}x \frac{M_{10}^{-2 l} }{ l!\pi^{l/4} } \left(\partial_{i_1}
\cdots \partial_{i_l} \phi \right)  X^{i_1} \cdots X^{i_l}
\, F_{\alpha \beta}^2 \right]\!.
\end{equation}
Here, $\smash{\langle \vec{X} \rangle}$ is the position of the brane stack. 
The dilaton $\phi$ and the $X^i$ are defined such that their kinetic
terms are canonically normalized. We ignore couplings to fermions, since they
are proportional to the fermionic equations of motion and thus give no
contributions to S-matrix elements~\cite{Gubser:1997yh}. Direct couplings
between the dilaton $\phi$ and the scalars $X^i$ are absent. Moreover, as can
be seen from Eq.~\eqref{coupling3}, couplings involving the $X^i$ as well as
$F_{\alpha \beta}$ are suppressed by extra factors
of $\smash{M_{10}^{-2 l}}$ and can therefore be ignored.

\section{Energy loss rate to flat 10d space}

Before we proceed, we should check whether a calculation in terms of
weakly coupled gauge fields is a good approximation in the strongly
coupled regime of the gauge theory. At zero temperature, this is
adequate due to the nonrenormalization theorem discussed at the end of
Section~\ref{3ba}. However, the gauge theory is at finite
temperature, which breaks supersymmetry. With supersymmetry being broken, this
nonrenormalization theorem cannot be expected to hold and it is not
immediately clear why to trust our calculation. Therefore, we analyse a simple
example in both the gauge  theory and the gravity picture and compare the
results. Namely, we consider a heated stack of D3-branes in flat 10d space
which is dual to a non-extremal black 3-brane and calculate the energy loss
rate in both pictures.

We model the heated, strongly-coupled gauge theory on the D3-brane
stack by a thermal plasma of free fields. In principle, one would have
to use finite temperature field theory for the calculation of the
energy loss rate. However, as we are only interested in the correct
order of magnitude, we can perform a zero-temperature calculation
using a thermal particle distribution in the initial state. Following
from Eq.~\eqref{coupling3}, the cross section for scattering of two
gauge bosons into one dilaton is 
\be
\label{csb}
\sigma \sim \frac{s^3}{M_{10}^8} 
\ee 
up to $\mathcal{O}(1)$
prefactors, where $\smash{\sqrt{s}}$ is the energy of the gauge bosons
in the center of mass frame. From Eq.~\eqref{csb}, we can calculate
the rate of energy loss per world-volume of the branes induced by this
scattering process. This is done by thermally averaging the product of
cross section and lost energy, in analogy to the standard calculations
of reaction rates in a hot plasma~\cite{swo,HMR}:
\begin{equation}
\label{rate}
\dot{\rho} = \frac{1}{2} \int d^3k_1 \, d^3k_2 \; f(\omega_1) \,
f(\omega_2) \, \sigma v \, (\omega_1 + \omega_2)\,.
\end{equation}
Here
\begin{equation}
f(\omega) = \frac{1}{4 \pi^3(e^{\omega/T} -1)}
\end{equation}
is the distribution function for the gauge bosons, $v$ is the relative
velocity of the colliding particles, and $T$ is the temperature of the
heated gauge theory. Inserting Eq.~\eqref{csb} into Eq.~\eqref{rate},
we get the energy loss rate due to scattering of one gauge boson
species. To get the total energy loss rate, we have to sum over all
species and polarizations.  In a U($N$) gauge theory there are $N^2$
gauge bosons. Thus, there is an extra factor of $2 N^2$ coming from
the summation. Using Eq.~\eqref{RN} and neglecting prefactors of order
one coming from the integration in Eq.~\eqref{rate}, we get 
\be
\label{sbel2}
\dot{\rho} \sim R^8 \, T^{13} \,, 
\ee 
where $R$ is the AdS scale of the corresponding black 3-brane.

Energy loss from the non-extremal black 3-brane is due to Hawking
radiation emitted by its black hole horizon. The corresponding rate
per brane world-volume $\dot{\rho}$ is given by a generalization of
the Hawking formula (see e.g.~\cite{Aharony:1999ti}). If we restrict
ourselves to the dilaton, we get 
\be 
\dot{\rho} = \int \frac{d^9k}{(2
\pi)^9} \, \frac{ v \, \omega \, \sigma_T(\omega)}{e^{\omega /T} -1} \,,
\ee 
where $v$ is the velocity of the emitted particles and $T$ is the
Hawking temperature of the horizon. The absorption cross section
of a dilaton by a non-extremal black 3-brane is given in
Eq.~\eqref{csNE}: $\sigma_T \sim \sigma_0 \,f(\omega/T)$, where
$\sigma_0 \sim \omega^3 R^8$ is the absorption cross section at zero
temperature and $f$ is some function. Using this result and performing the
integral, we have
\be
\label{sbel1}
\dot{\rho} \sim R^8 \, T^{13} \,.  
\ee 
Here we have neglected prefactors of order one which come in particular from
the integration over $f(\omega/T)$.

Both results for the energy loss rate, Eqs.~\eqref{sbel2} and
\eqref{sbel1}, agree up to $\mathcal{O}(1)$ factors. Accordingly, a
weak-coupling calculation in the gauge theory picture gives the right
order of magnitude.  The crucial ingredient is the fact that the
absorption cross section $\sigma_T$ of a dilaton by a non-extremal
black 3-brane differs from the zero-temperature absorption cross
section $\sigma_0$ only by a function of $\lambda \equiv \omega/T$. By
gauge/gravity duality, this means that the gauge boson-dilaton vertex
is corrected by a function of $\lambda$ at non-zero
temperature.\footnote{This is also the case if one takes
finite-temperature effects properly into account on the gauge theory
side.} Accordingly, the cross section for the process in
Fig.~\ref{scat} that we will calculate assuming weak coupling and zero
temperature has to be corrected by a function of $\lambda$. However,
inserting the corrected cross section into Eq.~\eqref{rate} and
performing the integral will just give a different ${\cal O}(1)$
prefactor, which we ignore anyway.

\section{Heat transfer rate to a different throat}
\label{etdt}
Let us now calculate the cross section for the process in
Fig.~\ref{scat}.  To this end, we need the KK expansion of the dilaton
in a 6d torus,
\begin{equation}
\label{expansion}
\phi(x,\langle \vec{X} \rangle) = \sum_{\vec{n} \in \mathbb{Z}^6}
\frac{1} {L^3} \, e^{2 \pi i \vec{n} \langle \vec{X} \rangle /L} \,
\Phi_{\vec{n}}(x) \,,
\end{equation}
where $L$ is the size of the torus and the expression is already
evaluated at the position $\smash{\langle \vec{X} \rangle}$ of one
brane stack. The mass of the $\vec{n}$th KK mode is
$\smash{m_{\vec{n}}=2 \pi |\vec{n}|/L}$.  Inserting
Eq.~\eqref{expansion} into Eq.~\eqref{coupling3}, one sees that the vertex
for the $\vec{n}$th KK mode in Fig.~\ref{scat} is
\begin{equation}
\label{vertex}
\sim\frac{s}{M_{10}^4 \, L^3} \, e^{2 \pi i \vec{n} \langle \vec{X}
\rangle /L} \,.
\end{equation}
Here the energy in the center of mass frame of the gauge bosons is
denoted by $\smash{\sqrt{s}}$. Let $\smash{\langle \vec{X_1} \rangle
}$ and $\smash{\langle \vec{X_2} \rangle }$ be the positions of the
two brane stacks inside the $T^6$.  If we denote the relative distance
of the stacks by $\smash{\vec{A} \equiv \langle \vec{X_2} \rangle -
\langle \vec{X_1} \rangle }$ and introduce the shorthand $\vec{a}
\equiv 2 \pi \vec{A} /L$, the matrix element corresponding to the
process in Fig.~\ref{scat} is given by
\begin{equation}
\label{Matrixelement}
\mathcal{M} \, \sim \, \frac{s^2}{ M_{10}^8 \, L^6} \, \sum_{\vec{n}
\in \mathbb{Z}^6 } \, \frac{ e^{ i \vec{n} \vec{a}}}{s -m_{\vec{n}}^2
+i \epsilon} \,.
\end{equation}
We have ignored prefactors of order one. For phenomenological
purposes, we can safely assume $\smash{\sqrt{s} < L^{-1}}$. Namely,
since the energy $\smash{\sqrt{s}}$ of the colliding gauge bosons is
determined by the temperature $T$ of the heated gauge theory, this
corresponds to $\smash{T < L^{-1}}$. If this were not the case, the
gauge theory would heat up the compact manifold and the geometrical
picture would be lost.  Following from $\smash{\sqrt{s} < L^{-1}}$,
one has $s < m_n^2$ for $n >0$ and the contribution of the energy
$\smash{\sqrt{s}}$ in the propagator can be neglected for all but the
zero mode. Thus, Eq.~\eqref{Matrixelement} simplifies to
\begin{equation}
\label{Matrixelement2}
\mathcal{M} \, \sim \, \frac{s^2}{ M_{10}^8 \, L^4} \, \sideset{}{'}
\sum_{\vec{n} \in \mathbb{Z}^6 } \, \frac{ e^{ i \vec{n} \vec{a}}}
{\vec{n}^2} + \frac{s}{ M_{10}^8 \, L^6} \,,
\end{equation}
where the prime denotes exclusion of $\smash{\vec{n} = \vec{0}}$ in
the sum. Since the 4d Planck scale is determined by $M_4^2\simeq
M_{10}^8 L^6$, the last term in Eq.~\eqref{Matrixelement2} simply
reflects the fact that the zero mode interacts with gravitational
strength. The sum, which would be UV divergent in absence of the
exponential factor, is dominated by terms with large
$\smash{\vec{n}}$.  It can therefore be approximated by an integral:
\begin{equation}
\label{sum}
\int d^6 n \, \frac{ e^{ i \vec{n} \vec{a}}}{\vec{n}^2} \sim
\frac{1}{a^4} \,.
\end{equation}
The r.h. side of Eq.~\eqref{sum} results from the fact that the
exponential function oscillates quickly for $|\smash{\vec{n}| \gtrsim
a^{-1} }$ ($\smash{a \equiv |\vec{a}|}$), effectively cutting off the
integral.\footnote{ One can see in particular that the sum in
Eq.~\eqref{Matrixelement2} is effectively cut off before the geometry
of the throats becomes relevant, justifying our flat-space
approximation.  } More precisely, we evaluate a similar but more
general integral, which we will need in
Section~\ref{decayingaugepicture2}, in Appendix~\ref{integral}.
Equation~\eqref{sum} follows from this integral in a particular limit,
which is displayed in Eq.~\eqref{lc}.

Inserting Eq.~\eqref{sum} into Eq.~\eqref{Matrixelement2}, we find
\begin{equation}
\label{Matrixelement3}
\mathcal{M} \, \sim \, \frac{s^2}{ M_{10}^8 \, A^4} + \frac{s}{
M_{10}^8 \, L^6}\,,
\end{equation}
where $\smash{A \equiv |\vec{A}|}$. For an order-of-magnitude
calculation, we can neglect the interference term in
$\smash{|\mathcal{M}|^2}$. The cross section for the process in
Fig.~\ref{scat} then reads
\begin{equation}
\label{sigma}
\sigma \, \sim \, \frac{s^3}{M_{10}^{16} A^8} + \frac{s}{M_{10}^{16}
L^{12}} \qquad\text{for}\qquad\sqrt{s} < L^{-1}\,.
\end{equation}
Inserting this cross section into Eq.~\eqref{rate}, we get the energy
loss rate due to scattering of one particle species into another
particle species.  To get the total energy loss rate, we have to sum
over all initial and final state species and polarizations. Let us
denote with $N_1$ and $N_2$ the number of colors of the heated gauge
theory and the gauge theory that is being heated, respectively. The
summation then gives extra factors of $2N_1^2$ and $2N_2^2$ and we
get, again neglecting prefactors of order one coming from the
integration in Eq.~\eqref{rate},
\begin{equation}
\label{etNN}
\dot{\rho} \sim \frac{N_1^2 N_2^2}{M_{10}^{16} A^8} \, T^{13} +
\frac{N_1^2 N_2^2}{M_{10}^{16} L^{12}} \, T^9 \,.
\end{equation}
Using Eq.~\eqref{RN}, this can be written in a slightly more compact
form. Denoting by $R_1$ and $R_2$ the AdS scales of the corresponding
throats, we arrive at the main result of this chapter:
\begin{equation}
\label{el}
\dot{\rho} \sim \frac{R_1^8 R_2^8}{A^8} \, T^{13} + \frac{R_1^8 R_2^8}
{L^{12}} \, T^9 \,.
\end{equation}
The applicability of this heat transfer rate to more general throat
geometries and embedding manifolds will be discussed in
Section~\ref{applicability}.

\chapter{Decay of KK modes between throats from a 10d
perspective}\label{KKdecay}
\section{The glueball decay vertex}\label{decayingaugepicture}

We will now calculate the decay rate of KK modes
which localized in one throat to a different throat. We will thus determine
the same quantity as in Chapter~\ref{et5d}. This time, though, we
will perform the calculation for the 10d
setup from the last chapter, two AdS$_5 \times$S$^5$ throats in a
6-dimensional torus. Again, we will use the dual gauge
theory picture and consider D3-brane stacks instead of the AdS$_5 \times$S$^5$
throats. 

\begin{figure}[t]
\begin{center}
\vspace{10mm} \includegraphics[scale=0.6]{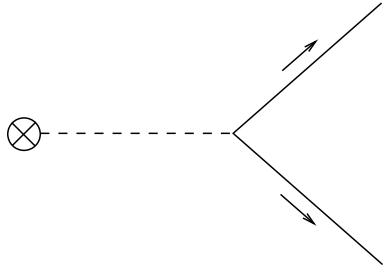}
\vspace{6.5mm}
\caption{Feynman diagram for the decay of a glueball in one throat
into fields in another throat.\label{dec}}
\end{center}
\end{figure}

In this picture, we have to calculate the decay rate of glueballs on
one brane stack into two gauge fields on another brane stack. The Feynman
diagram for this process is shown in Fig.~\ref{dec}. Due to the
nonrenormalization theorem described in the introduction, we do not have to
care whether the decay products will arrange into one or more glueballs. The
vertex for this part of the diagram is simply the one already derived in
Eq.~\eqref{vertex}. However, the other vertex between a dilaton and a
glueball can not so easily be read off from the Lagrangian. Therefore,
we first calculate the decay rate for a simpler situation in the gravity
picture. From this we determine the vertex by demanding that this decay rate
agree with the gauge theory picture.

Namely, we consider a single AdS$_5 \times$S$^5$ throat which is embedded into
flat 10d space. As we have discussed in Section~\ref{branes}, this is the
geometry of an extremal black 3-brane. Excitations in the throat region of
this 3-brane correspond to excitations in the dual gauge theory. The state
dual to a glueball on a D-brane stack is therefore a wave packet which
is
localized in the throat. Due to the different time evolution of its
constituent modes, this wave packet will decohere after a certain time
(cf.~Section~\ref{OtherWays}). Hence, excitations will show up in the
asymptotically flat region as well. This is the analogue of the decay of a
glueball on a D-brane stack to supergravity fields in the embedding
flat space.\footnote{As opposed to the `derivation' of the AdS/CFT
conjecture in Section~\ref{AdS/CFT}, we keep $M_s$ and $M_{10}$ finite. In
this
case, the asymptotically flat region does not decouple from the throat region
of a black 3-brane. Excitations in the two regions will thus mix with each
other. This reflects the fact that, since we do not send $M_{10}$ to infinity,
the gauge theory will interact with the supergravity fields in the embedding
space (cf.~Eq.~\eqref{coupling2}).}

Thus, we want to determine the decay rate of a dilaton which is localized
in the
throat into flat space. To this end, we assume the throat to be sharply
cut off somewhere in the IR. Such an AdS$_5 \times$S$^5$ throat with an IR
cutoff might not exist as a solution to supergravity, but it can serve as a
simple toy model capturing the relevant information. In
Chapter~\ref{modifications}, we will show how
to extend our results to realistic finite throats such as the
KS throat. On the gauge theory side,
the cutoff corresponds to a deformation by a relevant operator, in which case
the gauge theory has a discrete set of glueball states.

The tunneling probability $\mathcal{P}$ of a wave from the throat to
the asymptotically flat space is given in Eq.~\eqref{tp} (see also \cite{CT}).
Although this result has been derived for a throat which is infinite
in the IR direction, we can still use it for a finite throat, as long
as the mass $m$ of the wave is not too small. To see this, consider the
dilaton equation of motion in Schr\"odinger form, Eq.~\eqref{se1} with
$\omega = m$. In the
region $\smash{z \gg m^{-1}}$, the constant term in the potential is
dominant and the equation is approximately solved by plane waves:
\be
\label{pw}
\psi(z) \, \simeq \, A \cos m z+ B \sin m z \,.
\ee
We denote the position of the IR cutoff of the throat by $z_\IR$. The
boundary condition
for $\psi$ or its first derivative at $z_\IR$ lead to a quantization of the
mass in units of $m_\IR \equiv \smash{z_\IR^{-1}}$ such that $m \sim n \,
m_\IR$ with $n$ an integer. 

The approximation by a plane wave is valid in the region $\smash{z_\IR \geq
z \gg m^{-1} \sim z_\IR/ n}$, where we have used the mass quantization
condition.
As long as $n$ is sufficiently larger than 1,
the wave is a plain wave (in the parametrization of Eq.~\eqref{se1})
in a large portion of the throat. This plain wave tunnels to the
asymptotically flat space with the probability given in
Eq.~\eqref{tp}. Moreover, the analysis at the end of
Section~\ref{decayrate5d}
applies to this situation as well. The
relation of the tunneling probability to the decay rate is thus given by
Eqs.~\eqref{dr} and \eqref{TunnelFlux} and the decay rate follows as
\begin{equation}
\label{Gamma}
\Gamma \sim m_\IR (m R)^{8+4 l} \,.
\end{equation}

Now that we have the decay rate in the gravity picture, we need to
define a vertex $V$ in the gauge theory picture which reproduces this
result. We model the coupling by a term 
\be 
{\cal L_{\rm
10d}}\,\,\supset\,\, V\,\delta^{(6)}(\vec{X}-\langle\vec{X}
\rangle)\,\phi(x,\langle\vec{X}\rangle)\,{\cal G}(x) 
\ee 
in the 10d
Lagrangian, where $\mathcal{G}$ denotes the glueball state with
canonically normalized 4d kinetic term. Compactifying the 6 dimensions
perpendicular to the brane on a torus of size $L$ for the moment and
using the KK mode decomposition of Eq.~\eqref{expansion}, we get the
effective 4d Lagrangian
\begin{equation}
\label{effectiveL}
\mathcal{L}_{\rm 4d} \, \supset \, \sum_{\vec{n} \in \mathbb{Z}^6}
\left( - \frac{1}{2}\partial_{\mu} \Phi_{\vec{n}} \, \partial^{\mu}
\Phi_{\vec{n}} - \frac{1}{2} m_{\vec{n}}^2 \, \Phi_{\vec{n}}^2 + e^{2
\pi i \vec{n} \langle \vec{X} \rangle /L} \,\frac{V}{L^3} \,
\Phi_{\vec{n}}(x) \, \mathcal{G}(x) \right)\,.
\end{equation}
From this, the total decay rate of a glueball into KK modes of the
dilaton follows:
\begin{equation}
\Gamma = \frac{1}{2 \omega_i} \frac{1}{L^6} \sum_{\vec{n} \in
\mathbb{Z}^6} \int \frac{d^3 p_f}{(2 \pi )^3} \,\frac{1}{2 \omega_f}
\, (2 \pi)^4 \, \delta^{(4)}(p_f -p_i) \, |V|^2 \,.
\end{equation}
In this formula, $p_f=p_{f_\parallel}$ is a 4-vector characterizing
the momentum of the final-state dilaton parallel to the brane, while
$\omega_i$ and $\omega_f$ are the energies of the initial and final
state.  The 4-momentum of the decaying glueball is denoted by
$p_i$. Introducing the dilaton momentum in the compact dimensions as
$\smash{\vec{p}_{f_\perp} = 2 \pi \,\vec{n} /L}$, we can replace the
sum by an integral when we go back to $L \rightarrow \infty$:
\begin{equation}
\frac{1}{L^6} \, \sum_{\vec{n} \in \mathbb{Z}^6} \; \longrightarrow \;
\int \frac{d^6p_{f_\perp}}{(2 \pi)^6}\,.
\end{equation}
The decay rate of a glueball into a dilaton is then given by
\begin{equation}
\label{dr2}
\Gamma = \frac{1}{2 \omega_i} \int \frac{d^6 p_{f_\perp}}{(2 \pi )^6}
\frac{d^3 p_{f_\parallel}}{(2 \pi )^3} \,\frac{1}{2 \omega_f} \, (2
\pi)^4 \, \delta^{(4)}(p_{f_\parallel }-p_i) \, |V|^2\,.
\end{equation}
Since the dilaton is massless, $\smash{\omega_f=
\sqrt{|\vec{p}_{f_\perp}|^2+ |\vec{p}_{f_\parallel}|^2}}$. Going to
the rest frame of the glueball, $\smash{\vec{p_i}} = 0$, and
performing the momentum integrations, we arrive at
\begin{equation}
\Gamma = \frac{1}{2 \omega_i} \int \frac{d^6 p_{f_\perp}}{(2 \pi )^6}
\,\frac{1}{2 \omega_f} \, (2 \pi) \, \delta(\omega_f - \omega_i) \,
|V|^2 \sim \omega_i^3 \, |V|^2\,,
\end{equation}
where we have used $\smash{\omega_f=|\vec{p}_{f_\perp}|}$ and
neglected prefactors of order one. In its rest frame, $\omega_i$ is
simply the mass $m$ of the glueball. Comparing with Eq.~\eqref{Gamma},
we get
\begin{equation}
\label{vertex2}
V \sim \sqrt{m_\IR m}\,\,\,m^{2+ 2l}R^{4+2 l} \,.
\end{equation}

\section{Decay rate calculation in the gauge theory picture}
\label{decayingaugepicture2}
With the effective vertex $V$ at hand, calculating the decay rate of
one glueball into gauge fields living on a different brane stack is
straightforward. Following from Eqs.~\eqref{effectiveL} and
\eqref{vertex2}, the vertex between a glueball and a KK mode of the
dilaton is
\begin{equation}
\label{vertex4}
\frac{V}{L^3} \, e^{2 \pi i \vec{n} \langle \vec{X} \rangle /L} \,.
\end{equation}
The other vertex in the diagram is still given by
Eq.~\eqref{vertex}. Summing over all intermediate KK modes, we arrive
at an expression very similar to Eq.~\eqref{Matrixelement}:
\begin{equation}
\label{Matrixelement5}
\mathcal{M} \, \sim \, \frac{\sqrt{m_\IR m} \, ( m R )^{4+ 2 l}}{
M_{10}^4 \, L^6} \, \sum_{\vec{n} \in \mathbb{Z}^6 } \, \frac{ e^{ i
\vec{n} \vec{a}}}{m^2 -m_{\vec{n}}^2 +i \epsilon} \,.
\end{equation}
Compared to Eq.~\eqref{Matrixelement}, the only difference is the
prefactor and the substitution of the energy $\smash{\sqrt{s}}$ of the
colliding gauge bosons by the mass $m$ of the glueball.

We will analyse Eq.~\eqref{Matrixelement5} in two different regimes,
namely for $m^{-1}>L$ and for $m^{-1}\ll~L$. The former case is the most
interesting one from a phenomenological viewpoint. As we have argued in
Chapter~\ref{energytransfer}, we can assume that the reheating
temperature $T_{\text{RH}}$ in early cosmology is smaller than
$L^{-1}$. Accordingly, the mass $m$ of any relic KK modes is also
restricted by $m<L^{-1}$. The latter case, on the other hand, can be
easily analysed in the gravity picture as well. We will perform this
cross-check in Section~\ref{comparison}.

For $m^{-1}>L$, we can make the same simplifications as in
Eq.~\eqref{Matrixelement2} and use Eq.~\eqref{sum} for the sum. The
decay rate of a glueball into a pair of gauge bosons follows from the
standard 4d formula: 
\be 
\Gamma\sim m^{-1}|{\cal M}|^2\,.\label{4dg}
\ee 
To get the total decay rate, we have to sum over the $N^2$ final
state gauge bosons. If we denote by $R_1$ and $R_2$ the AdS scale of
the throat containing the initial and the final state, respectively,
and use Eq.~\eqref{RN}, we find
\begin{multline}
\label{DecayRate}
\Gamma \quad \sim \quad \frac{N_1^{2+l} N_2^2}{M_{10}^{16+4l} A^8} \, m_\IR
\,
m^{8+4 l}+
\frac{N_1^{2+l} N_2^2}{M_{10}^{16+4l} L^{12}} \,  m_\IR \, m^{4 + 4 l} \\
   \sim \quad  \frac{R_1^{8+4 l} R_2^8}{ A^8} \, m_\IR \, m^{8+4 l}+
\frac{R_1^{8+4 l} R_2^8}{L^{12}} \, m_\IR \, m^{4 + 4 l} \,.
\end{multline}
The applicability of this decay rate to more general throat geometries and
embedding manifolds will be discussed in Chapter~\ref{modifications}.

Let us now consider the case $m^{-1}\ll L$. We will also assume $A\ll
L$ for simplicity. Recalling that $\smash{m_{\vec{n}}=2 \pi
|\vec{n}|/L}$ and $\vec{a} = 2 \pi \vec{A} /L$, we can approximate the
sum in Eq.~\eqref{Matrixelement5} by an integral,
\begin{equation}
\label{propagator}
\frac{1}{L^6} \, \sum_{\vec{n} \in \mathbb{Z}^6 } \, \frac{ e^{2 \pi i
\vec{A} \, \vec{n}/L}}{m^2 -(2 \pi)^2 \vec{n}^2/L^2 +i \epsilon}
\;\;\; \longrightarrow \;\;\; \int \frac{d^6 \rho}{(2 \pi)^6} \,
\frac{ e^{i \vec{A} \, \vec{\rho}}}{m^2 -\vec{\rho}^2 +i \epsilon} \,,
\end{equation}
where $\vec{\rho} \equiv 2 \pi \,\vec{n} /L$. The resulting expression
is just the propagator of a massless particle in a mixed,
energy-configuration-space representation, with the `energy' $m$
characterizing the invariant 4-momentum. This is of course expected in
the large $L$ limit, where the torus goes over to flat space and the
infinite KK tower is replaced by the underlying higher-dimensional
dilaton field. The integral is evaluated in Appendix~\ref{integral}, the
outcome
being
\begin{equation}
\label{HankelFunktion}
\int \frac{d^6 \rho}{(2 \pi)^6} \, \frac{ e^{ i \vec{A} \,
\vec{\rho}}}{m^2 -\vec{\rho}^2 +i \epsilon} \sim \frac{m^2}{A^2} \,
H_2^+( m A) \,,
\end{equation}
where $H_2^+(x)=J_2(x)+i\, Y_2(x)$ is a Hankel function and we have
neglected prefactors of order one. Using the asymptotic forms of the
Bessel functions for large and small arguments,
Eq.~\eqref{HankelFunktion} can be simplified as follows:
\begin{equation}
\label{lc}
\frac{m^2}{A^2} \, H_2^+( m A) \sim
\begin{cases} 
\frac{m^{3/2}}{A^{5/2}} \, e^{i \, m A} & \text{for } m^{-1}\ll A \\
\frac{1}{A^4} & \text{for } m^{-1}\gg A\,.
\end{cases}
\end{equation}
Inserting these results in Eq.~\eqref{Matrixelement5}, we get the
matrix elements $\mathcal{M}$ for these two cases. The corresponding
partial decay rates follow from Eq.~\eqref{4dg}. Summing over all
final state species, we find
\begin{equation}
\label{DecayRate2}
\Gamma \sim
\begin{cases} 
\frac{R_1^{8+4 l} R_2^8}{A^{5}} \, m_\IR \, m^{11+4 l} & \text{for }
m^{-1}\ll A \\ \frac{R_1^{8+4 l} R_2^8}{A^8} \, m_\IR \, m^{8+4 l} &
\text{for } m^{-1}\gg A
\end{cases}\,.
\end{equation}
As a consistency check, we should examine, whether the appropriate limiting
cases of Eqs.~\eqref{DecayRate} and~\eqref{DecayRate2} coincide. The regions
of validity of the two calculations have a common border for $A\ll
m^{-1}\sim L$. Indeed, for this choice of parameters the first term in
Eq.~\eqref{DecayRate} dominates and the result agrees with the second
line of Eq.~\eqref{DecayRate2}.

\section{Some calculations in the gravity picture and relation to earlier
work}
\label{comparison}
As in the sections before, we consider two AdS$_5 \times$S$^5$ throats
embedded in a 6-dimensional torus of uniform size $L$. The geometry is
that of a multi-centered black 3-brane, the metric being
(cf.~Section~\ref{verlinde})
\begin{equation}
\label{multicenter}
ds^2 = H^{-1/2} \,  \eta_{\mu \nu} dx_\shortparallel^\mu
dx_\shortparallel^\nu  + H^{1/2}
\,  dx_\perp^{\textcolor{white}{.} i} dx_{\perp
 \textcolor{white}{.} i}^{\textcolor{white}{0}} 
\end{equation}
with
\begin{equation}
H(\vec{x}_{\perp}) = 1 + \sum_{\vec{n} \in \mathbb{Z}^6 } \left(
\frac{R_1^4}{|\vec{x}_{\perp} -\vec{A}_1 + \vec{n} L|^4} +
\frac{R_2^4}{|\vec{x}_{\perp} -\vec{A}_2 + \vec{n} L|^4} \right)\,.
\end{equation}
The positions of the two throats are denoted by $\vec{A}_1$ and
$\vec{A}_2$, their AdS scales by $R_1$ and $R_2$. The
$x_\shortparallel$ are coordinates along the 4 uncompactified
dimensions and the $\smash{x_\perp}$ refer to coordinates in
the torus. The sum in the warp factor $\smash{H(\vec{x}_{\perp})}$ is
due to mirror effects in the torus. Again, this is not a consistent
compactification. Including O-planes, for example, would give extra
contributions to the warp factor (see~\cite{Verlinde:1999fy}). We try to
calculate the transition of a dilaton between different throat
regions, which is the gravity counterpart to the gauge theory
calculation in Sections~\ref{decayingaugepicture}
and~\ref{decayingaugepicture2}. The equation of motion for the dilaton
is given in Eq.~\eqref{LE}. Inserting Eq.~\eqref{multicenter} in
Eq.~\eqref{LE} and using $\smash{\sqrt{g}=H^{1/2}}$, one gets
\begin{equation}
\partial_n \, \partial^n \, \phi + H(\vec{x}_{\perp}) \,
\partial_{\mu} \, \partial^{\mu} \, \phi = 0 \,.
\end{equation}
The indices $\mu$ and $n$ run from 0 to 3 and from 4 to 9,
respectively. Using the 4d Klein-Gordon equation, one arrives at
\begin{equation}
\label{se3}
\partial_n \, \partial^n \, \phi + H(\vec{x}_{\perp}) \, \, m^2 \,
\phi = 0 \,,
\end{equation}
where $m$ is the kinetic energy perpendicular to the branes. Like
Eq.~\eqref{se1}, this has the form of a Schr\"odinger
equation. Contrary to Eq.~\eqref{se1}, however, there is no potential
barrier separating the throat region and asymptotically flat space,
since the potential $\smash{V = - m^2 H(\vec{x}_{\perp})}$ is strictly
negative. The difference comes from using cartesian coordinates
perpendicular to the branes in Eq.~\eqref{multicenter} rather than
spherical coordinates in Eq.~\eqref{threebrane}. Still, a wave in the
throat region, moving away from the horizon, is reflected to a large
part before entering asymptotically flat space. In cartesian
coordinates, however, this is due to the steepness of the potential
well.

To determine the transition probability $\mathcal{P}$ of a dilaton
between two throat regions, one has to solve Eq.~\eqref{se3} with
appropriate boundary conditions. Then $\mathcal{P}$ is the ratio of
incoming flux in one throat and outgoing flux in the other throat. In
general, the corresponding calculation is difficult. However, if the
torus is very large ($L \rightarrow \infty$) and the throats are
sufficiently far apart ($A \gg \smash{m^{-1}}$), the problem
effectively splits into two simpler calculations. Namely, the latter
condition means that the de Broglie wavelength of the particle is
small compared to the distance of the throats. A transition between
two throats can then be described as a two-step process. For
simplicity, we take the initial state in the first throat to be an
s-wave. Only a small fraction of the outgoing flux reaches the
asymptotically flat region, the probability being given in Eq.~\eqref{tp}
(with $l=0$ and $\omega = m$)
\begin{equation}
\mathcal{P}_1 \sim \left(m R_1 \right)^8 \,.
\end{equation}
In between the two throats, one has a free spherical wave,
approximating a plane wave near the second throat. The absorption
cross section (per brane world-volume) for such a plane wave is given in
Eq.~\eqref{cs1}:
\begin{equation}
\sigma_2 \sim m^3 R_2^8 \,.
\end{equation}
Near the second throat, the incoming flux will be diluted by a factor
of $\smash{A^{-5}}$, since the free spherical wave is expanding in
6-dimensional flat space. The absorption probability by the second
throat thus is
\begin{equation}
\mathcal{P}_2 \sim \frac{\sigma_2}{A^5} \sim \frac{m^3 R_2^8}{A^5} \,.
\end{equation}
The transition probability between the two throats is just the product
$\mathcal{P}_1 \mathcal{P}_2$. If we denote by $m_\IR$ the mass gap in
the first throat, using Eqs.~\eqref{dr} and~\eqref{TunnelFlux} the decay
rate from the gravity calculation follows as
\begin{equation}
\Gamma \sim \frac{R_1^8 R_2^8}{A^5} \, m^{11} m_\IR \,.
\end{equation}
This is precisely what we found in Eq.~\eqref{DecayRate2} for $A \gg
m^{-1}$ and $l=0$.  The crucial ingredient is the $A^{-5}$
dependence. That it agrees in both calculations is, however, not too
surprising. In the gauge theory calculation, it came from the
propagator in a mixed energy-configuration-space representation
(cf. Eq.~\eqref{propagator}). The same is of course true in the above
gravity calculation, although we have not stated it explicitly.

As we have discussed in Section~\ref{OtherWays}, there is yet another
situation in which the decay rate of KK modes between two throats is
comparatively easy to
obtain. To this end, we assume that the AdS scale $R_1$ of the first
throat is of the same order as the size $L$ of the embedding
space, i.e.~$L \sim R_1$. We then have $R_1 \gtrsim R_2$, where $R_2$
is the AdS scale of the second throat. In this situation, we can approximate
the first throat by a
RS model, whereas the second throat corresponds to some degrees of freedom
which live on the UV brane of this RS model. Furthermore, we restrict
ourselves
to s-waves in the throats, i.e.~$l = 0$. The resulting decay rate
Eq.~\eqref{Gamma5d2} can be compared
with Eq.~\eqref{DecayRate} from the gauge theory calculation. The
distance $A$ between the two throats cannot be smaller than their AdS
scales $R_1$ and $R_2$. Since we also have $L \sim R_1$ and
$\smash{m \ll R_1^{-1}}$, the second term in
Eq.~\eqref{DecayRate} is
dominant. For $L \sim R_1$ and $l=0$, this term gives the
same result as Eq.~\eqref{Gamma5d2}!

Tunneling in a compact 10d setup with throats was also considered in~\cite{
Firouzjahi:2005qs,CT}. For the case $m^{-1} > L$, a decay rate of
$\smash{\Gamma \sim (m R)^{16} \, m_\IR}$ was derived, assuming that the
particle has to tunnel through two barriers described by the potential in
Eq.~\eqref{se1}. We see a conceptual problem with this approach since we do
not
know how to justify a 1-dimensional quantum-mechanical picture (this 1
dimension being the radial coordinate) in the two-throat case. But even if
we
accept this description for the moment, there are further issues related to
the two-barriers assumption: The barriers extend to values of $r \sim
m^{-1}$ as can be seen from Fig.~\ref{pot}. Since $m^{-1} \gg R$ and
$r$ measures the physical distance for $r \gg R$
(cf.~Eq.~\eqref{threebrane}), the width of each barrier is given by
$m^{-1}$. This just reflects the fact that a particle with mass $m$
has a de Broglie wavelength of $m^{-1}$. Accordingly, the particle has
to tunnel through two entire barriers only if the distance $A$ between
the two throats is $\sim 2 m^{-1}$. Indeed, from Eq.~\eqref{DecayRate}
for $l=0$ and since $L > A$, we get a decay rate of $\smash{\Gamma
\sim (m R)^{16} \, m_\IR}$ in this case, in agreement
with~\cite{Firouzjahi:2005qs,CT}. However, if $A$ is smaller than
$\sim 2m^{-1}$, the particle has to tunnel through a smaller barrier.
Correspondingly, the decay rate becomes larger, as can be seen from
Eq.~\eqref{DecayRate}.

The case $m^{-1} \lesssim L$ (without assuming $m^{-1} \ll L$) was also
considered in~\cite{Firouzjahi:2005qs,CT}. It was found that the decay rate
can be much larger than $\smash{\Gamma \sim (m R)^{16} \, m_\IR}$ if a
certain resonance condition is fulfilled. We have not determined the decay
rate for this case and therefore have no result to compare with.\footnote{Note
that we have assumed that $L \gg A,m^{-1}$ in deriving Eq.~\eqref{DecayRate2}.
This result is therefore not suitable to compare with the results from~\cite{
Firouzjahi:2005qs,CT} where the limit of extremely large $L$ was not
taken.} It would be interesting, though, to evaluate
Eq.~\eqref{Matrixelement5} for $m^{-1} \lesssim L$ and to see whether one can
reproduce the results from~\cite{Firouzjahi:2005qs,CT} as well as their
resonance condition.

\chapter{Modifications in more realistic setups}
\label{modifications}
In the last chapters, we have calculated the heat transfer rate and the
decay rate of KK modes for a simple geometry -- two AdS$_5 \times$S$^5$
throats embedded in a 6d torus. In this chapter, we will argue, that these
rates are also applicable to more general throat geometries and embedding
manifolds. A complication arises, though, for the decay rate of KK modes
in a KS throat. The 3-form flux, which is present in these throats, mixes
field fluctuations in a complicated way. The determination of the glueball
vertex along the lines of Section~\ref{decayingaugepicture} is therefore in
general difficult. We will show, however, that the glueballs decay to a
certain lightest glueball on very short timescales. It is therefore sufficient
to determine the decay rate to other throats only for this lightest glueball
which, as we will argue, decays again with the vertex derived in
Section~\ref{decayingaugepicture}. We will also see that an extra suppression
can arise for decay rates in flux compactifications because certain fields
which mediate the decay get high masses. This suppression is roughly
compensated, on the other hand, if the corresponding KK mode mixes
with tachyonic fields in the throat. Finally, we will generalize our
results to processes involving the standard model.

\section{Applicability to other geometries}
\label{applicability}
An apparent limitation of our analysis in Chapters~\ref{energytransfer}
and~\ref{KKdecay} is the assumption of a simple toroidal geometry for the
embedding space. This assumption was used to
determine the spectrum and the couplings of higher KK modes (which
determine the first term in Eqs.~\eqref{Matrixelement2}, \eqref{el}
and~\eqref{DecayRate}). By contrast, the coupling of the zero mode (which
determines the second term in Eqs.~\eqref{Matrixelement2},~\eqref{el}
and~\eqref{DecayRate}), depends only on the size of the embedding manifold
and not on its geometry. To see the relative importance of the terms
more clearly, we focus on the heat transfer rate and rewrite Eq.~\eqref{el}
as
\begin{equation}
\label{el2}
\dot{\rho} \sim \frac{R_1^8 R_2^8}{A^8} \, T^{13} \left( 1 + \left(
\frac{A}{L} \right)^8 \left( L \, T \right)^{-4} \right) \,.
\end{equation}
If the throat-to-throat distance is large, $A\sim L$, the second term
dominates (recall that $LT<1$) and the precise geometry is
irrelevant. By contrast, for small throat separation, $\smash{A \ll
L(LT)^{1/2}}$, the contribution of the KK modes is dominant. In this
case, the precise geometry of the embedding manifold may in principle
be relevant. However, it is then natural to assume that the curvature
scale in the region between the throats is smaller than
$1/A$. Furthermore, as we have already pointed out in Section~\ref{etdt}, the
sum in Eq.~\eqref{Matrixelement2} is dominated by contributions with
$|\vec{n}|\sim L/A$, corresponding to masses $\smash{m_{\vec{n}} \sim
A^{-1}}$. Such modes are only sensitive to the geometry at distance
scales $A$ in the vicinity of the two throats, which we just argued to
be approximately flat. Thus, the order of magnitude of our heat transfer rate
will
remain correct in most relevant cases, even if the overall geometry is
very different from that of a torus. 

Similar conclusions follow for the applicability of the decay rate in
Eq.~\eqref{DecayRate} to other embedding geometries. In particular, we see
that O-planes and further D-brane stacks will not change our results as long
as they are not too close to the two throats. In
Section~\ref{modificationsDecay}, we will see, however, that a suppression of
the decay rate can arise due to the stabilization of certain moduli in flux
compactifications. 

We can also apply our heat transfer rate to situations with one KS throat and
one
AdS$_5 \times $S$^5$ throat or with two KS throats as long as the curvature
scale of the space in between the two throats is not much larger than
$1/A$. In particular, one can easily see from the gravity picture that the
finite length of the KS throats will not change the heat transfer rate
qualitatively. This is obvious for the heated throat since the black hole
horizon hides the IR region.\footnote{Otherwise, if the temperature of the
throat is lower than the IR scale, it contains a non-relativistic gas of KK
modes whose decay rate to the other throat will be discussed in
Section~\ref{modificationsDecay}.} Energy transfer from the heated throat is
due to Hawking
radiation which is absorbed by the other throat. But only the geometry in
the UV region of the latter throat is important for the absorption of the
Hawking radiation. In particular, due to the latter fact, the relevant AdS
scales in
Eq.~\eqref{el} are those at the UV ends of the throats.

For the derivation of the decay rate, we have assumed that the decaying field
fulfills the equation of motion of a minimally coupled, massless scalar field
in the throat. This is no longer obvious for the dilaton in a KS throat
because 3-form flux mixes field fluctuations in a complicated way. We
will discuss the corresponding decay rate in
Section~\ref{modificationsDecay}. There is a different field which fulfills
the aforementioned equation of motion in the throat: the
graviton~\cite{Csaki:2000fc,Firouzjahi:2005qs}. Let us outline how to
determine the decay rate of graviton KK modes in a KS throat:

Away from the bottom of the throat at $r=r_s$, the metric of a KS throat is
well approximated by Eq.~\eqref{threebraneC} with the warp factor which is
given in Eq.~\eqref{lw}:
\begin{equation}
\label{KSwf}
H(r) = 1 +\frac{R_\IR^4 \, \ln(r/r_s)}{r^4} \,.
\end{equation}
For $R_\IR\gg r \gg r_s$, which defines the throat region, the warp factor is
approximately $H \simeq R_\IR^4 \ln(r/r_s)/r^4$. For $r\gg R_\IR$, where
the geometry is asymptotically a cone over $T^{1,1}$, we have $H \simeq 1$.
Near $r=r_s$, the geometry differs considerably from
Eqs.~\eqref{threebraneC}
and~\eqref{lw} and the throat is cut off by the KS region. For an order of
magnitude estimate, we can neglect the logarithmic $r$ dependence of the
warp factor away from $r=r_s$ and approximate the KS region by a sharp cut
off~\cite{Brummer:2005sh}. Thus, the tunneling probability
from the throat into the conical region can be (approximately) calculated
from the effective Schr\"odinger equation, Eq.~\eqref{se1}. Note,
however, that we have to replace the eigenvalues $l (l+4)$ of the angular
Laplacian on $S^5$ by the corresponding eigenvalues on $T^{1,1}$ for a KS
throat. Furthermore, since tunneling is mainly determined by the geometry
in the UV, we should use the AdS scale $R_\UV$ at the UV end of the throat
in Eq.~\eqref{se1}. For an AdS warp factor and a sharp cut off, the
incoming flux is as before given by Eq.~\eqref{TunnelFlux}. From
Eq.~\eqref{dr}, we can then determine the decay rate to the conical region
and match the glueball vertex such that this decay rate is reproduced. 

Once we have the glueball vertex, we can redo the steps which led to
Eq.~\eqref{DecayRate}. The decay rate of graviton KK modes which are
s-waves with respect to the $T^{1,1}$ in a KS throat is thus again given
by Eq.~\eqref{DecayRate}. For higher partial waves, the dependence on the
angular quantum number is different from that displayed in
Eq.~\eqref{DecayRate}.

\section{Some remarks on the spectrum of the Klebanov-Strassler theory}
\label{spectrum}
A number of
papers~\cite{Krasnitz:2000ir,Caceres:2000qe,Amador:2004pz,Firouzjahi:2005qs,
Noguchi:2005ws,Dymarsky:2006hn,Berg:2006xy,Dymarsky:2007zs,Benna:2007mb} 
have calculated parts of the bosonic glueball spectrum of the KS gauge
theory. In~\cite{Berg:2006xy}, masses of KK towers of $7$ coupled scalar
fields and the graviton polarized parallel to the uncompactified dimensions
were determined. In particular, several scalar states lighter than the lowest
spin-2 state were found. 
In~\cite{Caceres:2000qe}, the mass of the lowest KK mode of the dilaton
was 
calculated using some approximations in the geometry. Again, it was found to 
be lighter than a spin-1 and a spin-2 state~\cite{Caceres:2000qe, 
Amador:2004pz}. 
In the light of these findings, we expect the lightest state in the bosonic 
sector to be a scalar glueball. 

The KS gauge theory has $\mathcal{N}=1$ supersymmetry and the lightest scalar 
glueball has a spin-$\smash{\frac{1}{2}}$ superpartner. In a 
phenomenologically viable setup, supersymmetry is broken and the masses of
the scalar and the spin-$\smash{\frac{1}{2}}$ glueball are no longer
degenerate. Let us estimate the resulting mass splitting: 

To this end, note that we expect the throat to be sequestered from the
rest of the compact space in the sense of~\cite{Randall:1998uk}. To explain
this in more detail, let us consider a chiral multiplet $X$ from the throat
sector and another chiral multiplet $Y$ from somewhere in the rest of the
compact space. The Lagrangian can be written in standard ${\cal N}=1$
supergravity form
\be
\label{sugra}
\mathcal{L} \, = \, \int d^4 \theta \; \varphi \bar{\varphi} \; \Omega + 
\left(\int d^2 \theta \; \varphi^3 \; W + \text{ h.c.} \right)\,,
\ee
where $\varphi = 1 +\theta^2 F_\varphi$ is the chiral compensator, $\Omega$ 
is the kinetic function and $W$ is the superpotential. The sectors $X$ and $Y$
are said to be sequestered if~\cite{Randall:1998uk}
\begin{equation}
\begin{split}
\label{sequestered}
\Omega(X, \bar{X}, Y, \bar{Y}) \; & = \; \Omega(X, \bar{X}) \, + \, 
\Omega(Y, \bar{Y}) \\
W (X, Y) \; & = \; W(X) \, + \, W(Y) \,.
\end{split}
\end{equation}
In this case, supersymmetry breaking is communicated to the $X$-sector only
by the $F$-term of the chiral compensator $\varphi$ and not by
the $Y$-sector.

The sequestering assumption in our setup follows from the interpretation of a
Calabi-Yau orientifold with a long throat as supersymmetric RS
model~\cite{Brummer:2005sh}. In this 5d framework~\cite{Randall:1998uk},
this assumption has been widely accepted and has also been used in the context
of type IIB models with strongly warped regions (see
e.g.~\cite{Choi:2006bh,Brummer:2006dg} as well as the detailed discussion
of~\cite{Kachru:2007xp} and references therein). 

We thus assume that supersymmetry breaking is communicated to the lightest
glueball multiplet $X$ only by the $F$-term vev $\langle F_\varphi \rangle$
of the chiral compensator.\footnote{ Actually, the situation might be more
complicated since
the lightest glueball multiplet couples strongly to heavier glueballs,
which are themselves affected by supersymmetry breaking and which might
therefore influence the mass splitting of the lightest multiplet in a
non-negligible way. We therefore view the present calculation only as a
reasonable first guess.
}
The relevant part of the effective Lagrangian Eq.~\eqref{sugra} is 
\be
\label{susybreaking}
\mathcal{L} \, \supset \, \int d^4 \theta \; \varphi \bar{\varphi} \, X 
\bar{X} + \left( \int d^2 \theta \; m_\IR X^2 \varphi^3 + \text{ h.c.}
\right) \,.
\ee
We have to discuss two cases separately. For $\smash{\langle F_\varphi
\rangle \ll m_\IR}$, one can expand $X$ in components and split the lowest
component of $X$ into real and imaginary parts. Inserting this expression
in Eq.~\eqref{susybreaking} and diagonalizing the resulting mass matrix for
the real and imaginary part, one finds two scalar eigenstates with masses
\be
m_{1,2}^2 \, = \, 4 \, m_\IR^2 \pm \, 2 \, m_\IR \, |\langle F_\varphi
\rangle| \,.
\label{split}
\ee
Moreover, the mass of the fermion is $2 m_\IR$ and receives no contribution
from $\smash{\langle F_\varphi \rangle}$. Therefore, one scalar glueball is
lighter than its former spin-$\smash{\frac{1}{2}}$ superpartner and the
mass splitting is $|\langle F_\varphi \rangle|/2$.

For $\smash{\langle F_\varphi \rangle \gg m_\IR}$, Eq.~(\ref{split}) is
obviously
not applicable. In this case, one can analyse the situation from the
perspective of a chiral superfield with vanishing mass, i.e. one considers
the limit $m\to 0$. The theory then possesses a chiral symmetry which
ensures the
masslessness of the fermion even in the presence of supersymmetry breaking.
Thus, in
analogy to the matter superfields of the minimal supersymmetric standard
model, we expect that the scalar glueballs will be heavier than the
fermions if supersymmetry breaking in the throat is a large effect relative
to the supersymmetric mass term. 

Finally, we note that we can expect the F-term vev $\smash{\langle
F_\varphi \rangle}$ of the chiral compensator to be of the same order of
magnitude as
the gravitino mass $m_{3/2}$. In the following, we will therefore refer to
the
two aforementioned cases as $\smash{m_{3/2} \ll m_\IR}$ and $\smash{m_{3/2}
\gg m_\IR}$, respectively. 

\section{Processes in the throat sector}
\label{processes}
In Section~\ref{modificationsDecay}, we will discuss modifications in the 
decay rates of glueballs (respectively KK modes) to other sectors which are
due to flux-induced masses. We can restrict the analysis in that section
to the decay rates of the lightest glueball and its superpartner
(a scalar and a fermion according to the last section):\footnote{
It may happen that the lightest fermionic glueball is not the superpartner
of the lightest bosonic glueball. Moreover, it may happen that the mass of
the
lightest bosonic glueball is larger than twice the mass of the lightest
fermionic glueball. The former could then decay to the latter via couplings
discussed below. We will not consider these possibilities in the
following.}
As we will now discuss, all heavier glueballs decay to these states on
timescales which are short compared to the timescales for decays to other
sectors. 

Below the confinement scale, the glueballs are described by an effective
field theory. Generically, tree-level couplings and loop effects induce
various cubic interactions. For example, for a scalar glueball
$\mathcal{G}$, a spin-1 glueball $\smash{\mathcal{A}_\mu}$ and a spin-2
glueball $\smash{\mathcal{H}_{\mu \nu}}$, we expect interactions of the
type 
\be
\label{cubiccouplings}
\partial_\mu \mathcal{G} \, \mathcal{A}_\nu \, \mathcal{H}^{\mu \nu}  + 
m_\IR \, \mathcal{A}_\mu \mathcal{A}^\mu \, \mathcal{G} + m_\IR^{-1} 
\partial_\mu \mathcal{G} \, \partial_\nu \mathcal{G} \, \mathcal{H}^{\mu \nu} 
+ \dots 
\ee
The coupling strengths follow on dimensional grounds up to possible factors
of $N_\IR$ (which do not follow from dimensional analysis). Also, there may be
more partial derivatives involved or they may act differently. 
Interactions of this type allow for the decay of heavy glueballs to a few
light states which cannot decay further for kinematic reasons. 

Note, however, that the KS gauge theory has a global SU(2)$ \times$SU(2)
symmetry which forbids certain couplings of the type of
Eq.~\eqref{cubiccouplings}. From the dual gravity
point of view, this symmetry is due to isometries of the $T^{1,1}$ inside
a KS throat. In a compactified setup, the KS throat is attached to a
Calabi-Yau manifold which breaks this isometry in the UV. This symmetry
breaking is mediated to the IR as discussed
in~\cite{Kofman:2005yz,Aharony:2005ez,Berndsen:2007my,Dufaux:2008br}. We
therefore expect that couplings of the type of Eq.~\eqref{cubiccouplings}
which violate the global symmetry are nevertheless present, albeit with a
possibly smaller coupling strength. In the following, we ignore the effects
of glueballs charged under the SU(2)$ \times$SU(2) symmetry. In particular,
from the gauge theory point of view, we expect that the lightest glueball
and its superpartner are singlets with respect to this symmetry. 

\begin{figure}[t]
\begin{center}
\begin{minipage}{6.2cm}
\begin{center}
\includegraphics[scale=0.5]{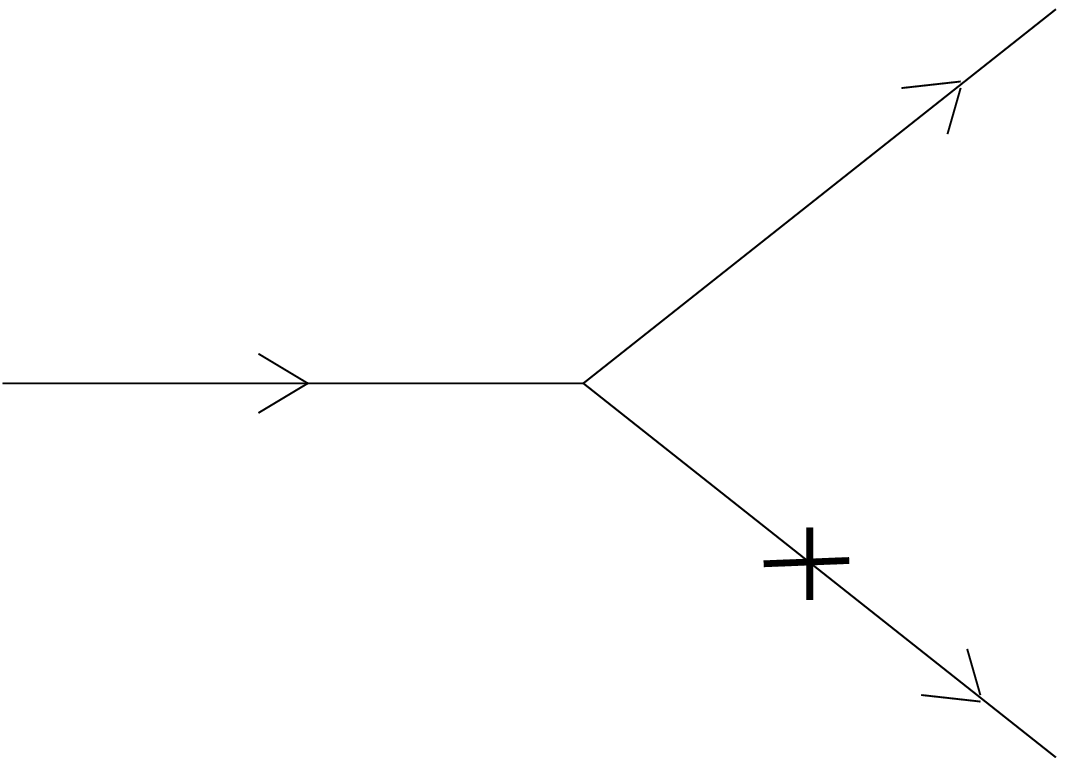}
\put(-23,-3){$h_{\mu \nu}$}
\put(-17,104){$\mathcal{G}$} 
\put(-67,31){$\mathcal{H}_{\mu \nu}$}
\put(-140,60){$\mathcal{B}$}
\caption{Decay of a bosonic glueball $\mathcal{B}$ into a bosonic
glueball 
$\mathcal{G}$ and a graviton $h_{\mu \nu}$ via a spin-2 glueball 
$\mathcal{H}_{\mu \nu}$.\label{fig:decay1}}
\end{center}
\end{minipage}
\hspace{1.4cm}
\begin{minipage}{6.2cm}
\begin{center}
\vspace{-.6cm}
\includegraphics[scale=0.5]{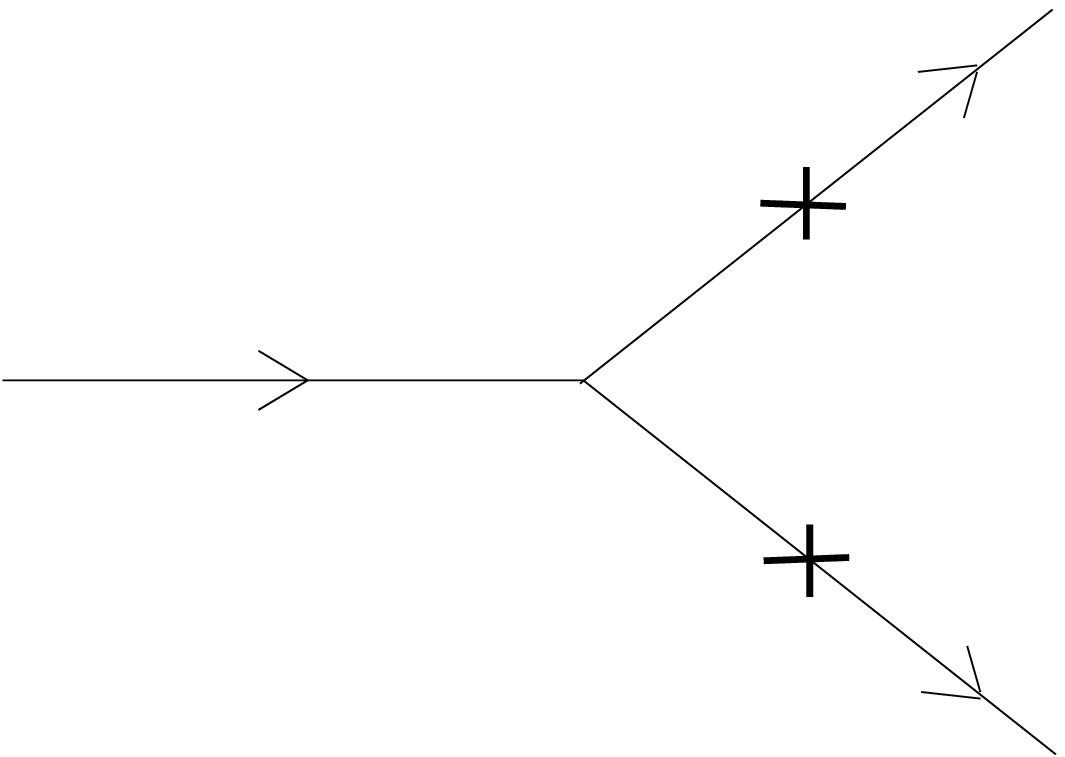}  
\put(-23,-3){$h_{\mu \nu}$} 
\put(-23,108){$h_{\mu \nu}$} 
\put(-67,75){$\mathcal{H}_{\mu \nu}$}
\put(-67,31){$\mathcal{H}_{\mu \nu}$}
\put(-140,60){$\mathcal{G}$}
\caption{Decay of a scalar glueball $\mathcal{G}$ into two 
gravitons $h_{\mu \nu}$ via spin-2 glueballs $\mathcal{H}_{\mu \nu}$.
\label{fig:decay2}}
\end{center}
\end{minipage}
\end{center}
\end{figure}

There are more induced interactions: Recall that Eq.~\eqref{vertex4} is the
vertex between a dilaton
KK mode in the embedding manifold and a scalar glueball. As we have
discussed in Section~\ref{applicability}, the steps leading
to this result are also valid for the mixing of graviton KK modes with
spin-2 glueballs. In particular, the mixing between the 4d graviton
(i.e.~the zero mode in the graviton KK tower, $\smash{\vec{n}=0}$ and $l =
0$) and a spin-2 glueball is
\be
\label{vertex5}
N_\UV \frac{m_\IR^{1/2} m^{5/2}}{M_4}\,.
\ee
We have used Eqs.~\eqref{RN} and \eqref{PlanckScale}. This mixing is similar
to the mixing between the photon and the $\rho$ meson known from QCD and was
also observed in~\cite{Batell:2007jv} for the gauge theory dual of a 5d RS
model.

The mixing in Eq.~\eqref{vertex5} combined with interactions of the type 
of Eq.~\eqref{cubiccouplings} allow for processes such as that shown in 
Fig.~\ref{fig:decay1}: A bosonic glueball $\mathcal{B}$ decays to the lightest
bosonic glueball $\mathcal{G}$ and, via a virtual spin-2 glueball
$\mathcal{H}_{\mu \nu}$, to a 4d graviton $h_{\mu \nu}$. Using
Eq.~\eqref{vertex5}, the decay rate is (for $m \sim m_\IR$ and up to an
unknown factor of $N_\IR$ which may result from the three-glueball vertex)
\be
\label{dr1}
\Gamma \, \sim \, N_\UV^2 \, \frac{m_\IR^3}{M_4^2} \,.
\ee
Similarly, fermionic glueballs decay to the lightest
fermionic glueball and a 4d graviton with the same rate.

The couplings in Eqs.~\eqref{cubiccouplings} and \eqref{vertex5} allow for
another process, shown in Fig.~\ref{fig:decay2}: The decay of the lightest
scalar glueball into two 4d gravitons. The decay rate is (again up to an
unknown factor of $N_\IR$)
\be
\label{dr3}
\Gamma \, \sim \, N_\UV^4 \, \frac{m_\IR^5}{M_4^4} \,.
\ee

By supersymmetry, there are processes in which one graviton in
Figs.~\ref{fig:decay1} and \ref{fig:decay2} is replaced by a gravitino and
one bosonic glueball is replaced by a fermion. For $m_{3/2} \gg m_\IR$,
the corresponding decays are kinematically not allowed. The case $m_{3/2}
\ll m_\IR$ is more involved:  As we have discussed in Section~\ref{spectrum},
we expect the lightest scalar glueball to be lighter than its fermionic
superpartner in this case. A process of the type shown
in Fig.~\ref{fig:decay1} may thus lead to the decay of the lightest
fermionic glueball to the lighter scalar superpartner. The
mass splitting between the superpartners, though, is just of the order of
magnitude of the gravitino mass (cf.~Eq.~\eqref{split}). It thus depends on
the precise relation between the gravitino mass and the mass splitting
whether this decay is kinematically allowed or not. The process of the type in
Fig.~\ref{fig:decay2}, on the other hand, allows for the decay of
fermionic glueballs to a graviton and a gravitino if $m_{3/2}
\ll m_\IR$.

Finally, let us summarize the decay channels discussed in this section. Due
to the interactions in Eq.~\eqref{cubiccouplings} and processes of the type
in Fig.~\ref{fig:decay1}, all glueballs (neglecting the glueballs charged
under the $R$-symmetry) decay to the lightest glueball and its
superpartner. This typically happens on cosmologically short timescales.
For example, for a sector with $\smash{m_\IR \sim
10^7}\hspace{+.09cm}\text{GeV}$ and $N_\UV
\sim\hspace{-.07cm}10$, this process takes $\smash{10^{-10}}$ s. For the case
$m_{3/2} \gg m_\IR$, the only other possible decay (of the channels
discussed in this section) is the decay of scalar glueballs to 4d
gravitons. Due to the suppression by four powers of the 4d Planck scale in
the corresponding rate, this decay happens on timescales much longer than
those of the decays discussed before (e.g.~$\smash{10^{10} \text{ s}}$ for
$m_\IR \sim\hspace{-.07cm}10^7\hspace{+.09cm}\text{GeV},
N_\UV\hspace{-.07cm}\sim\hspace{-.07cm}10$). In particular, the lightest
spin-$\smash{\frac{1}{2}}$ glueballs are stable if $m_{3/2} \gg
m_\IR$.

\section{Decay of scalar and fermionic KK modes to other throats}
\label{modificationsDecay}
Of the processes described in the last section, only the decay to two
gravitons (or to graviton and gravitino, if the gravitino is light enough)
can be sufficiently 
slow to be relevant for late cosmology. The other processes have a time
scale much shorter than the age of the universe for all relevant choices
of parameters. Thus, in late cosmology, the energy density in the throat
sector is completely in the form of the lightest glueball and its
superpartner. It is therefore sufficient to discuss the decay to other throats
only for these lightest states which we expect to be a scalar and
a spin-$\smash{\frac{1}{2}}$ fermion (cf.~Section~\ref{spectrum}). 

The vertex in Eq.~\eqref{vertex2} was derived
for a field which fulfills the equation of motion of a minimally coupled
scalar in the effective 5d description of the throat. An example of such a
field is the dilaton in an AdS$_5 \times $S$^5$ throat. In a KS throat, the
situation is more complicated because 3-form flux mixes the field
fluctuations in a complicated way. Nevertheless, Eq.~\eqref{vertex2}
for an s-wave (i.e.~$l=0$) applies to the coupling of
scalar glueballs from a KS sector to fields in the embedding manifold. To
see this, we again consider the dilaton $\phi$, whose equation of motion is
\be
\label{dilatonEOM2}
\nabla^2 \phi \, = \, \frac{1}{12} \,e^\phi \, \tilde{F}_{MNP}
\tilde{F}^{MNP} 
- \frac{1}{12} \, e^{- \phi} H_{MNP} H^{MNP} \,.
\ee 
Here, $\tilde{F_3}= F_3 - C H_3$ and $F_3 = d C_2$ and $H_3 = d B_2$ are
the field strengths of the RR 2-form $C_2$ and the NS 2-form $B_2$,
respectively. We have taken the RR scalar $C$ to be constant in
Eq.~\eqref{dilatonEOM2}. In a background with imaginary self-dual 3-form
flux~\cite{Giddings:2001yu}, the flux fulfills 
\be
\label{ISDflux}
H_{MNP} H^{MNP} = e^{2 \phi} \, \tilde{F}_{MNP} \tilde{F}^{MNP}
\ee
for the background value of $\phi$ and the right-hand side of 
Eq.~\eqref{dilatonEOM2} vanishes. This is no longer the case if one shifts 
the background value of $\phi$ while keeping $B_2$ and $C_2$ fixed. 
If one simultaneously shifts $B_2$ in such a way that 
Eq.~\eqref{ISDflux} remains fulfilled, however, the right-hand side of 
Eq.~\eqref{dilatonEOM2} still vanishes. In other words, there exists a
flat direction in the 5d effective theory which one can parameterize, e.g.,
by the value of the dilaton. The corresponding field fulfills the equation
of motion of a minimally coupled, massless scalar in 5d. Such a field
couples with the s-wave vertex from Eq.~\eqref{vertex2}.\footnote{The
angular momentum with respect to the S$^5$ in an AdS$_5 \times $S$^5$
throat acts as a mass term in the 5d effective theory. A massless 5d scalar
is the s-wave part of a 10d scalar in such a geometry.} A
light scalar KK mode localized at the bottom of the throat, i.e. a scalar 
glueball, will generically mix with this flat direction in the upper part of
the
throat~\cite{Giddings:2005ff, Frey:2006wv}. Thus, scalar glueballs couple to
fields in the embedding manifold with the previously derived s-wave vertex
given in Eq.~\eqref{vertex2}. Note, though, that stronger
couplings arise for KK modes mixing with fields of the 5d effective
theory which have tachyonic mass. We will analyse this effect in
the next section.

In a phenomenologically viable setup, the scalar fields in the embedding
manifold, which mediate the decay of scalar glueballs to other throats,
all have to be massive. This fact can lead to a suppression of decay
rates. Note that the heat transfer between different sectors can proceed
via the massless graviton. The heat transfer rate in Eq.~\eqref{el} is
therefore not suppressed by flux-induced mass terms. We now discuss three
classes of mediating fields separately:

\vspace{-.4cm}
\paragraph{Dilaton and complex structure moduli} In flux compactifications
\`a la GKP, the lowest KK mode of the dilaton and the complex
structure moduli get a mass
\be
\label{mtau}
m_\tau \, \sim \,  \frac{M_{10}^2}{M_4} \,, 
\ee
where we have assumed that $g_s \sim 1$. Redoing the calculation leading to 
Eq.~\eqref{DecayRate} with a massive instead of a massless propagator for the 
mediating field, we get an extra factor of 
\be
\label{suppressionfactor}
\left( \frac{m^2}{m^2-m_\tau^2} \right)^2 \, \sim  \,
\left( \frac{m}{m_\tau} \right)^4
\ee
in the second term. In the last step, we have assumed that the dilaton,
respectively the complex structure modulus, is much heavier than the
decaying
glueball. In this case, Eq.~\eqref{suppressionfactor} suppresses the decay
rate of scalar glueballs. Note that the second term in
Eq.~\eqref{DecayRate} always dominates if $m < m_\tau$ (since $A^{-1} <
M_{10}$). We then have
\be
\label{DecayRate3}
\Gamma \sim \frac{N_1^2 N_2^2}{M_{10}^8} \, m_\IR \, m^8 \,.
\ee
This is the decay rate of scalar glueballs (which are
s-waves, i.e.~singlets, with respect to the $R$-symmetry of the theory) to
other sectors. By
supersymmetry, it also applies to their spin-$\smash{\frac{1}{2}}$
superpartners.\footnote{This is certainly true for unbroken supersymmetry
in which case the total decay rates of two superpartners are related by a
supersymmetry transformation. For broken supersymmetry, one has to check
that the relevant vertices agree (up to $\mathcal{O}(1)$ prefactors) and
that the suppression by the mediating-field propagator is the same for both
superpartners. We expect this to be the case.}

\vspace{-.4cm}
\paragraph{K\"ahler moduli}
The K\"ahler moduli do not become massive by 3-form flux, and perturbative
or
non-perturbative corrections have to be taken into account in order to achieve
their stabilization (cf.~Section~\ref{GKP}). In particular, the K\"ahler
moduli
can be much lighter than the flux-stabilized complex structure moduli and
the dilaton. This is the case e.g.~in the KKLT scenario. One may therefore
expect that decays mediated by the K\"ahler moduli have a higher rate than
Eq.~\eqref{DecayRate3}. 

We believe that this is not the case. To explain this in more detail, we
restrict ourselves to the universal K\"ahler modulus. The crucial point is
that in the 5d effective theory of a KS throat, the universal K\"ahler
modulus is localized on the UV brane and is thus sequestered from the
bottom of the throat \cite{Brummer:2005sh,Brummer:2006dg,Choi:2006bh}. This
is the situation discussed in Section~\ref{spectrum}: If we denote the
universal K\"ahler modulus by $X$ and the glueball by $Y$ (more precisely,
the corresponding
superfields), the kinetic function $\Omega(X, \bar{X}, Y, \bar{Y})$ and
the superpotential $W(X, Y)$ fulfill Eq.~\eqref{sequestered}. Terms mixing
the two superfields appear neither in the kinetic part nor in the
superpotential. Thus, since the universal K\"ahler modulus does not mix
with the glueballs, it cannot mediate their decays to other sectors. Even
if the sequestered form of Eq.~\eqref{sequestered} turns out to be
violated, we expect that the cross-couplings are much more suppressed than
the mixing vertex of Eqs.~\eqref{vertex2} and \eqref{vertex4} between
glueball and dilaton or complex structure modulus. The effect
of the K\"ahler moduli in mediating glueball decays is then still
negligible.

\vspace{-.4cm}
\paragraph{Gravitino} Also the gravitino can be considerably lighter than the
dilaton and the complex-structure moduli and may therefore play an
important role in mediating decays. Unfortunately, we have not completely
settled this issue.
There are two relevant processes: The gravitino may mix with the fermionic
glueballs and may thus mediate their decays to other sectors. In addition, the
heavier superpartner may decay to the lightest glueball by the emission of a
gravitino. This process follows from the process shown in
Fig.~\ref{fig:decay1} by replacing the virtual spin-2 glueball by a virtual
spin-$\smash{\frac{3}{2}}$
glueball, the outgoing graviton by a gravitino and one of the bosonic
glueballs by a fermionic glueball. If the gravitino mass is large, $m_{3/2}
> m_\IR$, the gravitino is off-shell and must in turn decay to another
sector. It is not immediately clear how strongly the gravitino propagator
suppresses the decay rate for these two processes, i.e. with which power
the gravitino mass enters. In the following, we will assume that there is a
limit of large gravitino mass $m_{3/2} \gg m_\IR $ for which decays
mediated by the gravitino are subdominant. 

\section{Decay of KK modes via tachyonic fields in the throat}
\label{tachyon}
As we have mentioned in the last section, scalar KK
modes which mix with tachyons in the effective 5d description of the
throat couple to supergravity in the embedding space with a stronger vertex
than that in Eq.~\eqref{vertex2}.
The reason is that the profile of a tachyon, i.e. a scalar with
a negative mass squared, is less suppressed than the profile of the dilaton if
one moves from the IR to the UV
end of the approximate AdS$_5$ geometry. 

To see this in more detail, we determine the decay rate of KK modes of such a
tachyon between two throats. For simplicity, we assume that both throats
have the same AdS scale $R$ and that one throat can well be approximated by
the RSI model, whereas the other throat can be accounted for by fields
which live on the UV brane of this RSI model. As we have discussed in
Section~\ref{OtherWays}, this approximation is valid if the size of
the embedding space is of the same order as the AdS scales of the throats. A
tachyonic
scalar $\phi$ in the RSI model is described by the action
\be
\label{tachyonic}
S_{\text{5d}}\, = \, \int d^4x \int_{-\ell}^\ell dy \, \sqrt{g} \,
\frac{1}{2} 
\left( g^{MN} \partial_M \Phi \partial_N \Phi + m_{\text{5d}}^2 \Phi^2
\right) \,,
\ee
where $m_{\text{5d}}^2<0$ is the negative mass squared of the scalar. As in
Section~\ref{RSmodels}, we view the RSI model
as an $S^1/\mathbb{Z}_2$ orbifold and use the same parametrization of AdS$_5$
as in
Eqs.~\eqref{ansatz} and~\eqref{solution}. 

In the full AdS$_5$-space (i.e. without branes),
tachyonic scalars do not lead to instabilities if the masses fulfill the 
Breitenlohner-Freedman bound  $m_{\text{5d}}^2 \geq -4
k^2$~\cite{Breitenlohner:1982bm}, where $k$ is the AdS scale. Let us consider
a scalar $\Phi$
in the RSI model which just satisfies this bound, i.e.~$m_\text{5d}^2 = -4
k^2$.
The KK expansion of the scalar is
\be
\label{TachyonKK}
\Phi(x,y) \, = \, \sum_n  \chi_n(x) \phi_n(y) \, ,
\ee
where the $\chi_n(x)$ are eigenmodes of the 4d d'Alembertian with eigenvalues
$m_n^2$. 

\begin{figure}[t]
\begin{center}
\includegraphics[scale=0.4]{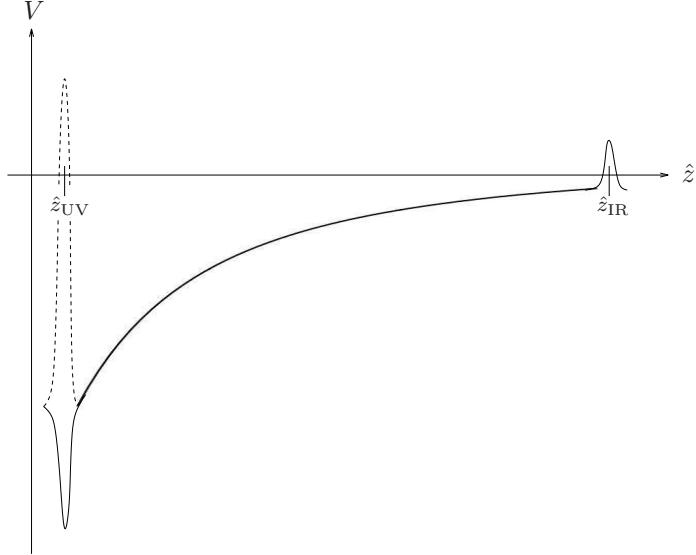}  \put(-245,203){\footnotesize $V$} 
\put(4,141){\footnotesize $\hat{z}$} 
\put(-28,130){\scriptsize $\hat{z}_\IR$} 
\put(-235,130){\scriptsize $\hat{z}_\UV$} 
\caption{Schematic plot of the potential in the effective Schr\"o\-din\-ger
equation for a tachyon in a RSI model. The dotted $\delta$-peak appears when a
large mass term on the UV brane is switched on.\label{potTachyon}}
\end{center}
\end{figure}

We define a new coordinate $\hat{z} \equiv \sgn(y) \, k^{-1}
( e^{k |y|} -1)$.\footnote{This coordinate is related to the coordinate $z$,
which appeared in earlier sections, by a shift: $|\hat{z}|=|z| - k^{-1} $.}
In the orbifold, positive and negative $\hat{z}$ corresponding to positive
and negative $y$ are identified. For a rescaled field 
$\psi_n\hspace*{-.01cm}\equiv\hspace*{-.01cm}(1+ |\hat{z}|
k)^{-3/2}\hspace*{-.01cm}\phi_n$, the equation of motion that follows from
Eq.~\eqref{tachyonic} can be written as
\be
\label{se4}
- \frac{d^2}{d \hat{z}^2} \psi_n \,+\, \left(3 k \,e^{-k \ell}
\delta(|\hat{z}|-\hat{z}_\IR) - 3 k
\,
\delta(\hat{z}) - \frac{1}{4 \, (k^{-1}+ |\hat{z}|)^2} \right) \psi_n
\, = \, m_n^2 \psi_n \,,
\ee 
where $\hat{z}_\IR = k^{-1} (e^{k \ell}-1)$ denotes the
position of the IR brane. The position of the UV brane is $\hat{z}_\UV = 0$.
This equation has the form of a Schr\"odinger equation with
`energy' $E=m_n^2$ and a potential $V$ which is given by the term in
brackets. We have plotted this potential for positive $\hat{z}$ in
Fig.~\ref{potTachyon}. From the equivalent quantum-mechanical problem with
that potential, we expect a mode with negative `energy' $E=m_n^2$ and
accordingly a mode with tachyonic mass in the 4d KK
spectrum. Such a mode is indeed contained in the KK spectrum and the
Breitenlohner-Freedman bound is
no longer sufficient to ensure stability in a RS model~\cite{Ghoroku:2001pi}. 

The absence of
tachyonic directions in the 4d theory can be achieved by
switching on a localized mass term on the UV brane for the scalar $\Phi$.
To see this in more detail, we consider the corresponding term in the action,
\be
\label{UVaction}
S_\UV \, \supset \, \int d^4x \int_{- \ell}^\ell dy \, \sqrt{g} \, \lambda k
\, \delta(y) \,,
\Phi^2 
\ee
where the dimensionless parameter $\lambda$ measures the mass in
units of the AdS scale $k$. If we redo the steps leading to Eq.~\eqref{se4}
taking Eq.~\eqref{UVaction} into account, we find an additional term in the
potential:
\be
\label{UVMassPotential}
V \, = \, 3 k \, e^{-k \ell} \delta(|\hat{z}|-\hat{z}_\IR)+
\Bigl(2 \, \lambda- 3 \Bigr) \, k \,
\delta(\hat{z}) - \frac{1}{4 \, (k^{-1}+ |\hat{z}| )^2} \,.
\ee
We have plotted the resulting $\delta$-peak at the UV brane for $\lambda$
much larger than $3/2$ schematically as a dotted curve in
Fig.~\ref{potTachyon}. As one can see, this $\delta$-peak
has changed the sign as compared to the setup without a mass term at the UV
brane and now leads to a repulsive instead of an attractive contribution to
the potential. Due to this fact, as we will discuss in more detail in
Appendix~\ref{TachyonAppendix}, there are no longer modes with negative
`energy' $E=m_n^2$ in the KK spectrum for sufficiently large $\lambda$.
Instead, all the modes have a positive mass squared. 

Thus, a scalar with a tachyonic bulk mass in a RS model must have an
appropriate mass on the UV brane in order to avoid tachyonic
KK modes. Furthermore, the Breitenlohner-Freedman bound has no
particular meaning in a
RS model. Even tachyonic scalars with bulk masses $m_\text{5d}^2 < -4 k ^2$ do
not lead to instabilities if they are stabilized by sufficiently masses
on the UV brane. Nevertheless, we still consider a tachyon
which just satisfies the Breitenlohner-Freedman bound, i.e.~with
$m_\text{5d}^2= -4 k^2$, in the following. The reason is that
such scalars appear in the 5d effective theory of the KS throat: The geometry
of the KS throat in the UV is approximately AdS$_5 \times$T$^{1,1}$ and
scalars with negative mass squared down to the Breitenlohner-Freedman bound
are contained in the spectrum of type IIB supergravity on
AdS$_5 \times$T$^{1,1}$~\cite{Ceresole:1999zs}. 

For each tachyonic bulk mass, there exists a value of the mass on the UV
brane which lifts the tachyonic mode in 4d to a massless mode. As we will
derive in
Appendix~\ref{TachyonAppendix}, for a scalar with $m_\text{5d}^2 = -4 k^2$,
this happens for $\lambda \simeq 2$. The resulting massless scalar KK mode,
however, would be in conflict with observations, as we have discussed in
Section~\ref{introduction}. In the following, we
therefore assume that $\lambda$ is somewhat larger than $2$, say
$\lambda \approx 3$. This lifts the tachyonic mode in 4d to a very massive
mode.

We now want to determine the decay rate of KK modes of such a tachyon
between two throats. We replace first throat by the RSI model and the second
throat by a gauge theory with $\sim
N^2$ degrees of freedom which lives on the UV brane of the RSI model. Here,
$N$ is related to the AdS scale $R$ of the throat by Eq.~\eqref{RN}. We assume
that the 10d scalar leading to the tachyon in 5d effective theory couples to
the operator $F_{\mu \nu}^2$ of the gauge
theory in the same way as the dilaton couples to that operator in the DBI
action. This gives the effective 5d coupling
\be
\label{TachyonicCoupling}
S_\UV \, \supset \, \frac{1}{  M_{10}^4 \, R^{5/2}} \int d^4x \int_{-
\ell}^\ell dy \, \sqrt{g} \, F_{\mu \nu} F^{\mu \nu} \, \Phi \, \delta(y) \, .
\ee
The prefactor can be understood as follows: We consider a setup in
which the AdS scales $R$ of the throats are of the same order as the size
$L$ of the embedding space. A reduction from 10d to 5d then gives a factor
of $L^5 \sim R^5$ in the kinetic term of the scalar. This factor
can be absorbed into a field redefinition and leads to the factor $R^{-5/2}$
in
Eq.~\eqref{TachyonicCoupling} in addition to the factor $M_{10}^{-4}$ from
the DBI action.

We consider a field which is even under the orbifold $\mathbb{Z}_2$-action,
i.e.~$\psi_n(\hat{z}) = \psi_n(-\hat{z})$. The $\delta$-function peaks in the
potential can be rewritten as boundary
conditions on the branes. Furthermore, it is
sufficient to consider the field only for positive $\hat{z}$. It
is then more convenient to use a
shifted and rescaled variable $\bar{z} \equiv m_n(\hat{z} +
k^{-1})$.\footnote{This coordinate is related to the coordinate $z$,
which appeared in earlier sections, by a rescaling: $\bar{z}= m_n \, z $.}
For an even field, a Schr\"odinger-equation with the
potential Eq.~\eqref{UVMassPotential} is equivalent to the equation
\be
\label{se5}
\left( \frac{d^2}{d \bar{z}^2} \, + \, \frac{1}{4 \, \bar{z}^2} + 1
\right) \psi_n \,
= \, 0 
\ee
with the boundary conditions
\be
\label{bc}
\bar{z}_\UV \frac{d}{d\bar{z}} \psi_n(\bar{z}_\UV) \, = \,
\Bigl(\lambda-\frac{3}{2} \Bigr) \,
\psi_n(\bar{z}_\UV) \qquad \text{and} \qquad
\bar{z}_\IR \frac{d}{d\bar{z}} \psi_n(\bar{z}_\IR)\, =
\,-\frac{3}{2} \psi_n(\bar{z}_\IR)\, .
\ee
Here, $\bar{z}_\UV=m_n k^{-1}$ and $\bar{z}_\IR= m_n k^{-1} e^{k \ell}$
are the positions of the UV brane and the IR brane,
respectively. The wave functions and the spectrum, which follow from
Eqs.~\eqref{se5} and~\eqref{bc}, are determined in
Appendix~\ref{TachyonAppendix}. Here, a simplified analysis will be
sufficient:

Since Eqs.~\eqref{se5} and~\eqref{bc} correspond to a quantum-mechanical
problem in a box of size $\bar{z}_\IR - \bar{z}_\UV$, it is clear that the
masses are quantized approximately in units of $\smash{(\bar{z}_\IR -
\bar{z}_\UV)^{-1} \simeq k e^{-k \ell}}$, such that $m_n \sim n \, k \,
e^{-k \ell}$ with $n \in
\{1,2,3,\dots \}$. In the following, we consider a KK mode with a mass
somewhat larger than the minimal mass and correspondingly $n$ somewhat larger
than $1$. For such a
KK mode, we can solve Eq.~\eqref{se5} approximately in two regions: For
$\smash{\frac{1}{2} \ll \bar{z} \leq \bar{z}_\IR \sim n}$, we can neglect
the
$z$-dependent term in the potential in Eq.~\eqref{se5}. Similarly, for
$\smash{\frac{1}{2} \gg \bar{z} \geq \bar{z}_\UV \sim n \, e^{-k \ell} }$,
we can neglect the constant term. Approximate solutions to Eq.~\eqref{se5} in
these regions are
\be
\label{approx}
\psi_n(\bar{z}) \, \simeq \, 
\begin{cases}
\frac{1}{N_n} \, \bigl( \bar{z}^\frac{1}{2}  \, + \, A_n \,
\bar{z}^\frac{1}{2} \ln \bar{z} \bigr) & \text{for} \quad \bar{z}_\UV
\leq  \bar{z} \ll  \smash{\frac{1}{2}} \\
\frac{1}{N_n}\, \bigl( \cos \bar{z} \, + \, B_n \, \sin \bar{z} \bigr) &
\text{for} \quad \smash{\frac{1}{2}}  \ll \bar{z} \leq  \bar{z}_\IR \; .
\end{cases}
\ee
The constant $A_n$ can be determined from the boundary condition on the UV
brane in Eq.~\eqref{bc}. We find
\be
\label{A_n}
A_n \, = \, \frac{\lambda - 2}{1- (\lambda-2) \, \ln \bar{z}_\UV} \, .
\ee
We choose the constant $N_n$ such that the kinetic term of the 4d field
$\chi_n$ is canonically normalized. With the relations $\smash{\bar{z} =
\frac{m_n}{k} e^{k \ell}}$ and $\smash{\phi_n = (\frac{\bar{z}
k}{m_n})^{3/2} \psi_n}$ as well as with Eqs.~\eqref{tachyonic}
and~\eqref{TachyonKK}, one can check that this is fulfilled if
\be
m_n^{-1} \int_{\bar{z}_\UV}^{\bar{z}_\IR} d\bar{z} \, \psi_n^2 \,
\sim \, 1 \,,
\ee
up to $\mathcal{O}(1)$ prefactors. We can split this integral and use the
approximate solutions in
Eq.~\eqref{approx} in the regions $\smash{\bar{z}_\UV
\leq  \bar{z} \lesssim \frac{1}{2}}$ and $\smash{\frac{1}{2} 
\lesssim \bar{z} \leq \bar{z}_\IR}$, respectively. Since $A_n < 1$ and we
have assumed that
$\bar{z}_\IR \sim n \gg 1$, the integral is dominated by the
latter region. In Appendix~\ref{TachyonAppendix}, we find that the constant
$B_n$, which is
determined by the boundary condition at the IR brane in Eq.~\eqref{bc} and
the precise relation for the masses $m_n$,
is also small, i.e.~$B_n < 1$. We then find
\be
\label{TachyonNormalization}
N_n \, \sim \,  \sqrt{\frac{\bar{z}_\IR}{m_n}} \, . 
\ee

Inserting the KK expansion Eq.~\eqref{TachyonKK} into the coupling
Eq.~\eqref{TachyonicCoupling} and using the approximate solution near
the UV brane in Eq.~\eqref{approx} together with
Eqs.~\eqref{A_n} and~\eqref{TachyonNormalization}, the coupling of the n-th KK
mode $\chi_n$ to the operator $F_{\mu \nu}^2$ follows as:
\be
\label{TachyonicCoupling2}
S_\text{4d} \, \supset \, \frac{g_n}{ M_{10}^4 \, R^3} \int d^4x \,
F_{\mu \nu}
F^{\mu \nu} \, \chi_n  \,, \qquad \text{where} \quad g_n \, \sim \, \sqrt{m_n
\, m_\IR} \frac{R^2}{\ell} \,.
\ee
We have used $\lambda \approx 3$ and the fact that $\log \bar{z}_\UV =
\log(m_n/k) \sim -k \, \ell = -\ell /R$. This factor
is approximately the logarithm of the generated hierarchy of the throat and
therefore typically at most of the order 10. Up to $\mathcal{O}(1)$
prefactors, our simplified analysis reproduces the coupling in
Eqs.~\eqref{AppendixCoupling1} and~\eqref{AppendixCoupling2}
from the rather lengthy KK decomposition in Appendix~\ref{TachyonAppendix}. 

The resulting decay rate of a KK mode into $\sim N^2$ degrees of freedom on
the UV brane, and thus into another throat, is
\be
\label{TachyonDecayRate}
\Gamma \, \sim \, (m_n R)^4 \, m_\IR \, \Bigl(\frac{R}{\ell}\Bigr)^2 \,,
\ee
where we have used Eq.~\eqref{RN}. Since typically $\ell /R =
\mathcal{O}(10)$, this decay rate is only slightly smaller than the
corresponding decay rate of graviton KK modes Eq.~\eqref{Gamma5d}. 

We want
to compare this decay rate with the flux-suppressed decay
rate Eq.~\eqref{DecayRate3} that we have derived in
Section~\ref{modificationsDecay}. Using Eq.~\eqref{RN} and
assuming that both throats have the same
number of degrees of freedom, i.e.~$N_1 = N_2$, Eq.~\eqref{DecayRate3} can be
written as
\be
\label{Flux-Suppressed}
\Gamma \, \sim \, N^2 (m_n R)^8 \, m_\IR \, .
\ee
Since the mass $m_n$ is
exponentially smaller than the AdS scale $R^{-1}$, whereas typically $\ell
/R =\mathcal{O}(10)$ and $\smash{n \lesssim 10^4}$
(cf.~Section~\ref{SingleThroat}), the relative factor $(m_n R)^4 N^2 (\ell
/R)^2 $ between Eqs.~\eqref{TachyonDecayRate} and~\eqref{Flux-Suppressed} is
usually very small. Decays of KK modes which
are mediated by a tachyon in the KS throat therefore have a much
higher rate than decays which are mediated by the flat direction in
the KS throat that we have discussed in Section~\ref{modificationsDecay}.

\section{Processes involving the standard model sector}
\label{DecaytoSM}
The results from Chapters~\ref{energytransfer} and~\ref{KKdecay} were
derived for stacks with a large number of D-branes because these stacks are
dual to throats. These results, however, are also applicable to small
stacks of D-branes. In particular, we can consider a realization of the
standard model on some D3-
and/or D7-branes which live in the unwarped part of the compact
space.\footnote{In the literature, it is often assumed that the standard model
is realized in a throat. In this case, the standard model couples to
supergravity modes in that throat which in turn couple to other throats
according to Eqs.~\eqref{el} and \eqref{DecayRate}.} The heat transfer
rate between the standard model and a throat is then given by
Eq.~\eqref{etNN} if we set $N_1^2
= g$ to account for the standard model with
$g\hspace{-.07cm}\sim\hspace{-.07cm}100$ degrees of
freedom.
According to the discussion in Section~\ref{applicability}, $N_2$ is the
number $N_\UV$ of 5-form flux at the UV end of the throat.

Similarly, Eq.~\eqref{DecayRate3} with $N_1^2 = g$ and $N_2 = N_\UV$ is the
decay rate of scalar glueballs to the standard model.
Note, however, that this decay rate is not applicable to the inverse
process (a decay from the standard model to a throat sector) since our
derivation of the vertex in Eq.~\eqref{vertex2}
assumed a weakly curved gravity description. This is not fulfilled for a
small stack of D-branes. 

There is a subtlety concerning the decay rate of
spin-$\smash{\frac{1}{2}}$ glueballs to the standard model which arises for
large gravitino masses, $m_{3/2} \gg m_\IR$. Namely, we expect the
mass splitting between superpartners in the standard model sector to be of
the order of the gravitino mass. This means that the superpartners of
standard model particles are heavier than the decaying
spin-$\smash{\frac{1}{2}}$ glueballs. If $R$-parity is conserved, most
decay
channels involve such a superpartner as a final state and the corresponding
decays are therefore kinematically forbidden. A coupling that does not
involve a superpartner is
\be
\label{vertex3}
\lambda \, \bar{l} \, \psi \, H \,,
\ee
where $l$ is a lepton doublet, $H$ is the Higgs doublet and $\psi$ is a 
dilatino or any other modulino.\footnote{This coupling was already considered
in~\cite{Benakli:1997iu} since it also leads to a mixing between the modulino
and the neutrino.} The coupling strength $\lambda$ may be $\mathcal{O}(1)$ or 
it may be suppressed as $\lambda = m /M_4$ where $m$ is some
low mass scale. 

Note, however, that the coupling in Eq.~\eqref{vertex3} probably requires
$R$-parity violation. Namely, the corresponding coupling containing the
modulus instead
of the modulino generates a bilinear $R$-parity violating term for nonzero
modulus vev. For high-scale supersymmetry breaking, a large
coupling $\lambda$ may nevertheless be allowed. Moreover, even for
maximally broken $R$-parity, all other decay channels involve standard
model
superpartners which further decay into standard model particles. The
corresponding decay rates are therefore suppressed by the propagators of
the heavy superpartners and are smaller than the decay rate resulting from
Eq.~\eqref{vertex3}. Therefore, we focus on this coupling. Redoing
the steps leading to Eq.~\eqref{DecayRate3} with the vertex of
Eq.~\eqref{vertex3}, we find
\be
\label{fermionicdr}
\Gamma \, \sim \, \lambda^2 \, N^2_{\UV} \frac{m^6 m_\IR}{M_{10}^8 /M_4^2}
\ee
for the decay rate of spin-$\smash{\frac{1}{2}}$ glueballs to the standard
model if $m_{3/2} \gg m_\IR$.

Finally, we note that, in absence of the coupling in Eq.~\eqref{vertex3}
and of $R$-parity violation, the spin-$\smash{\frac{1}{2}}$ glueballs can
not
decay at all to the standard model sector. If, in addition, there is no
throat with lower IR scale (otherwise decays to this sector with the rate
in Eq.~\eqref{DecayRate3} are possible), the spin-$\smash{\frac{1}{2}}$
glueballs are absolutely stable.

\chapter{Sequestered Dark Matter: Thermal production}
\label{thermalproduction}
\section{Preliminaries}
\label{intro}
Dark matter is frequently assumed to consist of massive weakly interacting 
particles which are stable (or have a very long lifetime) because their 
decay is forbidden by some (approximate) symmetry. The observed dark
matter abundance is obtained if these particles were in thermal
equilibrium in the early universe and later fell out of equilibrium with an
appropriate freeze-out abundance.

But it is also well-known that dark matter may originate in a hidden sector
which is coupled to the standard model only via higher-dimension
operators, ensuring that dark matter has a sufficiently long lifetime (see
e.g.~\cite{Kolb:1985bf,Dimopoulos:2001ui,CT}). Since the
annihilation cross section of these particles is suppressed by a high
mass scale (proceeding via a higher-dimension operator), they may
overclose the universe today if they had been in thermal equilibrium in the
early universe. Instead, one often assumes that the hidden sector
particles were not in thermal equilibrium after reheating and that
they were produced by thermal
reactions in the hot standard model plasma.
The resulting abundance is typically much lower than the equilibrium
abundance since the relevant rates are again suppressed by a high mass scale.
Depending on the reheating temperature, the abundance can have the right
magnitude to account for the observed dark matter. A well-known particle which
falls into this class of dark matter candidates is (to a certain
extent\footnote{The gravitino would decay too quickly were it not (at least
partially, see~\cite{Bertone:2007aw,DecayingDarkMatter}) protected by
$R$-symmetry. Moreover, another well-known production mechanism of
gravitino
LSPs is by the decay of NLSPs which had an appropriate freeze-out
abundance.}) the gravitino. 

In the following sections, we present a new dark matter candidate with the
aforementioned properties (we discuss earlier related work in
Section~\ref{relation}): KK modes in a throat or, equivalently, glueballs of
the dual gauge theory. Due to the warping, these particles have highly
suppressed couplings to other sectors and can therefore be very stable. By
the
same token, they have redshifted and thus relatively low masses which allows
them to be produced thermally, even if the reheating temperature is not very
high.

The throats respectively the dual gauge theories would lead to too much
dark radiation during big bang nucleosynthesis and/or would overclose the
universe if
they were in thermal equilibrium after reheating. We therefore assume
that only the standard model is reheated initially and that the
throats receive no energy from the reheating process.\footnote{This is, for
example, not fulfilled for reheating after brane-antibrane inflation. In
this scenario (cf.~Section~\ref{motivation}), the throat in which inflation
takes place is heated
by the annihilation of the brane with the antibrane. This energy is
subsequently transferred to other throats and the standard model as
discussed in earlier sections. Aspects of KK dark matter in such a
scenario were discussed in~\cite{CT} (cf.~Section~\ref{relation}).} Glueballs
are then produced by thermal reactions in the hot standard
model plasma. Nevertheless, it is of course possible that the inflaton
interacts only very weakly with the throats respectively the dual gauge
theories, thereby producing an abundance of glueballs during reheating which
is not too high.
Since this abundance depends on the coupling of the inflaton and on
the model at hand, we can view our result for the produced amount of dark
matter, that we derive, as a model-independent lower bound. 

There is an important aspect of the production mechanism of our dark matter
candidate that can best be seen from the gauge theory point of view: 
Annihilating particles in the hot standard model plasma inject energy
into the gauge theory. The gauge theory states which are produced
in that way hadronize shortly afterwards (if the energy density in the
gauge theory sector is not above the confinement scale). As we will discuss
in the next section,
there are no jets in the large-$N$, large-$\lambda$ gauge theories which
are dual to throats. Instead, after hadronization, all the energy is in
the form of slow glueballs with masses and kinetic energies of the
order of $m_\IR$. These glueballs immediately scale like matter with the
expansion of the universe and thereby give an important contribution to the
total energy density at our epoch already for reheating
temperatures which are only moderately high.

For definiteness, we focus on scenarios in which the standard model is
realized on some D3- and/or D7-branes in the unwarped part of the compact
space (as discussed in Section~\ref{DecaytoSM}). The fact that the glueball
couplings to the standard model are highly suppressed ensures a long lifetime
of the dark matter particles. Nevertheless, it turns out that in many cases
decays mediated by the gravitino must be negligible because otherwise the
decay rate of glueballs would be too high. We therefore focus on scenarios
with high-scale supersymmetry breaking in which the gravitino is much heavier
than the glueballs, $m_{3/2} \gg m_\IR$. 
We comment on models with low-scale supersymmetry breaking in
Section~\ref{extra}.

\section{Energy transfer}\label{et}
As outlined in the last section, we assume that the throats have
received no energy from the reheating process, whereas the standard model
is heated to a temperature $T_{\RH}$ initially. Subsequently, energy will 
be transferred from the standard model to the throats. Note that this 
process is similar to the energy transfer from the hot brane to the bulk in 
RSII models~\cite{HMR} (see also~\cite{ 
Gubser:1999vj}). The AdS$_5$ bulk plays the role of the throat, which we 
however assume to be of finite length, with the KS region corresponding to
the IR brane.

We assume that the temperature of the standard model is smaller than the
compactification scale, i.e.~$\smash{T < L^{-1}}$. Furthermore, we consider
the generic situation that the distance between the two throats is of
the same order of magnitude as the size of the embedding manifold, i.e.~$A
\sim L$. The heat transfer rate is then given by Eq.~\eqref{etNN} and the
second term in that rate dominates.
We set $N_1^2 = g$ to account for the standard model sector with
$g\hspace{-.07cm}\sim\hspace{-.07cm}100$ degrees of freedom and $N_2 = N_\UV$
for the throat at hand.
Using Eq.~\eqref{PlanckScale}, the heat transfer rate can be written as
\be
\label{etr}
\dot{\rho} \, \sim \, g \, N^2_{\UV} \frac{T^9}{M_4^4} \,.
\ee
This rate is easily understood as being due to a gravitational strength 
coupling between a sector with $g$ degrees of freedom (the standard model) 
and a sector with $N_\UV^2$ degrees of freedom (the throat). Note that,
due to the existence of a confinement scale $m_\IR$, glueballs can only be
created if $T_\RH>m_\IR$. We expect these initial gauge theory states to
have masses up to $m \sim T_\RH$. 

The glueballs may decay back to the standard model. As we have
discussed in Section~\ref{applicability}, spin-2 glueballs have the highest
decay rate. From Eq.~\eqref{DecayRate}, we get
\be
\label{KK2SM}
\Gamma(m) \, \sim \, g \, N^2_{\UV} \frac{m^4 m_\IR}{M_4^4} \,.
\ee
On the other hand, glueballs can also decay to lighter glueballs within the 
same throat. We will have to discuss this process in some detail below. At 
the moment, it is sufficient to establish that the decay to lighter glueballs
wins over the possible decay back to the standard model. For this purpose, we
recall that we are dealing with a strongly coupled system with a dense 
spectrum. Thus, the initially created gauge theory state of mass $m$ will 
have a lifetime $\sim 1/m$. In the most conservative scenario, it will decay 
to 2 states of mass $m/2$. These states will in turn decay to states of mass 
$m/2^2$ after a time-interval $\sim 2/m$, and so on. Summing up the 
probabilities for the decay back to the standard model at each step of this 
cascade, we arrive at a total probability 
\be 
w\sim\sum_{n=0} \Gamma(m/2^n)\cdot\frac{2^n}{m}\,.\label{w}
\ee
This sum is of the same order of magnitude as the first term and hence very 
small in all cases of interest. Clearly, we could equally well have assumed 
that each glueball decays to $k_1$ lighter states with mass $m/k_2$, arriving 
at the same conclusion for any ${\cal O}(1)$ numbers $k_1$ and $k_2$. Thus, 
the relaxation to lighter states within the same throat always wins over the 
decay back to the standard model or to other throats. 

The heat transfer rate Eq.~\eqref{etr} is strongly temperature dependent, 
$\dot{\rho} \propto T^9$. Therefore, heat transfer is effectively finished 
soon after reheating and the corresponding time scale is
$\smash{|T/\dot{T}|}$ 
at $T=T_\RH$. The total energy density after reheating is dominated by the 
relativistic gas in the standard model sector with $\smash{\rho = g 
\frac{\pi^2}{30} T^4}$. We find
\be
\label{rts}
|T/\dot{T}| \, = \, H^{-1} \, \sim \,  \frac{M_4}{ g^{1/2}T^2} \,,  
\ee
where $H$ is the Hubble rate. Using
Eqs.~\eqref{etr} and \eqref{rts} at $T 
\sim T_\RH$, the energy density deposited in a throat directly after 
reheating is 
\be
\label{energydensity}
\rho \, \sim \, \dot{\rho} \, |T/\dot{T}| \, \sim \, g^{1/2} 
N^2_\UV \frac{T^7_\RH}{M_4^3} \,.
\ee

Before closing this section, we note that the heat transfer processes 
we consider compete with the unavoidable energy deposition in the throat 
sectors occurring during inflation. This can be understood by noting that 
de Sitter space has a temperature $T_{\rm dS} \sim 1/R_{\rm dS}\sim 
M_{\rm inf}^2/M_4$. We assume that inflation lasts long enough for 
the throats to be thermalized with this temperature. Furthermore, 
parameterizing the efficiency of reheating by an efficiency factor
$\epsilon\le 1$, we have $gT_{\rm RH}^4\sim \epsilon M_{\rm inf}^4$. Thus, 
all throats have a temperature $T_{\rm dS} \sim \sqrt{g/\epsilon}\,T_{\rm 
RH}^2/M_4$ at the time of reheating. Jumping ahead, we note that for 
typical long throats (where this effect is most relevant), we find 
initial throat temperatures $\sim\hspace{-.07cm}10^6\hspace{+.09cm}\text{GeV}$
for a reheating
temperature $\sim\hspace{-.07cm}10^{11}\hspace{+.09cm}\text{GeV}$. For such
throats, `de-Sitter heating' in
fact wins over the heating process analysed in this section if
$\epsilon<1$, allowing in principle for even more throat dark matter than
we find in our conservative analysis.

\section{Time evolution of the energy density}
\label{te}
The result in Eq.~\eqref{energydensity} is the energy density in the gauge
theory sector directly after reheating. During cosmological
evolution, this energy density is diluted by the expansion of the universe.
To determine the contribution of glueballs to the total
energy density at our epoch, it is important to know how their energy
density scales with the scale factor $a$ of the universe.

Let us first consider the case that the energy density is above the
critical value for a deconfinement phase transition directly after
reheating. The gauge theory thermalizes in this situation. To see this in
more detail, we view each initially created gauge theory state as a
localized excitation of a strongly coupled system with energy $\sim T_\RH$.
The localization assumption can be justified by recalling that, from the
D-brane perspective, the mediating bulk supergravity fields couple to local
gauge theory operators like $F_{\mu\nu}F^{\mu\nu}$. We model the further
evolution of this state as a ball of gauge theory plasma expanding with the
velocity of light.\footnote{ Note that this physical picture is equivalent
to the picture of a cascade decay used in the derivation of Eq.~(\ref{w})
if we assume that a glueball with mass $m/2^n$ fills out a volume
$(2^n/m)^3$. 
}
The number density of these balls is
\be
\label{nd}
n \, \sim \frac{\rho}{T_\RH} \, \sim \, g^{1/2} N^2_\UV 
\frac{T_\RH^6}{M_4^3} \,,
\ee
where we have used Eq.~\eqref{energydensity}. It follows that the balls
fill out the whole space after a time
\be
t \, \sim \, n^{-1/3} \, \sim \, \frac{M_4}{g^{1/6}N^{2/3}_\UV \, T_\RH^2}\,.
\ee 
Comparing this with the Hubble time Eq.~\eqref{rts} at $T=T_\RH$, we see that
the gauge theory plasma fills out the whole space before the Hubble expansion 
becomes relevant if $N_\UV^2 \gtrsim g$, which holds for all relevant throats.

The gauge theory dual to a KS throat has a logarithmically varying number of
degrees of freedom, corresponding to the logarithmic deviation of the KS
geometry from AdS$_5$. In the deconfined phase, the effective number of
colours $N_\eff$ of the gauge theory depends on the temperature
$\smash{\tilde{T}}$ of the plasma as
\be
\label{N_eff2}
N_\eff \, \sim \, N_\IR \, \ln \left( \frac{\tilde{T}}{m_\IR} \right) \,.
\ee
The deconfined phase of the gauge theory is dual to a throat with a
black hole horizon which replaces the IR end. The highest meaningful value
in Eq.~\eqref{N_eff2} is $N_\eff \sim N_\UV$. This corresponds to a
temperature where the black hole horizon reaches the UV end of the throat.
The energy density as a function of the plasma temperature
$\smash{\tilde{T}}$ is 
\be
\rho \, \sim \, N_\eff^2 \, \tilde{T}^4 \,.
\ee
Since the logarithmic variation of $N_\eff$ with $\smash{\tilde{T}}$ is small
compared to the variation of the $\smash{\tilde{T}^4}$-term, we will neglect 
it in the following. The deconfined phase can then be described by an 
approximate conformal field theory and the energy density correspondingly 
scales like radiation with $\smash{a^{-4}}$. 

When the energy density drops to $\rho \sim N^2_\IR m^4_\IR$, a confinement
phase transition begins which lasts until the energy density has reached $\rho
\sim \lambda \, m_\IR^4$, where $\lambda = g_s N_\IR$ is the 't Hooft
coupling.\footnote{The thermodynamics of large-$N$ gauge theories is
reviewed e.g.~in Appendix A in \cite{Aharony:2005bm}.} In the
transition region for $\rho$, space is divided into separate regions in
either the confined phase with $\rho < \lambda \, m_\IR^4$ or the still
deconfined phase with $\rho > N^2_\IR m^4_\IR$. At even lower energy
densities $\rho < m_\IR^4$ (assuming $\lambda >1$), a description in terms
of a nonrelativistic glueball gas is applicable and the energy density
correspondingly scales with $\smash{a^{-3}}$. We do not know the scaling of
$\rho$ with $a$ in the transition region $\smash{N^2_\IR m^4_\IR  > \rho >
m_\IR^4}$ though, since the equation of state during the phase transition
is unknown. Since we expect the scaling to be in between the two extremes
$\smash{\rho \propto a^{-3}}$ and $\smash{\rho \propto a^{-4}}$ in this
region, we will take $\smash{\rho \propto a^{-4}}$ for $\smash{\rho > N_\IR
m_\IR^4}$ and $\smash{\rho \propto a^{-3}}$ for $\smash{\rho < N_\IR
m_\IR^4}$ for simplicity.\footnote{
Using the intermediate value $\smash{\rho \sim N_\IR m_\IR^4}$ for the
distinction between the two behaviours, the error in $\rho$ is a factor of
$\smash{(N_\IR m_\IR^4 / N^2_\IR m_\IR^4)^{1/4} = N_\IR^{-1/4}}$ if
$\smash{\rho \propto a^{-3}}$ in the entire transition region or a factor of
$\smash{(N_\IR m_\IR^4 / m_\IR^4 )^{1/3} = N_\IR^{1/3}}$ if $\smash{\rho
\propto a^{-4}}$ in the entire transition region. In both cases, this factor
is typically $\mathcal{O}(1)$.}

Thus, the energy density in the gauge theory sector,
Eq.~\eqref{energydensity}, initially scales like radiation if it is larger
than $\rho \sim N_\IR m_\IR^4$. As a function of the standard model
temperature $T$, we have
\be
\label{ed}
\rho \, \sim \, g^{1/2} N^2_\UV \left( \frac{T_\RH}{M_4} \right)^3 T^4\,.
\ee
\begin{figure}[t]
\begin{center}
\includegraphics[scale=0.5]{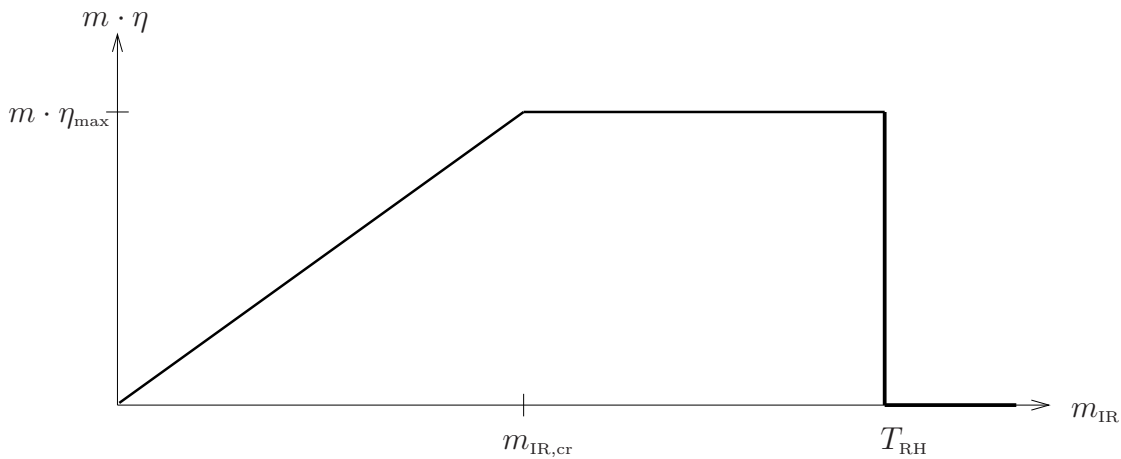} \put(-367,148){$m \cdot \eta$}
\put(-395,112){$m \cdot \eta_\maxx$}
\put(7,0){$m_\IR$} 
\put(-208,-12){$m_\IRc$}
\put(-65,-12){$T_\RH$}
\vspace{.4cm}
\caption{Schematic plot of $m \cdot \eta$ as a function of the IR
scale 
$m_\IR$ of the throat. Here, $m_\IRc $ is the IR scale for which $T_\pt \sim 
T_\RH$, i.e., for which the throat is heated precisely to its phase
transition 
temperature. See text for more details.}
\label{fig:plot}
\end{center} 
\end{figure}

We assume the scaling behaviour to change when the energy density has dropped 
to $\smash{\rho \sim N_\IR m^4_\IR}$. This happens when the standard model 
has a temperature
\be
\label{Tpt}
T_\pt \, \sim \, m_\IR \frac{ N_\IR^{1/4}}{ N_\UV^{1/2}}  \hspace{-1mm} 
\left( \frac{M_4}{T_\RH} \right)^{3/4} \hspace{-1mm},
\ee
where we have neglected a factor of $g^{1/8}$ which is close to 1. The energy 
density scales like matter afterwards and the ratio of energy density and 
entropy density, $ \rho/s = m \cdot \eta$, stays constant. Here $\eta=n/s$ 
is the glueball number density normalized by the entropy density. Using 
Eq.~\eqref{ed} and $\smash{s = g \frac{2 \pi^2}{45}  T^3}$ (dominated by the 
standard model sector) at $T=T_\pt$, we find the glueball mass density per
entropy density
\be
\label{energydensity2}
m \cdot \eta  \, \sim \,   \frac{N^2_\UV}{g^{1/2}} \left( \frac{T_\RH}{M_4} 
\right)^3 T_\pt \,.
\ee
The factor of $T_\pt$ is 
smaller than or equal to $T_\RH$ and reflects the fact that the energy
density undergoes a phase of $\smash{a^{-4}}$ dilution. The quantity 
$m \cdot \eta$ is useful because it determines the contribution of the 
glueballs to the total energy density in the late universe. We have plotted 
$m\cdot \eta$ as a function of $m_\IR$ schematically in Fig.~\ref{fig:plot}. 
The part corresponding to Eq.~\eqref{energydensity2} is the line which
grows linearly\footnote{
Note 
that this plot has to be read either at fixed $N_\UV$, in which case 
$N_\IR$ must be interpreted as function of $N_\UV$ and $m_\IR$, or at fixed 
$N_\IR$, in which case $N_\UV$ must be interpreted as function of $N_\IR$ and 
$m_\IR$. In both cases an extra logarithmic dependence of $m\cdot\eta$ on 
$m_\IR$ is introduced, which we however neglect. 
}
with the IR scale from $m_\IR = 0$ up to an $m_\IR$ such 
that $T_\pt$ in Eq.~\eqref{Tpt} is of the same order of magnitude as $T_\RH$. 
This is the maximal IR scale for which Eq.~\eqref{energydensity2} is valid 
because at this point the initial energy density in the gauge theory
sector, Eq.~\eqref{energydensity}, is of the same order of magnitude as the
critical energy density $\rho \sim N_\IR m_\IR^4$.

Dividing Eq.~\eqref{energydensity2} by $\smash{\rho_c/s_0 \simeq 2 \cdot 
10^{-9}}\hspace{+.09cm}\text{GeV}$, where $\smash{\rho_c}$ is the current
critical energy density 
for a flat universe and $s_0$ is the current entropy density, and using
$g\hspace{-.07cm}\sim\hspace{-.07cm}100$ as well as $M_4\hspace{-.07cm}\simeq
2 \cdot 10^{18}\hspace{+.09cm}\text{GeV}$, we have
\be
\label{Omega}
\Omega \, = \, \frac{\rho}{\rho_c} \,  \sim \, \left( \frac{T_\RH \,
N_\UV^{1/2}}{6 \cdot 10^{11} \, 
\text{GeV}} \right)^4 \, \left( \frac{T_\pt}{T_\RH} \right) \,.
\ee
This is the contribution of the gauge theory sector to the density
parameter. The second factor is smaller than or equal to 1 and again
reflects the fact that the corresponding energy density undergoes a phase
of $\smash{a^{-4}}$ dilution.

Let us now consider the case that the initial energy density in the gauge
theory sector, Eq.~\eqref{energydensity}, is smaller than $\rho \sim N_\IR
m_\IR^4$. Recall that the energy density is due to standard model
particles which annihilate via a KK mode of supergravity fields in the
embedding space into gauge theory states (cf.~Fig.~\ref{scat}). This is
similar to electron-positron annihilation via a photon into a quark and an
antiquark. In QCD, these partons subsequently radiate other partons in a
narrow cone along their trajectories, leading to two jets. The hadrons which
are formed in this process are ultrarelativistic and would initially scale
like radiation with $a^{-4}$ with the expansion of the universe. In
contrast, in the large-$N$, large-$\lambda$ gauge theories which are dual
to throats, no jets are
formed~\cite{Hofman:2008ar,Hatta:2008tx,Hatta:2008tn} (see
also \cite{Harling:2008px,Strassler:2008bv}). Instead, if the energy
density is below the critical energy density, the initial gauge theory
states hadronize to slow and light glueballs, i.e.~with masses and kinetic
energies $\sim m_\IR$. This is not surprising since the gauge theory is
strongly coupled on all scales.\footnote{The calculations
in~\cite{Hofman:2008ar,Hatta:2008tx,Hatta:2008tn} were mainly performed on
the gravity side of the duality.} Highly energetic partons which result
from the annihilation of standard model particles will therefore split into
more and more partons until the individual
energy of the partons reaches the confinement scale~\cite{Hatta:2008tx}.
This leads to the
aforementioned distribution of light and slow glueballs. These glueballs
immediately become nonrelativistic and then scale like matter with $a^{-3}$
with the expansion of the universe.

Note that Refs.~\cite{Hofman:2008ar,Hatta:2008tx,Hatta:2008tn}, in which
the hadronization in large-$N$, large-$\lambda$ gauge theories was
analysed in great detail, appeared after Ref.~\cite{Harling:2008px}, on which
this chapter is based. Ref.~\cite{Harling:2008px} additionally
analyzed the evolution of the energy density for the case that jets are
formed in gauge theories which are dual to throats. This analysis is now
obsolete.

Taking this scaling into account, the mass density over entropy density $m
\cdot \eta$ and the density parameter $\Omega$ are given by
\be
\label{Omega2}
m \cdot \eta_\maxx  \, \sim \,   \frac{N^2_\UV}{g^{1/2}}
\frac{T_\RH^4}{M_4^3} \,,
\qquad \qquad \Omega_\maxx \, = \,  \left( \frac{T_\RH \,
N_\UV^{1/2}}{6 \cdot 10^{11} \, 
\text{GeV}} \right)^4 \,.
\ee
Note that these expressions are just those in Eqs.~\eqref{energydensity2} and
\eqref{Omega} with $T_\pt$ replaced by $T_\RH$. This reflects the
fact that the energy density now scales like matter from the beginning
on. As a function of $m_\IR$, $m \cdot \eta$ is thus constant in this
case. Note, though, that throats with $m_\IR > T_\RH$ are not heated for
kinematic reasons. The mass density over entropy density $m \cdot
\eta$ is therefore zero in this region. Again, we have plotted the
corresponding parts of $m \cdot \eta$ as a function of $m_\IR$ in
Fig.~\ref{fig:plot}.

\chapter{Sequestered Dark Matter: Cosmological scenarios}
\label{scenarios}
\section{A single throat}
\label{SingleThroat}
Glueballs are an interesting dark matter candidate if they are stable until
our epoch. In this section, we analyse the prospects of this new kind of
dark matter for a compactification which has only a single
throat region. Setups with a large number of throats are discussed in
Section~\ref{ManyThroats}.

As one can see from Fig.~\ref{fig:plot}, the glueball mass density over
entropy density $m\cdot\eta$ or, equivalently, the contribution of glueballs
to the density
parameter $\Omega$ is maximized for throats with IR scales in the
range
$m_\IRc\hspace{-.05cm}\lesssim\hspace{-.05cm}m_\IR\hspace{-.05cm}
\lesssim\hspace{-.02cm}T_\RH$. This is therefore an
interesting region to look for glueball dark matter. 

Let us first consider throats at the lower end of the `optimal' interval,
i.e.~with IR scales of the order of $m_\IRc$. For this IR scale, the initial
energy density in the throat, Eq.~\eqref{energydensity}, is just the
critical energy density $\smash{\rho
\sim N_\IR m_\IR^4}$ of that throat. Solving for $m_\IR$, we find
\be
\label{mass}
m_\IRc \, \sim \, \left( \frac{  N_\UV^2 \, T_\RH^7}{N_\IR \, M_4^3} 
\right)^{1/4} \,,
\ee
where we have neglected a factor of $\smash{g^{1/8}}$ which is 
$\mathcal{O}(1)$. 

In order to evaluate Eq.~\eqref{mass} as well as other
relevant equations, we have to fix $N_\IR$ and $N_\UV$. These numbers
determine the warp factor which in turn is related to the IR scale of the
throat:\footnote{For the warp factor, we have used Eq.~\eqref{hierarchy}
and the fact that $N_\UV \sim K M$ and $N_\IR \sim g_s M^2$, where $K$ and
$M$ are the relevant numbers of $H_3$-flux and $F_3$-flux, respectively.
The mass quantization follows from Eqs.~\eqref{massesKK} and \eqref{RN} and
the fact that the masses of light KK modes are determined by the geometry
near the IR end of the throat.}
\be
\label{hm}
h \, \sim \, e^{2 \pi N_\UV / 3 N_\IR}  \qquad \qquad m_\IR \, \sim \,
h^{-1} N_\IR^{-1/4} M_{10} \,.
\ee
To simplify the discussion and to avoid uncertainties associated with
unknown factors of $N_\IR$ in the various glueball decay rates, we focus on
throats where $N_\IR=\mathcal{O}(1)$. In this case, $N_\UV$ is a function
of $m_\IR$ and $M_{10}$. For our purposes it will be sufficient to use the
typical values $N_\UV \approx 10$ for long throats (e.g. for $m_\IR\sim
10^6\hspace{+.09cm}\text{GeV}$ and $M_{10}\sim
10^{15}\hspace{+.09cm}\text{GeV}$) and $N_\UV \approx 4$ for shorter
throats (e.g. for
$m_\IR\hspace{-.07cm}\sim\hspace{-.07cm}10^{11}\hspace{+.09cm}\text{GeV}$).

Inserting $N_\UV \approx 10$ and $N_\IR \approx 1$ in Eq.~\eqref{Omega2},
we can determine the reheating temperature which leads to the right amount
of dark matter, i.e.~$\Omega \sim 1$. The mass of the dark matter candidate
subsequently follows from Eq.~\eqref{mass}. We find 
\be
\label{massglueball}
T_\RH \, \sim \, 10^{11}\hspace{+.09cm}\text{GeV} \qquad \text{and} \qquad
m_\IR \, \sim \,
10^6\hspace{+.09cm}\text{GeV} \,.
\ee
Thus, for a reheating temperature $\smash{\sim
10^{11}}\hspace{+.1cm}\text{GeV}$ and a throat
with IR scale $\smash{\sim\hspace{-.04cm}10^6}\hspace{+.09cm}\text{GeV}$, we
get the right amount of glueballs
to explain the observed dark matter. 

In the late universe, this kind of dark matter consists solely of the
lightest scalar glueball and its spin-$\smash{\frac{1}{2}}$ superpartner,
due to the processes discussed in Section~\ref{processes}. Let us consider
the fermionic glueballs for the moment. For reason that we have discussed in
Section~\ref{intro}, here and below, we focus on setups in which the gravitino
is much heavier than the glueballs,
$m_{3/2}\hspace{-.03cm}\gg\hspace{-.03cm}m_\IR$. 
The decay of spin-$\smash{\frac{1}{2}}$ glueballs to a
graviton and a gravitino is then kinematically forbidden. If the coupling
in Eq.~\eqref{vertex3} is present, the spin-$\smash{\frac{1}{2}}$ glueballs
decay to the standard model with the rate given in Eq.~\eqref{fermionicdr}.
The resulting lifetime depends on the coupling strength $\lambda$:
\be
\label{lt1}
\tau \, \sim \, 10^{26} \left(  \frac{  M_{10} \cdot \lambda^{-1/4}}{2 \cdot 
10^{16} \, \text{GeV}} \right)^8 \text{ s} \,.
\ee
It is not sufficient to make this lifetime just longer than the present age of
the universe (which is $\sim\hspace{-.07cm}10^{17}\hspace{+.09cm}\text{s}$):
The glueball decays produce
photons (e.g.~via hadronic showers) which contribute with a continuous
spectrum to the diffuse $\gamma$-radiation. The $\gamma$-ray flux
measured e.g.~by the EGRET experiment gives constraints on the lifetime of
unstable particles in dependence of their
mass density
(see e.g.~\cite{KribsRothstein,UP1,Bertone:2007aw,DecayingDarkMatter,
DecayingDarkMatter2}).
In particular, an unstable particle with the mass density of dark matter
has to live longer than $\sim\hspace{-.07cm}10^{26}\hspace{+.09cm}\text{s}$ to
comply with
observations \cite{KribsRothstein}.\footnote{Here we have used that the
hadronic branching ratio for decays via the coupling in Eq.~\eqref{vertex3}
is $\mathcal{O}(1)$. For decays exclusively to photons or leptons, the
constraints are less severe.} Thus, it depends on the two unknown
parameters $M_{10}$ and $\lambda$ whether the spin-$\smash{\frac{1}{2}}$
glueballs are a good dark matter candidate or not.

An interesting scenario is to have $\lambda=\mathcal{O}(1)$.\footnote{
The coupling in Eq.~\eqref{vertex3} leads to a mixing between the modulino
$\psi$ and a left-handed neutrino $\nu$ after electroweak symmetry
breaking. Since the modulino has a large mass $m_\tau$, the seesaw
mechanism results in a light mass eigenstate. For $M_{10} \sim
10^{16}\hspace{+.09cm}\text{GeV}$
(the minimal value allowed for $\lambda \sim 1$ according to
Eq.~\eqref{lt1}), Eq.~\eqref{mtau} gives $m_\tau \sim
10^{14}\hspace{+.09cm}\text{GeV}$. Using
this value and the mixing mass term for $\lambda \sim 1$ in the seesaw
formula, the resulting neutrino mass is $\sim 0.1$ eV. Interestingly, this
is precisely the mass range indicated by various experiments.
}  
To get a viable dark matter candidate, the 10d Planck scale has a rather
limited range in this case according to Eq.~\eqref{lt1}. This makes it more
probable that the lifetime of the glueballs is in a range that can be probed
by new $\gamma$-ray telescopes like GLAST.
If this scenario with $\lambda=\mathcal{O}(1)$ is realized in nature, one may
be able to see a signal in the near future.

The scalar glueballs from a throat with IR scale
$10^6\hspace{+.09cm}\text{GeV}$ decay to
gravitons after approximately $ 10^{15}\hspace{+.09cm}\text{s}$ according to
Eq.~\eqref{dr3}.
Note, however, that this lifetime is proportional to $\smash{m_\IR^{-5}}$.
The lifetime will thus be somewhat larger or smaller for IR scales slightly
different from $\smash{10^6\hspace{+.09cm}\text{GeV}}$. If the lifetime is in
the range of
$10^{17}\hspace{+.09cm}\text{s}$ (the present age of the universe) to
$10^{12}\hspace{+.09cm}\text{s}$ (the time of
matter-radiation equality), the resulting decrease in the dark matter density
may have interesting observable consequences. Note that, if the initial energy
density was slightly above the critical energy density, a gauge theory
with confinement
scale of the order of $m_\IRc$ was thermalized in the early universe. In
this case, part of the scalar glueballs (which we expect to be heavier than
their superpartners for $m_{3/2} \gg m_\IR$, cf.~Section~\ref{spectrum})
annihilate into their superpartners after the phase transition. The
remaining abundance of scalar glueballs is the freeze-out abundance for
this process which we determine in Appendix~\ref{additional}. 
Inserting the above values for $N_\UV$, $m_\IR$
and $T_\RH$ into the relation for the depletion
factor, Eq.~\eqref{dilutionfactor}, we see that the scalar glueballs
make up for only $\smash{10^{-2}}$ of the total dark matter abundance in
this case. The loss of mass density due to their decay is correspondingly
small. If the gauge theory was never thermalized, on the other hand, the
initial abundance of scalar glueballs and fermionic glueballs is equal and
the dark matter
mass density is halved due to the decay of
the scalar glueballs. It would be interesting to see whether such a large
decrease of the dark matter abundance may still be allowed by observations
and which observable consequences (e.g.~on structure formation) it might
have.

If the lifetime of scalar glueballs is large enough, a non-negligible
amount still exists at our epoch. Decays of the glueballs to the standard
model again produce $\gamma$-rays. The corresponding decay rate
is given by Eq.~\eqref{DecayRate3} for $N_1^2=g$ and $N_2=N_\UV$
(cf.~Section~\ref{DecaytoSM}). For
$m_\IR\hspace{-.07cm}\sim\hspace{-.07cm}10^6\hspace{+.09cm}\text{GeV}$, the
partial
lifetime of scalar glueballs with respect to decays to the standard model
is
\be
\label{lt5}
\tau \, \sim \, 10^{26} \, \left( \frac{M_{10}}{3 \cdot
10^{13}\hspace{+.09cm}\text{GeV}} 
\right)^8 \text{ s} \,.
\ee
If the scalar glueballs still make up an $\mathcal{O}(1)$ fraction of the
dark matter at our epoch, this partial lifetime has to be larger than
$\smash{\sim\hspace{-.07cm}10^{26}\hspace{+.09cm}\text{s}}$ to comply with
the EGRET measurements. If the current
abundance of scalar glueballs is reduced (by decays to gravitons or by
annihilation in a thermalized situation), the lower bound on the lifetime
becomes correspondingly weaker. 

In contrast to fermionic glueballs, scalar glueballs can decay directly to
two photons. Decays via this channel in the halo of our galaxy lead to a sharp
line in the $\gamma$-ray spectrum. The $\gamma$-rays at 
energies around $\smash{ 10^6\hspace{+.09cm}\text{GeV}}$ cannot be measured by
satellite-based
experiments like EGRET or GLAST. Ground-based $\gamma$-ray telescopes like
HESS have the necessary energy range, but a limited sensitivity due to the
cosmic ray background. At $10^6\hspace{+.09cm}\text{GeV}$, the measured flux
in the cosmic ray
spectrum is (see e.g.~\cite{Horandel:2004bd})
\be
F \, \sim \,   10^{-12} \, ( \text{m}^2 \text{ sr s GeV})^{-1} \,.
\ee
To be detectable against this background, the flux from the decaying
glueballs in the halo has to have the same order of magnitude. This flux is
emitted as a sharp line at energy $ m_\IR$ but smeared out by the
detector due to a finite energy resolution $\Delta E$. We model this effect by
replacing the $\delta$-function peak of the flux by a box of width $\Delta E$.
The flux is also inversely proportional to the mass $m_\IR$ and the lifetime
$\tau$ of the glueballs. Assuming that the scalar glueballs make up an
$\mathcal{O}(1)$ fraction of the dark matter at our epoch, we find (see
e.g.~\cite{Bertone:2007aw})
\be
F \, \sim \,  10^{-12} \, \left( \frac{10^5\hspace{+.09cm}\text{GeV}}{\Delta
E}
\right) \, \left( \frac{10^6\hspace{+.09cm}\text{GeV}}{m_\IR} \right) \,
\left( 
\frac{10^{26}\text{ s}}{\tau} \right)\, ( \text{m}^2 \text{ sr s GeV})^{-1}
\,.
\ee
For $m_\IR\hspace{-.04cm}\sim\hspace{-.04cm}10^6\hspace{+.09cm}\text{GeV}$ and
$\Delta E\hspace{-.03cm}\sim\hspace{-.03cm}10^{-1}
\hspace{-.03cm}\cdot\hspace{-.01cm}E\hspace{-.03cm}\sim\hspace{-.03cm}
10^5\hspace{+.09cm} \text{GeV}$
as quoted by the HESS collaboration, the partial lifetime of scalar
glueballs (for decays to the standard model) has to be less than
$\sim\hspace{-.07cm}10^{26}\hspace{+.09cm}\text{s}$ to be detectable against
the cosmic ray background. 
If the partial lifetime is somewhat larger, the $\gamma$-line may nevertheless
become detectable in the near future with an improved rejection of cosmic ray
events and a better sensitivity and energy resolution.

In summary, if an $\mathcal{O}(1)$ fraction of the dark matter at our epoch
are scalar glueballs and if their partial lifetime is not much larger than
$ 10^{26}$ s, two experiments may see a signal: The contribution of
glueball decays to the $\gamma$-ray spectrum below
$10^2\hspace{+.09cm}\text{GeV}$ may be
detected by GLAST. Moreover, the $\gamma$-line near
$10^6\hspace{+.09cm}\text{GeV}$ may be
seen by HESS. A lifetime of the order of $ 10^{26}\hspace{+.09cm}\text{s}$
follows if $M_{10}
\sim\hspace{-.07cm}10^{13}\hspace{+.09cm}\text{GeV}$ according to
Eq.~\eqref{lt5}. Such a low 10d Planck scale
may be realized in a large-volume compactification along the lines
of~\cite{largevolume}. Note that this scenario is incompatible with the
aforementioned scenario in which $\lambda=\mathcal{O}(1)$: According to
Eq.~\eqref{lt1}, $\lambda$ has to be very small (or zero) for such a low 10d
Planck scale.

Throats with IR scales smaller than $\smash{10^6}\hspace{+.09cm}\text{GeV}$
also lead to an
interesting dark matter candidate. According to Eqs.~\eqref{Tpt} and
\eqref{Omega}, $\Omega$ is proportional to $\smash{T_\RH^{9/4} m_\IR}$. In
order still to have the abundance of dark matter with $\Omega \sim 1$, we have
to increase the reheating temperature as $\smash{T_\RH \propto m_\IR^{-4/9}}$
if we lower the IR scale. For example, for a throat with IR scale
$\smash{10^4\hspace{+.09cm}\text{GeV}}$, a reheating temperature of
$\smash{10^{12\hspace{+.09cm}\text{GeV}}}$ would
give the right abundance (again assuming that $N_\UV\approx 10$). Since the
various glueball decay rates are proportional to $m_\IR$ to some positive
power, the glueballs become more stable for lower IR scales.

Let us now analyse throats with IR scales still in the `optimal'
interval between $m_\IRc$ and $T_\RH$ but much larger than the critical
value $m_\IRc$. For definiteness, we consider a throat with IR scale $\sim
T_\RH$ and take $N_\UV \approx 4$ in order to have
$N_\IR = \mathcal{O}(1)$. According to Eq.~\eqref{Omega2}, such a throat
again gives the right amount of dark matter for a reheating temperature
$\smash{ \sim\hspace{-.07cm}10^{11}\hspace{+.09cm}\text{GeV}}$. The mass of
this dark matter candidate
correspondingly is $\smash{ \sim
10^{11}\hspace{+.09cm}\text{GeV}}$.\footnote{Note that the
glueballs are never in thermal equilibrium for such short throats.
Therefore, the heavier superpartners do not annihilate into the lightest
glueball states and the initial abundance of scalar and
spin-$\smash{\frac{1}{2}}$ glueballs is equal.}
The scalar glueballs decay to gravitons already after $\smash{10^{-8}}$
s, according to Eq.~\eqref{dr3}. If the coupling in Eq.~\eqref{vertex3} is
present, the spin-$\smash{\frac{1}{2}}$ glueballs decay to the standard
model
after 
\be
\label{llt1}
\tau \, \sim \, 10^{27} \left(  \frac{  M_{10} \cdot \lambda^{-1/4} }{ 5
\cdot 
10^{20}\hspace{+.09cm}\text{GeV}}\right)^8 \text{ s} \,. 
\ee
Hadronic decays of particles in this mass range have been considered 
in~\cite{Birkel:1998nx} to explain events in the cosmic ray spectrum
beyond the GZK cutoff. Taking the measured flux in this energy range, claimed
by several collaborations, as an upper limit, a lifetime of at
least $10^{27}\hspace{+.09cm}\text{s}$ is required for a particle with mass
$10^{11}\hspace{+.09cm}\text{GeV}$. If
$\lambda = \mathcal{O}(1)$, the spin-$\smash{\frac{1}{2}}$ glueballs decay
too quickly since $M_{10}$ cannot be larger than
$M_4 \simeq 2 \cdot 10^{18}\hspace{+.09cm}\text{GeV}$. The coupling $\lambda$
can be much
smaller, though, and the spin-$\smash{\frac{1}{2}}$ glueballs may be
sufficiently stable for large enough $M_{10}$. 

Finally, let us comment on throats with higher 5-form flux numbers $N_\IR$
and $N_\UV$. Since $N_\IR$ is no longer of the order 1, we have only rough 
estimates of the glueball decay rates in these cases. In the following, we
ignore this issue and assume that the glueballs are sufficiently stable.
The number $N_\UV$ is constrained by the requirement that the Calabi-Yau
orientifold has enough negative charge to compensate for the flux. If one
considers the
orientifold limit of an F-theory compactification, then this amount of
negative charge is given by $\chi_4 /24$, where $\chi_4$ is the Euler number
of the underlying Calabi-Yau fourfold. Examples with $\chi_4 /24$ up to
$\smash{10^4}$ are known (see e.g.~\cite{Klemm:1996ts}) and we therefore
assume that $\smash{N_\UV\hspace{-.03cm}\lesssim\hspace{-.03cm}10^4}$.

It follows from Eq.~\eqref{Omega2} that throats with maximal
$N_\UV\hspace{-.07cm}\sim\hspace{-.07cm}10^4$ and with IR
scales in the range $m_\IRc \lesssim m_\IR \lesssim T_\RH$ can account for
the observed dark matter if the reheating temperature was $\sim
10^{10}\hspace{+.09cm}\text{GeV}$.
The mass of these dark matter candidates is between $\sim
10^5\hspace{+.09cm}\text{GeV}$ (using
Eq.~\eqref{mass}) and $\sim\hspace{-.07cm}10^{10}\hspace{+.09cm}\text{GeV}$.
Together with the results from the
first part of this section (where we have chosen the other extreme with $N_\IR
= \mathcal{O}(1)$) this gives the possible range of parameters in our scenario
if the 5-form flux number is varied from its minimal to its maximal value:
For a throat in the `optimal' range $m_\IRc \lesssim m_\IR \lesssim T_\RH$ to
account for the observed dark matter, the required reheating temperature is
between $10^{10}\hspace{+.09cm}\text{GeV}$ and
$10^{11}\hspace{+.09cm}\text{GeV}$. The IR scale and thus the
mass of the corresponding dark matter particles can vary between
$10^5\hspace{+.09cm}\text{GeV}$
and $10^{11}\hspace{+.09cm}\text{GeV}$.

\section{Many throats}
\label{ManyThroats}
As we have discussed in Section~\ref{statistics}, typical vacua in
the type IIB string theory landscape can have a large number
of throats. An estimate of the expected number of throats with a hierarchy $h$
larger than some $h_*$ for a Calabi-Yau orientifold with $K$ 3-cycles was
given in Eq.~\eqref{expectation}. Using this estimate and Eq.~\eqref{hm} and
neglecting a factor $\smash{N_\IR^{-1/4}}$ for simplicity, the expected number
of throats with IR scale in the range $\smash{m_\minn < m_\IR < m_\maxx}$
follows as
\be
\label{distribution}
\bar{n} (m_\minn < m_\IR < m_\maxx ) =  \frac{K/3}{\log
(M_{10}/ 
m_\maxx)} - \frac{K/3 }{\log (M_{10}/ m_\minn)} \,.
\ee
Here, we have assumed that $c=1$.

The function $m \cdot \eta$ (and correspondingly $\Omega$) is maximal for
throats with IR scales between $m_\IRc$ and $T_\RH$
(cf.~Fig.~\ref{fig:plot}). If many throats are present, we can expect that
the observed dark matter (or at least the dominant throat contribution to dark
matter) will come from throats with an IR scale in this region. As before, we
simplify the analysis by using $N_\UV \approx 10$ for long throats
($m_\IR\hspace{-.07cm}\sim\hspace{-.07cm}10^6\hspace{+.09cm}\text{GeV}$) and
$N_\UV \approx 4$ for short throats
($m_\IR\hspace{-.07cm}\sim\hspace{-.07cm}10^{11}\hspace{+.09cm}\text{GeV}$).
As we have discussed in the last section, a reheating
temperature of the order of $\smash{10^{11}\hspace{+.09cm}\text{GeV}}$ then
leads to the observed
amount of dark matter for all throats with IR scales between $m_\IRc \sim
10^6\hspace{+.09cm}\text{GeV}$ and $T_\RH \sim
10^{11}\hspace{+.09cm}\text{GeV}$. Note that, in general, the situation
might be more complicated: For example, a throat with an IR scale smaller than
the critical value $m_\IRc$ but with very large $N_\UV$ (recall that $N_\UV$
is fairly arbitrary if we do not insist on $N_\IR = \mathcal{O}(1)$) may
provide the dominant contribution to dark matter (cf.~Eqs.~\eqref{Omega} and
\eqref{Omega2}).

It is clear from Eq.~\eqref{distribution} that the expected number of
throats with IR scales in the aforementioned region grows with
the number of 3-cycles $K$. Moreover, one can easily check that this number
of throats also becomes larger for smaller 10d Planck scales $M_{10}$. We
begin the evaluation of Eq.~\eqref{distribution} with an optimistic
scenario in which
$K=200$ and
$M_{10}\hspace{-.07cm}\sim\hspace{-.07cm}10^{14}\hspace{+.09cm}\text{GeV}$.
The number of 3-cycles is a
moderately high but not untypical
value within the set of known Calabi-Yau
spaces~\cite{K}. The value of the 10d Planck scale, on the other hand, is
roughly the minimal value for which our analysis is valid: In deriving the
heat transfer rates in Eqs.~\eqref{el} and \eqref{etr}, we have assumed that
the reheating temperature is smaller than the compactification scale,
i.e.~$\smash{T_\RH < L^{-1}}$. Using Eq.~\eqref{PlanckScale}, we find that
the compactification scale $\smash{L^{-1} \sim
10^{12}}\hspace{+.09cm}\text{GeV}$ corresponds to the 10d Planck
scale
$M_{10}\hspace{-.04cm}\sim\hspace{-.04cm}10^{14}\hspace{+.09cm}\text{GeV} $.
Thus, since we consider a reheating
temperature of the order of $10^{11}\hspace{+.09cm}\text{GeV}$, the assumption
would no longer be
fulfilled for a much lower 10d Planck scale.

For the aforementioned values of $K$ and
$M_{10}$, the expected number of throats with IR
scales in the range $10^6\hspace{+.09cm}\text{GeV}\hspace{-.03cm}\lesssim
\hspace{-.04cm}m_\IR\hspace{-.04cm}\lesssim\hspace{-.04cm}10^{11}\hspace{
+.09cm}\text{GeV}$ follows
as\footnote{More precisely, we have evaluated Eq.~\eqref{distribution} for the
range
$5\hspace{-.04cm}\cdot\hspace{-.04cm}10^5\hspace{+.09cm}\text{GeV}\hspace{
-.04cm}<\hspace{-.04cm}m_\IR\hspace{-.04cm}<\hspace{-.04cm}5\hspace{-.04cm}
\cdot\hspace{-.04cm}10^{11}\hspace{+.09cm}\text{GeV}$. The mean number of
throats in other ranges
of IR scales
have been calculated in a similar way.}
\be
\label{number}
\bar{n} \left( 10^6\hspace{+.09cm}\text{GeV} \lesssim m_\IR \lesssim
10^{11}\hspace{+.09cm}\text{GeV} \right) \, \simeq \, 9.1 \,.
\ee
For definiteness, we restrict ourselves to setups which have a throat
with IR scale of the order of $\hspace{-.07cm}10^6\hspace{+.09cm}\text{GeV}$.
This is the case in a large fraction of
models, since the mean number of throats with this IR scale
is\footnote{We have used the interval $5 \cdot 10^5\hspace{+.09cm}\text{GeV} <
m_\IR <
5 \cdot 10^6\hspace{+.09cm}\text{GeV}$ to estimate the number of throats with
IR
scale $\sim\hspace{-.07cm}10^6\hspace{+.09cm}\text{GeV}$ from
Eq.~\eqref{distribution}.}
\be
\bar{n} \left(
m_\IR\hspace{-.07cm}\sim\hspace{-.07cm}10^6\hspace{+.09cm}\text{GeV}
\right) \,
\simeq \, 0.5 \,.
\ee
Certain partial lifetimes of glueballs with mass $\sim
10^6\hspace{+.09cm}\text{GeV}$ have been
determined in the last section. In particular, we see from
Eq.~\eqref{lt5} (with $M_{10}\sim10^{14}\hspace{+.09cm}\text{GeV}$) that the
scalar glueballs may
lead to interesting observable signatures if they do not decay too quickly to
two gravitons. The fermionic glueballs cannot decay to a graviton
and a gravitino for kinematic reasons since we still assume that $m_{3/2}
\gg m_\IR$. They can therefore account for the
observed dark matter even if the scalar glueballs decay to gravitons already
at an early epoch. According to Eq.~\eqref{lt1} (with $M_{10}\sim10^{14}$
GeV), the fermionic glueballs must not couple too strongly to the standard
model sector, i.e.~$\lambda$ must not be too large. In addition, we now
expect
\be
\bar{n} \left(
m_\IR\hspace{-.01cm}\lesssim\hspace{-.01cm}10^5\hspace{+.12cm}\text{GeV}
\right) \, \simeq \, 3.5
\ee
throats with IR scales smaller than $\smash{10^6\hspace{+.09cm}\text{GeV}}$
which provide another 
decay channel for the glueballs. These throats can have a large
number $\tilde{g}$ of degrees of freedom. We want the glueballs from the
throat at $\sim\hspace{-.07cm}10^6\hspace{+.09cm}\text{GeV}$ to account for
the observed dark matter and we
therefore have to check whether their lifetime is still larger than the
current age of the universe. If we denote the 5-form flux number at the UV
end of the $i$th throat by $N_i$, we have
\be
\label{dof}
\tilde{g} \, = \,  \sum_i N^2_i \,,
\ee
where the sum runs over all throats with IR scales smaller than 
$\smash{10^6}\hspace{+.09cm}\text{GeV}$. Using Eq.~\eqref{DecayRate3} with
$N^2 =
\tilde{g}$, the partial lifetime of the dark matter glueballs for decays to
these throats is
\be
\tau \, \sim \, \tilde{g}^{-1} \, 10^{32} \text{ s} \,.
\ee
According to the discussion at the end of the last section, we
expect $\tilde{g}$ to be somewhere in the range of $\smash{10^2}$ to
$\smash{10^8}$. Even for maximal $\tilde{g}$ this partial lifetime
is thus much larger than the current age of the universe and the glueballs
are still a good dark matter candidate. 

There are also throats with IR scales larger than the IR scale
of the dark matter throat (cf.~Eq.~\eqref{number}). Glueballs from these
throats have shorter lifetimes than the dark matter glueballs. The
abundance of particles which decay to the standard model with a lifetime
in the range of $\smash{10^{-2}}\hspace{+.09cm}\text{s}$ and
$\smash{10^{12}}\hspace{+.09cm}\text{s}$ is
severely constrained by nucleosynthesis~\cite{UP2,UP1,Kawasaki:2004qu}. For
lifetimes larger than $10^{12}$ s, bounds from the diffuse $\gamma$-radiation
are again important~\cite{KribsRothstein}.
Therefore, we have to check whether the decaying glueballs fulfill these
observational constraints. 

\begin{figure}[t]
\begin{center}
\includegraphics[scale=0.3]{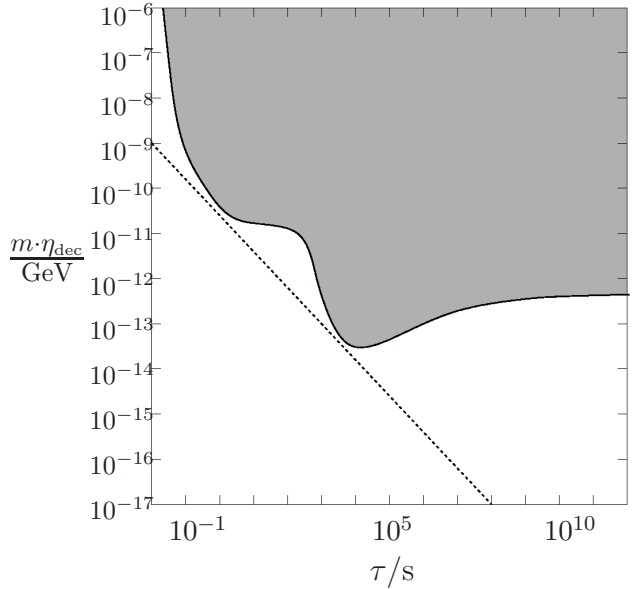}  \put(-236,90){\large $\frac{m \cdot
\eta_\dec}{\text{GeV}}$} 
\put(-201,182){\footnotesize $10^{-6}$}
\put(-201,165){\footnotesize $10^{-7}$}
\put(-201,148){\footnotesize $10^{-8}$}
\put(-201,131){\footnotesize $10^{-9}$}
\put(-205,114){\footnotesize $10^{-10}$}
\put(-205,97){\footnotesize $10^{-11}$}
\put(-205,80){\footnotesize $10^{-12}$}
\put(-205,63){\footnotesize $10^{-13}$} \put(-205,46){\footnotesize
$10^{-14}$} \put(-205,29){\footnotesize $10^{-15}$}
\put(-205,12){\footnotesize $10^{-16}$} \put(-205,-5){\footnotesize
$10^{-17}$}
\put(-99,-28){$\tau /\text{s}$} 
\put(-33,-13){\small $10^{10}$} 
\put(-98,-13){\small $10^5$} 
\put(-175,-13){\small $10^{-1}$} 
\vspace*{.5cm} 
\caption{Nucleosynthesis constraints on the mass density $m
\cdot \eta_\dec$ of the fraction of particles that have
decayed to the standard model in dependence of their lifetime $\tau$ (the
shaded region is excluded, schematic
plot based on results from \cite{Kawasaki:2004qu}). The
corresponding function for the glueballs is also shown (the dotted
curve). See text for more details.\label{nu}}
\end{center}
\end{figure}

We analyse only decays of scalar glueballs. The discussion can be easily
extended to include the fermionic glueballs. Since fermionic glueballs
decay to the standard model via the operator in Eq.~\eqref{vertex3},
nucleosynthesis or the diffuse $\gamma$-radiation may give a bound on the
coupling strength $\lambda$. 
Furthermore, we assume that the aforementioned throats have 
IR scales of at least $10^7\hspace{+.09cm}\text{GeV}$. One can check that the
corresponding
glueballs decay already after $10^{10}\hspace{+.09cm}\text{s}$ or earlier.
These lifetimes are too
short (i.e.~shorter than $10^{12}\hspace{+.09cm}\text{s}$) to give relevant
constraints from the
diffuse $\gamma$-radiation. We will therefore restrict our analysis to bounds
from nucleosynthesis. Scalar glueballs have three important decay
channels: They decay to gravitons, to throats with lower IR scales and to
the standard model. The total decay rate is the sum of the three
corresponding decay rates:\footnote{In deriving the decay rate in
Eq.~\eqref{DecayRate3}, we have assumed that the mass $m_\tau$ of the
modulus which mediates the decay is larger than the mass $m_\IR$ of the
decaying glueball. According to Eq.~\eqref{mtau}, we have $m_\tau \sim
10^{10}\hspace{+.09cm}\text{GeV}$ for $M_{10} \sim
10^{14}\hspace{+.09cm}\text{GeV}$. For a throat with IR scale
larger than $10^{10}\hspace{+.09cm}\text{GeV}$ we would therefore have to use
the unsuppressed
decay rate in Eq.~\eqref{DecayRate} instead of Eq.~\eqref{DecayRate3}. We
will not have to consider such heavy glueballs in the following.}
\be
\label{totaldr}
\Gamma_{\rm \text{total}} \, \sim \, \tilde{g} \, N_\UV^2 \frac{
m_\IR^9}{M_{10}^8}+g \, N_\UV^2 \frac{ m_\IR^9}{M_{10}^8} + N_\UV^4 \frac{
m_\IR^5}{M_4^4} \,.
\ee
Here, $\tilde{g}$ is the combined number of degrees of freedom of all the
throats with IR scales smaller than the IR scale $m_\IR$ of a given throat.

Since $\tilde{g} \gtrsim g$, we can neglect the decay rate to the standard
model (the $g$-dependent term) in Eq.~\eqref{totaldr}. Furthermore, we can
see from Eq.~\eqref{totaldr} that for sufficiently large glueball lifetimes
(i.e.~small
$m_\IR$), the total decay rate is dominated by decays to gravitons
(the last term in Eq.~\eqref{totaldr}). To be conservative, we take
$\tilde{g}\hspace{-.07cm}\sim\hspace{-.07cm}10^2$ in the following. Higher
values of $\tilde{g}$
would give a smaller branching ratio for decays to the standard model and
would thus make the constraints from nucleosynthesis easier to fulfill. One
can then check that, for
$M_{10}\hspace{-.07cm}\sim\hspace{-.07cm}10^{14}\hspace{+.09cm}\text{GeV}$ and
$N_\UV \approx 4$
to $10$, the decay rate to gravitons dominates for glueballs which live
longer than $10^{-2}$ s. Since nucleosynthesis gives no constraints for
particles which decay earlier than $10^{-2}$ s, we can restrict our
analysis to these glueballs. We thus have
\be
\label{totaldr2}
\Gamma_{\rm \text{total}} \, \sim \,  N_\UV^4 \frac{
m_\IR^5}{M_4^4} \,.
\ee

We denote the mass density over entropy density of the fraction of
glueballs
that have decayed to the standard model by $ m\cdot \eta_\dec$ and the
branching ratio for decays to the standard model by $br$:
\be
m \cdot \eta_\dec \, = \, br \cdot  m \cdot \eta \,.
\ee
Dividing the decay rate to the standard model by the total decay rate in
Eq.~\eqref{totaldr2}, we find
\be
br \, = \, \frac{\Gamma_{\rm \text{SM}}}{\Gamma_{\rm \text{total}}} \,
\sim
\, \frac{g\, M_4^4 \, m_\IR^4} {N_\UV^2\, M_{10}^8} \,.
\ee
Using again Eq.~\eqref{totaldr2}, this branching ratio and correspondingly $m
\cdot \eta_\dec$ can be expressed as a function of the lifetime $\smash{\tau =
\Gamma_{\rm \text{total}}^{-1}}$. Furthermore, we know from
Fig.~\ref{fig:plot}
that the decaying glueballs initially have the same mass density as dark
matter, i.e.~$\smash{m \cdot \eta \approx 10^{-9}}\hspace{+.09cm}\text{GeV}$. 

In Fig.~\ref{nu}, we have plotted $m \cdot \eta_\dec$ as a function of
$\tau$ as a dotted curve. The shaded region in this plot is the parameter
space that is excluded by the requirement of successful nucleosynthesis. Note,
however,
that these constraints were calculated in \cite{Kawasaki:2004qu} for a
decaying particle of mass $10^4\hspace{+.09cm}\text{GeV}$ (and with an
$\mathcal{O}(1)$ hadronic
branching ratio). Using
Eq.~\eqref{totaldr2}, the glueball masses which correspond to
the range of lifetimes in Fig.~\ref{nu} are between
$10^7\hspace{+.09cm}\text{GeV}$ and
$10^9\hspace{+.09cm}\text{GeV}$. This is much heavier than the particle mass
$10^4\hspace{+.09cm}\text{GeV}$ for
which the constraints were actually determined. On the other hand, in
\cite{Kawasaki:2004qu},
constraints were also calculated for lighter particles with masses $10^2$
GeV and $10^3\hspace{+.09cm}\text{GeV}$. The results differ only slightly from
the constraints
for particle mass $10^4\hspace{+.09cm}\text{GeV}$. We therefore believe that
it is a
reasonable approximation to extrapolate the bounds to particles with
masses in the range of $10^7\hspace{+.09cm}\text{GeV}$ to
$10^9\hspace{+.09cm}\text{GeV}$. Assuming that this is the
case, we see from Fig.~\ref{nu} that glueball decays from throats with
IR scales larger than $10^6\hspace{+.09cm}\text{GeV}$ do not destroy
nucleosynthesis.

Up to now, we have considered a scenario in which $K=200$ and
$M_{10}\hspace*{-.05cm}\sim\hspace*{-.05cm}10^{14}\hspace{+.09cm}\text{GeV}$.
This choice of parameters led to a relatively large number of
throats in the interesting region of IR scales between
$10^6\hspace{+.09cm}\text{GeV}$ and
$10^{11}\hspace{+.09cm}\text{GeV}$. Let us finally analyse a more conservative
scenario in which
we choose $K=60$ and
$M_{10}\hspace{-.05cm}\sim\hspace{-.05cm}10^{18}\hspace{+.09cm}\text{GeV}$.
The number of 3-cycles is, for example, roughly the minimal value
consistent with fine-tuning of the cosmological constant in the KKLT
construction~\cite{Hebecker:2006bn}. According to Eq.~\eqref{distribution},
the expected number of throats with IR scales between
$\smash{10^{6}\hspace{+.09cm}\text{GeV}}$
and 
$\smash{10^{11}\hspace{+.09cm}\text{GeV}}$ then is
\be
\bar{n} \left( 10^6\hspace{+.09cm}\text{GeV} \lesssim m_\IR \lesssim
10^{11}\hspace{+.09cm}\text{GeV}
\right) \, \simeq \, 0.7\,.
\ee
Thus, a significant fraction of vacua has a throat in this range of IR
scales and a reheating temperature of the order of
$\smash{10^{11}\hspace{+.09cm}\text{GeV}}$
would give the right amount of dark matter for such a throat. 

For definiteness, we assume that there is a single throat with IR scale $\sim
10^{11}\hspace{+.09cm}\text{GeV}$ in the aforementioned range. This choice of
IR scale
reflects our knowledge of the distribution of throats: As we have discussed
above, the
expected number of throats in a given interval, Eq.~\eqref{distribution},
becomes smaller if the 10d Planck scale becomes larger. By the same token,
the mean number of throats in the interval also becomes smaller if the
interval is shifted to smaller IR scales. This means that it is more
probable to find a throat at large IR scales.

Certain partial lifetimes of the glueballs from a throat with IR scale
$\sim\hspace{-.07cm}10^{11}\hspace{+.09cm}\text{GeV}$ have been discussed in
the last section. In
addition, there may be throats with lower IR scales which provide new decay
channels for the glueballs. Their mean number is 
\be
\bar{n}\left(m_\IR < 10^6\hspace{+.09cm}\text{GeV} \right) \, \simeq \, 0.7
\,.
\ee
We denote the number of degrees of freedom of this sector by $\tilde{g}$. As
before, we have to check that the dark matter glueballs do not decay too
quickly to this sector. Using Eq.~\eqref{DecayRate3}, the partial lifetime for
this decay channel is
\be
\label{lt2}
\tau \sim \tilde{g}^{-1} \, 10^{20} \text{ s} \,.
\ee
Thus, $\tilde{g}$ must not be larger than $\sim\hspace{-.07cm}10^3$ for the
dark matter
glueballs still to live longer than the current age of the universe. If
$\tilde{g}$ is larger than that, on the other hand, a throat with a lower
IR scale would be required (cf.~Eq.~\eqref{DecayRate3}). 

Obviously, many other cases, including more extreme choices of parameters,
are conceivable. However, an exhaustive study of the parameter space 
is beyond the scope of this thesis.

\section{Scenarios with low-scale supersymmetry breaking}
\label{extra}

In this section, we consider the limit of light gravitinos or, equivalently, 
low-scale supersymmetry breaking. As we have discussed
in Section~\ref{processes}, the fermionic glueballs can decay to a 
graviton and a gravitino if the gravitino is lighter than the glueballs.
The scalar glueballs decay to two gravitons with the same rate. As we have
seen 
in Section~\ref{SingleThroat}, scalar glueballs with mass 
$\smash{10^6\hspace{+.09cm}\text{GeV}}$ have a lifetime of 
\be
\tau \, \sim \, 10^{15} \text{ s} \,.
\ee
This is shorter than the current age of the universe. This lifetime now also
applies to the fermionic glueballs which accordingly decay too quickly to be
a good dark matter candidate.
On the other hand, the partial lifetime for this decay channel is
proportional to $m_\IR^{-5}$ according to Eq.~\eqref{dr3}. Thus, we obtain a
longer lifetime if the IR scale is somewhat lower than 
$10^6\hspace{+.09cm}\text{GeV}$.
In this case, throat dark matter is still possible even if the
supersymmetry breaking
scale is low. 

There are additional decay channels which are relevant in the limit of
light gravitinos: Recall that, in Section~\ref{spectrum}, we have found that a
scalar glueball is lighter than its fermionic superpartner in the limit of
low-scale supersymmetry breaking. 
Depending on the precise relation between the F-term vev of the chiral
compensator and the gravitino mass, the fermionic glueball can or can not
decay to this lighter scalar glueball with the emission of a gravitino
(cf.~Fig.~\ref{fig:decay1}). If this decay is kinematically not 
allowed, the fermionic glueball can still decay via a gravitino to standard
model particles and the lighter scalar glueball.
The gravitino propagator gives no suppression of the decay rate in this 
case and the decay rate is given by Eq.~\eqref{DecayRate}. For a throat with
$\smash{m_\IR\hspace{-.07cm}\sim\hspace{-.07cm}10^6\hspace{+.09cm}\text{GeV}}$
and $N_\IR\sim {\cal O}(1)$, the partial
lifetime of the fermionic glueballs for this decay channel is again
\be
\tau \, \sim \, 10^{15} \text{ s} \,.
\ee
Since decays of this kind are constrained by diffuse $\gamma$-ray
measurements, a partial
lifetime larger than $\sim\hspace{-.07cm}10^{26}\hspace{+.09cm}\text{s}$ is
actually required. Since the partial
lifetime is again proportional to $m_\IR^{-5}$ (cf.~Eq.~\eqref{DecayRate}),
the glueballs can as before be made sufficiently stable if the IR scale is
somewhat smaller than $10^6\hspace{+.09cm}\text{GeV}$. As we have explained
in Section~\ref{SingleThroat}, in order to obtain the observed amount of
 dark matter, a higher reheating temperature is then required.

\section{Relation to earlier work}
\label{relation}
The fact that dark matter can come from a hidden (or more precisely 
conformally sequestered) sector realized by a KS throat has
already been emphasized in~\cite{CT}. This paper focuses on scenarios
where 
a KS throat is heated by the annihilation of a brane with
an antibrane at the end of inflation. Subsequently, energy is transferred from
this throat to other throats which may be present in the compact space.
Throat-localized KK modes, which are produced in this way,
can be the observed dark matter if they are sufficiently long-lived and if
certain
parameters which determine their relic density are tuned. More
precisely, three types of dark matter candidates are discussed in
\cite{CT}:
KK modes which are localized in the throat where inflation took
place and those localized in the throat where the standard model lives have
to carry an (approximately) conserved angular momentum in the throat in order
to be sufficiently stable. KK modes which are localized in other
throats can, by contrast, be long-lived also without such an angular 
momentum.\footnote{
In addition, \cite{CT} also discusses particles on D-branes in these throats
as a dark matter candidate.}

We have approached the possibility of throat dark matter from a different and,
to a certain extent, more general perspective: We have not assumed
that reheating is due to brane-antibrane annihilation in a warped region.
Instead, we have only relied on the fact that the standard model has a
certain reheating temperature after 
inflation ends. In our approach, throats are not a `model building feature'
introduced to realize inflation, uplifting, etc. Instead, we view the
presence of (a potentially large number of) throats as a prediction of the
type IIB string theory landscape. Accordingly, we have considered scenarios in
which various throats are present. The amount of dark matter in these
throats that we have derived can be viewed as a generic prediction of the
type IIB landscape. 

We have focused in particular on the phenomenological importance of fermionic
KK modes in the throat. To the best of our knowledge, this point has so far
not received sufficient attention in the literature. As we have seen, 
fermionic KK modes play a central role in the phenomenology of sequestered
dark matter. Furthermore, we have used our values for the heat transfer rate
and the decay rates between throats from Chapters~\ref{energytransfer}
to~\ref{modifications}. For 
temperatures and Kaluza-Klein masses smaller than the compactification
scale, our rates differ from the values used in~\cite{CT}. We have restricted
our ana\-lysis to this case, for the following reasons: On the
one hand, a large compactification scale is required to make the KK modes
sufficiently stable against decay to the standard model or other
throats.\footnote{
An exception is the decay rate of fermionic KK modes to
the standard model sector which can be made small even for small
compactification scales (cf.~Section~\ref{DecaytoSM}).} 
On the other hand,
temperatures above the compactification scale would imply that also the
unwarped part of the manifold is heated up. The resulting gas of KK modes
in the compact space may then destabilize the volume modulus, making the
analysis of the early cosmological evolution much more difficult.
Furthermore, the suppression of the decay rates due to the flux-stabilization
of certain moduli was not taken into account in~\cite{CT} and our decay rates
differ also in this respect.

\chapter{Conclusions}
\label{conclusions} 

\section{General review}
In this thesis, we have studied throats in the early, hot universe. We have
reviewed various aspects of throats and of their dual gauge theories in
Chapters~\ref{throats} and~\ref{StringRealizations}. A
throat is a region in higher-dimensional space with a strong
gravitational potential along a certain direction. Energy scales of physical
processes, that are localized at different positions along this direction,
are exponentially redshifted or blueshifted relative to a fixed observer.
In that way, large hierarchies of scales can be generated. 

The Randall-Sundrum (RS) I model~\cite{Randall:1999ee} realizes this
generation of hierarchies in a simple geometry: A slice of 5-dimensional
anti-de-Sitter space (AdS$_5$) which is cut off by two branes. As Randall and
Sundrum have shown, the 4-dimensional graviton is localized towards one of the
branes, called the UV brane, in this geometry. Energy scales
of physical process that are localized near the other brane, called the IR
brane, are exponentially redshifted relative to the 4-dimensional Planck
scale. In particular, the RSI model allows for a solution to the hierarchy
problem of the standard model.

In string theory, a throat is, for instance, realized near a black
3-brane. This background solution of type IIB supergravity can be viewed as an
AdS$_5 \times$S$^5$ throat which is embedded into flat 10d space. As Verlinde
has shown~\cite{Verlinde:1999fy}, a string
realization of the RSII model~\cite{Randall:1999vf} (without IR brane) can be
obtained from such a black 3-brane if one compactifies the 6 dimensions
perpendicular to the brane on a torus. 

Black 3-branes are equivalent to stacks of D3-branes of type IIB string
theory. This equivalence can e.g. be tested by comparing the absorption cross
sections of particles for both objects. We
have reviewed the corresponding calculation for incident dilatons in
Section~\ref{3ba}. The fact that the absorption cross sections
agree exactly is evidence that black 3-branes and D3-brane stacks are just two
descriptions of the same object. 

A string realization of the RSI model can be obtained if one
places D3-branes and fractional D3-branes on a conifold singularity of a
Calabi-Yau orientifold~\cite{Giddings:2001yu}. The backreaction of these
branes on the geometry creates a Klebanov-Strassler (KS)
throat~\cite{Klebanov:2000hb} which is embedded into the compact space.
Alternatively, the KS throat in this setup can be viewed as the result of the
backreaction of 3-form flux on the geometry. In the landscape of type IIB flux
compactifications, 3-form flux is
distributed in various ways over the 3-cycles of the compact space and it
turns out that throats are commonly created from the backreaction of this
3-form flux. In Section~\ref{statistics}, we have reviewed how the expected
number of throats in a given Calabi-Yau orientifold can be determined.

\section{Throats in the early universe}

Since throats are a common feature of type IIB string compactifications, it
is interesting to think about observable signals which can result from their
presence in the early universe. In particular, as high temperatures have been
involved at this stage of the cosmological evolution, the throats can have
been heated to a certain temperature. If the energy density in a throat is
above its IR scale (which is related to the length of the throat), the
backreaction of the thermal plasma on the
geometry creates a black hole horizon which replaces the IR end of the
throat. This black hole horizon emits Hawking radiation which in turn can heat
up other throats and the standard model. The rate of this process is an
important quantity for the further cosmological evolution.

In Chapter~\ref{energytransfer}, we have
calculated this heat transfer rate a simple setup -- a heated AdS$_5
\times$S$^5$ throat and another AdS$_5 \times$S$^5$ throat embedded into a 6d
torus. Due to the warping, the Hawking radiation from the black hole
horizon in the heated throat has to tunnel through an effective energy barrier
before it can reach the other throat. The corresponding tunneling probability
determines the rate of heat transfer. The calculation of this probability,
however, is a multi-dimensional tunneling problem which is difficult to solve.
Therefore, we have chosen a different approach and replaced the AdS$_5
\times$S$^5$ throats by equivalent stacks of D3-branes. These D3-brane stacks
carry gauge theories on their world-volume. Correspondingly, we have referred
to the description in terms of D-branes as the gauge theory picture as opposed
to the gravity picture in terms of throats.

In particular, the heated throat is
equivalent to a D3-brane stack with a heated world-volume gauge theory. After
a KK expansion of supergravity fields in the embedding torus, our 10d setup
can be described by a 4d theory: A heated gauge
theory which is coupled by a tower of KK modes to another gauge theory. 
Heat transfer is in this description due to the annihilation of fields in the
thermal plasma of the heated gauge theory into KK modes which in turn decay to
the other gauge theory. The calculation of the
heat transfer rate is then a straightforward exercise in quantum field
theory. In particular, it is much simpler than the calculation involving the
tunneling problem. 

If the temperature drops below the IR scale of a throat, a phase
transition takes place. During this phase transition, the black hole
horizon is replaced by the IR end of the throat and a nonrelativistic
gas of throat-localized KK modes is formed. These KK modes decay to other
throats and the standard model with a certain rate. Again, we have calculated
this rate, which is another important quantity for the cosmology of throats,
in the gauge theory picture. 

A KK mode which is localized in a throat corresponds to a glueball of the dual
gauge theory. The coupling of such a glueball to supergravity fields in the
embedding space, however, can not be read off from any Lagrangian. Therefore,
in Chapter~\ref{KKdecay}, we have first calculated the decay rate
in a simpler setup using the gravity picture: The decay rate of a dilaton KK
mode from the AdS$_5 \times$S$^5$ part of a black 3-brane to the
asymptotically flat part. Using this result, we have determined the coupling
between the glueball and supergravity fields in the embedding space from the
requirement that the decay rates from the gravity picture and the gauge theory
picture agree. Redoing the calculation from Chapter~\ref{energytransfer} using
this coupling, we have found the decay rate of glueballs between the two gauge
sectors or, equivalently, of KK modes between the two throats.

The heat transfer rate and the decay rate that we have found depend on the
distance between the throats and on the size of the embedding
compact space. The rates become smaller for larger distances and larger
embedding spaces. In addition, the decay rate depends on the angular quantum
number of the decaying mode with respect to the S$^5$ in the AdS$_5
\times$S$^5$ throat containing the initial state. Modes with angular momentum
have a much lower decay rate than modes which are s-waves with respect to the
S$^5$.

If the size of the embedding space is of the same order as the AdS scales of
the throats, a compact space with two throats can be approximated by two
RS models which are glued together at a common UV
brane. As we have reviewed in Chapter~\ref{et5d}, the decay rate of
graviton KK modes in this situation follows from a straightforward calculation
of the tunneling probability between the two RS models. The decay rate
that we have found in our setup with two AdS$_5 \times
$S$^5$ throats in a torus should reproduce this result in the aforementioned
limit of small embedding space. In Section~\ref{comparison}, we have shown
that this is indeed the case. Furthermore, in Chapter~\ref{et5d}, we have
also presented an alternative derivation of the decay rate of KK modes between
two RS models. To this end, we have replaced one RS model by its dual gauge
theory. From the point of view of the remaining RS model, this gauge theory
lives on the UV brane. The decay rate of graviton KK modes in the
remaining RS model into gauge fields on the UV brane reproduces the decay
rate of graviton KK modes between two RS models.

As we have discussed in Chapter~\ref{modifications}, our results for the heat
transfer rate and the decay rate of KK modes can be applied to other
geometries of the throats and the embedding space. In particular, the heat
transfer rate is applicable to two KS throats if the curvature
of the space connecting them is not larger than the inverse distance.
Furthermore, the decay rate can be applied to the decay of graviton KK
modes between such throats. For KK modes of other supergravity fields, on the
other hand, the decay rate is in general difficult to determine since the
3-form flux in a KS throat mixes field fluctuations in a
complicated way. In order to determine the vertex between glueballs and
supergravity fields in the embedding space as discussed above, we would have
to solve the resulting complicated equations of motion of the field
fluctuations. We have shown, however, that the glueballs in a given sector
decay to a lightest scalar glueball and its fermionic superpartner
on cosmologically short timescales. For the purpose of cosmology, it is then
sufficient to determine only the decay rate of these glueballs. Furthermore,
we have found a flat direction for the supergravity fields in a KS throat.
The scalar which parameterizes this flat
direction fulfills the same equation of motion as the dilaton in an AdS$_5
\times$S$^5$ throat. Accordingly, it couples to supergravity in the embedding
space with the previously derived vertex. A scalar KK mode in general mixes
with this flat direction in a KS throat and therefore also decays via the
previously derived vertex. 

A suppression of the decay rate can arise, on the other hand, since 
certain fields which mediate decays get high masses in flux
compactifications. As we have discussed in Section~\ref{tachyon}, this
suppression is roughly
compensated by a stronger decay vertex if the decaying KK mode mixes with
tachyons in the 5d effective description of the throat. The reason is that
tachyons, i.e.~scalars with a negative mass squared, are less suppressed in
the UV direction than scalars with a vanishing or
positive mass. To analyse this effect in more detail, we have calculated the
decay rate of KK modes of a tachyon
between two RS models. For the calculation, we have replaced one RS model by
its dual gauge theory which then lives on the UV brane of the remaining RS
model. In the full AdS$_5$-space (without branes), scalars with negative mass
squared down to the Breitenlohner-Freedman bound~\cite{Breitenlohner:1982bm}
do not lead to instabilities. In a RS model, however, the tachyon must have a
large mass term on the UV brane in order to avoid tachyonic KK modes. Similar
to the mass of mediating fields in flux compactifications, this UV-localized
mass leads to a suppression of the decay rate. The stronger decay vertex of KK
modes of a tachyon, however, compensates this suppression. These KK modes and
KK modes which mix with a tachyon therefore decay roughly with the
previously derived rate of graviton KK modes.

\section{Dark matter in throats}

We have presented an application of the heat transfer rate and the decay
rate in Chapters~\ref{thermalproduction} and~\ref{scenarios}: KK modes
whose wavefunctions are localized in a throat are an interesting dark matter
candidate. These KK modes have redshifted masses, allowing for their
production after reheating in the standard model, even if the reheating
temperature is not very high. In addition, their decay rates 
are highly suppressed, potentially resulting in a very long lifetime. 

We have considered scenarios in which the standard model lives in the
unwarped part of a compact space, which in addition has a certain number of 
throats. To be conservative, we have assumed that the throats receive
no energy from the reheating mechanism and that the reheating mechanism only
heats up the standard model. Even under this minimal assumption, the throats
are heated up by energy
transfer from the hot standard model plasma. 

From the dual gauge theory point of view, the resulting energy density in a
throat is initially in the form of gauge theory states with energy of the
order of the reheating temperature. As we have shown, if the total
energy
density is above the critical energy density for a deconfinement phase
transition, the gauge theory thermalizes. Accordingly, the energy density
initially scales like radiation with the expansion of the universe until a
confinement phase transition takes place. Afterwards, the energy density is in
the form of nonrelativistic glueballs whose energy density correspondingly
scales like matter. In Section~\ref{te}, we have determined the resulting
contribution of glueballs to the total energy density at our epoch.

If the energy density is always below the critical energy density, on the
other hand, the initial gauge theory states hadronize to a large number of
nonrelativistic glueballs. As we have explained in Section~\ref{te}, this is
due to the fact that the gauge
theory (which is dual to a throat) is strongly coupled on all scales. In an
asymptotically free theory such as QCD, jets with ultrarelativistic particles
would instead be formed during hadronization. The energy density in the
throat sector thus scales like matter immediately after reheating and the
resulting contribution to the total energy density at our epoch
is relatively large already for reheating temperatures which are only
moderately high.

We have plotted the contribution of a throat 
to the total energy density for fixed reheating temperature 
$T_\RH$ and as a function of the IR scale $m_\IR$ in Fig.~\ref{fig:plot}.
As one can see, this
function has a maximum between IR scales $ m_\IRc$ and $T_\RH$. Here,
$m_\IRc$ is the IR scale for which the dual gauge theory thermalizes precisely
to the phase transition temperature. Since throats are a common feature in the
landscape of type IIB string theory and a given compact space typically has
several throats, it is probable to have throats with IR
scales in the `optimal' range between $m_\IRc$ and $T_\RH$. We can therefore
expect that dark matter is due to these throats.

As we have discussed in Section~\ref{processes}, scalar glueballs decay to
two gravitons with a certain rate. Similarly, fermionic
glueballs can in principle decay to a graviton and a gravitino. In order to
obtain a stable dark matter candidate, we have focused on scenarios in which
such decays are kinematically forbidden due to a heavy gravitino. This
requires that the supersymmetry breaking scale
is larger than the mass of the glueball. It may also mean that the
superpartners of standard model particles are heavier than the
glueballs. If $R$-parity is conserved, most decay channels
of fermionic glueballs to the standard model involve
such a superpartner and are therefore kinematically forbidden. In
Section~\ref{DecaytoSM}, we have identified an operator which does not involve
a superpartner and which would allow the decay of fermionic
glueballs to a Higgs and a lepton. If present, this operator would give the
dominant decay channel even for maximally broken $R$-parity. 

In Section~\ref{SingleThroat}, we have first discussed scenarios with a
single throat in the `optimal' range of IR scales between $m_\IRc$ and
$T_\RH$. We have found that, in many cases, KK modes in 
throats with IR scale $T_\RH$ have a decay rate which is too high for them 
to be a good dark matter candidate. However, if the gravitino is very heavy 
(high-scale supersymmetry breaking) and the aforementioned operator is
suppressed, the
lightest fermionic glueballs may nevertheless survive and play the role of
dark matter. The more promising case is that of throats with lower IR scales.
For definiteness, we have analysed a throat with IR scale $m_\IRc$, which is
the lower end of the `optimal' range of IR scales. We have found that a
reheating temperature of $10^{10}\hspace{+.09cm}\text{GeV}$ to
$10^{11}\hspace{+.09cm}\text{GeV}$ leads to the
right amount of glueballs to account for the observed dark matter. The
critical IR scale $m_\IRc$ is a function of $T_\RH$. After having fixed 
$T_\RH$, we find a mass for the dark matter candidate in this scenario which
is between
$\smash{10^5\hspace{+.09cm}\text{GeV}}$ and $10^6\hspace{+.09cm}\text{GeV}$. 

Our dark matter scenario may lead to some interesting observable signatures. 
The dark matter glueballs decay to the standard model with a very low,
but non-negligible rate. The decays produce photons to which experiments like 
GLAST or HESS may be sensitive. It turns out that the decay rates depend on
two parameters: The 10d Planck mass enters via the flux
stabilized mass of fields which mediate the decay. Moreover, the
decay rate of fermionic glueballs depends on the dimensionless coupling
strength $\lambda$ of the aforementioned operator which allows their decay to
a lepton and a Higgs. In Section~\ref{SingleThroat}, we have identified two
interesting scenarios: 

If $\lambda$ is of the order 1, the 10d Planck scale has to be very large in
order to get a sufficiently stable dark matter candidate. Namely, for a lower
10d Planck scale, the fermionic glueballs decay to the standard model
with a rate which is in conflict with measurements of the diffuse
$\gamma$-radiation. On the other hand, the 10d Planck scale cannot be larger
than the 4d Planck scale. This makes it more probable that the lifetime of
fermionic glueballs is in a range that can be probed with
new experiments like GLAST. If the scenario with $\lambda
=\mathcal{O}(1)$ is realized in nature, one can hope to detect a signal from
the decaying glueballs in the near future.

If $\lambda$ is much smaller than 1, a lower 10d Planck scale still leads to
sufficiently stable fermionic glueballs. For a
low 10d Planck scale, also the decay of scalar glueballs can become relevant
for detection.
In contrast to fermionic glueballs, scalar glueballs can
decay directly to two photons. This decay channel leads to a sharp
$\gamma$-line at an energy of
$10^5\hspace{+.09cm}\text{GeV}$ to $10^6\hspace{+.09cm}\text{GeV}$ which could
be detected with experiments like HESS.
In addition, the scalar glueball decays also produce a continuous
$\gamma$-ray spectrum to which e.g.~GLAST may again be sensitive. If the 
scalar glueballs make up an $\mathcal{O}(1)$ fraction of the dark matter at
our epoch, a 10d Planck scale of the order of
$10^{13}\hspace{+.09cm}\text{GeV}$
would allow for a detection of both signals in the near future. Such a
10d Planck scale corresponds to a compactification radius
of the order of just 50 times the string length, which is not extremely
large. 

Finally, in Section~\ref{ManyThroats}, we have considered scenarios
with a large number of throats, using the result for the expected number of
throats in a given Calabi-Yau orientifold that we have reviewed in
Section~\ref{statistics}. We have found that there are setups in which the
probability of having a throat with IR scale in the aforementioned `optimal'
range is large. The only free parameter which has then to be fixed in order to
obtain the observed amount of dark matter is the reheating temperature. The
lifetime of the glueballs strongly depends on their mass and therefore on the
IR scale of the corresponding throat. In a scenario with several throats
with various IR scales, it is therefore possible to have glueballs
which decay already at early epochs while other glueballs are sufficiently
stable to account for the observed dark matter. Decays to the standard model
at early epochs are severely constrainted by nucleosynthesis. We have checked
in a specific example that these constraints can be fulfilled.

\section{Outlook}

Our results for the heat transfer rate and the decay rate depend on the
distance between the throats and the size of the embedding compact space.
They generalize the aforementioned decay rate of graviton KK modes between
two RS models. This dependence on distance
and size is, for instance, relevant for the analysis of reheating after
brane-antibrane inflation. At the end of inflation in this scenario, the brane
annihilates with the antibrane and the energy which is released in this
process subsequently thermalizes. If the standard model is realized in another
throat, at least part of this energy has to be channeled to this throat in
order to allow for the reheating of the standard model.

We find that, as long as the embedding manifold is not of minimal
size, our decay rate is considerably lower than the decay rate
of graviton KK modes between two RS models which was used in previous
analyses~\cite{Barnaby:2004gg,Kofman:2005yz,Frey:2005jk,Chialva:2005zy} of
this reheating mechanism. Given our
results, it will be interesting to reconsider reheating after brane-antibrane
inflation. 

Apart from reheating after brane-antibrane inflation and our dark matter
scenario, one can apply our
results to several cosmological scenarios where reheating takes place either
in the standard model or in a throat and the
standard model resides either at the bottom of a throat or somewhere in the
rest of the Calabi-Yau orientifold. The heat transfer rate and the 
decay rates that we have calculated can then be used in a set of
Boltzmann equations to determine the evolution of energy densities of
the standard model and the throats.

For our dark matter scenario in Chapters~\ref{thermalproduction}
and~\ref{scenarios}, we have assumed that the dark matter glueballs decay via
the vertex from Chapter~\ref{KKdecay} for a dilaton in an AdS$_5 \times$S$^5$
throat. The decay rate of these glueballs is suppressed relative to the decay
rate from Chapter~\ref{KKdecay}, however, since certain fields which
mediate the decay get high masses in flux compactifications. 

On the other hand, we have found in Section~\ref{tachyon} that KK modes
mixing with tachyons in the throat decay with a
considerably stronger vertex than that from Chapter~\ref{KKdecay}. Such
tachyons also appear in the 5d effective description of the KS throat. We
expect that the KK modes dual to the dark matter glueballs mix
with these tachyons in the KS throat. The resulting stronger decay vertex of
these particles would then roughly compensate the suppression of the
decay rate due to flux-induced masses of mediating fields. In particular, the
dark matter glueballs would decay with a higher rate than that used in
Chapters~\ref{thermalproduction}
and~\ref{scenarios}. It will be interesting to work out the consequences of
such a higher decay rate for our dark matter scenario. The resulting shorter
lifetimes of the dark matter may eliminate a large portion of the
phenomenologically viable region in parameter space.
On the other hand, the remaining region in parameter space may be so
small that our dark matter scenario could either be excluded or verified by
experiments in the near future.

In Chapter~\ref{thermalproduction}, we have mentioned an alternative
production mechanism of the dark matter. This mechanism is operative
during inflation and is due to the fact that de Sitter space has a
temperature. It would be interesting to investigate this mechanism and its
influence on our dark matter scenario in more detail. Furthermore, we have
mainly neglected KK modes which are charged under global (approximate)
symmetries in the throat. As we have seen in Chapter~\ref{KKdecay}, in
an AdS$_5 \times$S$^5$ throat these charged KK modes decay with a
considerably lower rate than KK modes which are s-waves. It would be
interesting to determine whether such a suppression also arises in a KS
throat or whether the KK modes again mix with a tachyon in such a throat. Part
of the global symmetries in some throats are only present in the
high-temperature phase and are broken at the phase transition during which the
black hole horizon is replaced by the IR end of the throat. As we have seen in
Chapter~\ref{KKdecay}, long throats are heated to the black-hole phase for a
sufficiently large reheating temperature. They subsequently cool with the
cosmological expansion until the aforementioned phase transition takes place.
The breaking of global symmetries during the phase transition can then
lead to the formation of topological defects. It would be interesting to
investigate this possibility in more detail.

\appendix

\chapter{Kaluza-Klein expansion of the graviton in a Randall-Sundrum model}
\label{KKmodes}
In the following, we perform the KK expansion of the 5d graviton $h_{MN}$ in
the RS background discussed in Section~\ref{RSmodels} and determine the
couplings of the
KK modes to field theories on the UV brane and the IR brane. The results in
this section
are taken from \cite{Randall:1999vf,Davoudiasl:1999jd,Chung:2000rg}. 

Due to the orbifold symmetry $y \rightarrow -y$, there is no zero mode of
the graviphoton $h_{\mu N}$. The massless spectrum thus consists of the 4d
graviton $\smash{h^{(0)}_{\mu \nu}}$ and the radion. In addition, there is a
tower of
massive spin-2 fields $\smash{\smash{h^{(n)}_{\mu \nu}}}$. By counting the
number of
degrees of freedom, it becomes clear that there are no massive vectors or
scalars in the spectrum. In a phenomenologically viable theory, the radion has
to be stabilized, e.g. with the Goldberger-Wise
mechanism~\cite{Goldberger:1999uk}. In the following, we assume that this
has been achieved and we do not consider the radion any more. A convenient
parametrization for the spin-2 fluctuations is
\be
\label{fluctuation}
ds^2 \, = \, e^{-2 k|y|} \, \left( \eta_{\mu \nu} + h_{\mu \nu}(x,y)
\right) \, dx^\mu dx^\nu + dy^2 \,,
\ee
where $h_{\mu \nu}$ is chosen to be transverse and traceless. With this
parametrization, the equation of motion for $h_{\mu \nu}$ is
just a Laplace equation in the background geometry
\cite{Csaki:2000fc,Firouzjahi:2005qs}:
\be
\label{LE2}
\partial_M \left( \sqrt{g} g^{MN} \partial_N h_{\mu \nu} \right) = 0 \,.
\ee
Next, we perform the KK expansion of the 5d field $h_{\mu \nu}$:
\be
\label{KKexpansion}
h_{\mu \nu}(x,y) \, = \, \sum_{n=0}^\infty \, h^{(n)}_{\mu \nu}(x) \,
\chi^{(n)}(y) \,.
\ee
Using this expansion and denoting the mass of the n-th KK mode by $m_n$, the
equation of motion for the profile $\smash{\chi^{(n)}}$ of the n-th KK
mode along the 5th dimension reads
\be
\label{EOM}
\frac{d}{d y} \left( e^{-4 k |y|} \frac{d \chi^{(n)}}{d y} \right) \,+\, m_n^2
\, e^{-2 k |y|} \, \chi^{(n)} \,= \,0 \, .
\ee
For a field which is even under the orbifold $\mathbb{Z}_2$-action $y
\rightarrow -y$, the boundary conditions at the two branes follow from
Eq.~\eqref{EOM} as
\be
\label{BC}
\frac{d}{d y} \chi^{(n)}(0) \, = \, 0 \quad \text{and} \quad \frac{d}{d y}
\chi^{(n)}(\ell) \, = \, 0 \,.
\ee
For $m_n=0$, the solution to Eqs.~\eqref{EOM} and \eqref{BC} is
$\chi^0  = \text{const}$. The corresponding KK mode is the 4d graviton. The
wave functions $\chi^{(n)}$ have to be normalized according to 
\be
\label{nn}
\int_{-\ell}^\ell  dy \, e^{-2 k |y|} \left(\chi^{(n)}(y)\right)^2 \,
\overset{!}{=} \, 1
\ee 
in order to obtain a canonically normalized kinetic term for the corresponding
KK mode. The norm of the wavefunction $\chi^0$
is dominated by the region near the UV brane. This means that 4d
gravity is localized near the UV brane and is the reason for the appearance of
the unrescaled metric $\smash{g^{(4)}}$ in Eq.~\eqref{gravityaction}. 

Solutions to Eqs.~\eqref{EOM} and \eqref{BC} for $m_n \neq 0$ are
\be
\label{BesselFunctions}
\chi^{(n)} \, = \,  \frac{1}{N_n} e^{2 k |y|} \left[ J_2\left(
\frac{m_n}{k} e^{k |y|} \right) +B_n Y_2\left(
\frac{m_n}{k} e^{k |y|} \right)     \right] \,,
\ee
where $J_2$ and $Y_2$ are Bessel functions. The constant $B_n$ is
determined by the boundary conditions. Using the boundary condition at the UV
brane in Eq.~\eqref{BC}, we find
\be
B_n \, = \, -\frac{J_1\left(
\frac{m_n}{k}\right)}{Y_1\left( \frac{m_n}{k}\right)} \,.
\ee
Alternatively, we can determine $B_n$ from the boundary condition at the IR
brane in Eq.~\eqref{BC}. Since both results for $B_n$ have to agree, we get a
condition on the masses $m_n$:
\be
\frac{J_1\left(
\frac{m_n}{k}\right)}{Y_1\left( \frac{m_n}{k}\right)} \, = \,
\frac{J_1\left(\frac{m_n}{k}e^{k \ell}\right)}{Y_1\left(
\frac{m_n}{k}e^{k \ell}\right)} \, .
\ee
We consider only light modes with masses $m_n \ll k$. Furthermore, we
assume that $m_n \gg k \, e^{-k \ell}$. This assumption will be justified
in a moment. Using the asymptotic forms of the Bessel functions for large
and small arguments, we find that the mass spectrum is determined by
\be
\label{masses}
J_1\left(\frac{m_n}{k}e^{k \ell} \right)\, \simeq \, 0 \quad \Rightarrow \quad
m_n \, \simeq \, \Bigl( n +\smash{\frac{1}{4}} \Bigr) \pi \,  e^{-k \ell} \, k
\quad \text{for} \quad n \in \mathbb{N} \,.
\ee
We see that the aforementioned assumption can be justified. The last
approximation becomes better for larger $n$ but works already well
for the lowest mass eigenvalue (which to a higher precision is $m_1 \simeq
1.22 \, \pi \, e^{-k \ell} \, k$).

The normalization constants $N_n$
follow from Eq.~\eqref{nn} and are given by
\be
N_n \, = \, \frac{1}{\sqrt{k}} \, \left[ e^{2 k \ell} Z^2_2\left(
\frac{m_n}{k} e^{k \ell} \right) - Z^2_2\left( \frac{m_n}{k} \right)
\right]^{1/2} \, ,
\ee
where $Z_2(x) \, \equiv \, J_2(x) - B_n  \, Y_2(x)$. The norm of
the wavefunctions Eq.~\eqref{BesselFunctions} is dominated by the
region near the IR brane. Intuitively, this is the reason why the masses,
Eq.~\eqref{masses}, are quantized in units of the warped curvature scale,
i.e.~$\smash{e^{-k \ell} k}$. 

Using the results for the wavefunctions $\smash{\chi^{(n)}}$, we can determine
the couplings of the KK modes to theories on the UV brane and the IR brane.
The 4d action for the tower of KK modes
reads~\cite{Randall:1999vf,Davoudiasl:1999jd,Chung:2000rg}:
\begin{multline}
\label{5daction2}
S = \int d^4 x \;  \frac{1}{2} \sum_{n=0}^\infty \bigl( \partial_{\alpha}
h_{\mu \nu}^{(n)} \, \partial^{\alpha} h^{\mu \nu (n)} +\, m_n^2
\, h_{\mu \nu}^{(n)} \, h^{\mu \nu (n)} \bigr) \\
+ \frac{1}{\sqrt{2}} \left( \frac{1}{M_4} T^{\mu \nu}_\UV + \frac{1}{M_4}
T^{\mu \nu}_\IR \right) h^{(0)}_{\mu \nu} 
+ \frac{1}{\sqrt{2}} \, \sum_{n=1}^\infty \left(\frac{g_n}{M_4}
\, T^{\mu \nu}_\UV + \frac{1}{e^{-k \ell} M_4} T^{\mu \nu}_\IR
\right) h^{(n)}_{\mu \nu} \,.
\end{multline}
Here, $T^{\mu \nu}_\UV$ and $T^{\mu \nu}_\IR$ are the energy-momentum tensors
on the UV and IR brane, respectively. In particular, note that the coupling of
the massive KK modes to the IR brane is only suppressed by the warped Planck
scale. The other coupling constants are
\be
\label{gn2}
g_n = \left( \left(
\frac{Y_1\left(\frac{m_n}{k}\right)}{Y_1\left(\frac{m_n}{k} e^{k \ell}\right)}
\right)^2 -1
\right)^{-1/2} \simeq \; \sqrt{\frac{n}{2}} \,\pi \, e^{-k \ell} \,.
\ee
In the last step we have used the asymptotic forms of the Bessel function
$Y_1$. The approximation is valid for $\smash{n \ll e^{k \ell}}$. As one can
see,
the coupling of massive KK modes to the UV brane has an extra exponential
suppression factor. 

\chapter{Kaluza-Klein expansion of a tachyon in a
Randall-Sundrum model}
\label{TachyonAppendix}

In Section~\ref{tachyon}, we have considered a scalar with a tachyonic mass in
the RSI model. We have seen that, in order to avoid a tachyonic KK mode,
a sufficiently large mass term on the UV brane has to be switched on. The
action of such a scalar then reads
\be
\label{5dAction}
S\, = \, \int d^4x \int_{-\ell}^\ell dy \, \sqrt{g} \,
\left( \frac{1}{2}  \, g^{MN} \partial_M \Phi \partial_N \Phi \, +\, 
\frac{1}{2}  \, m_{\text{5d}}^2 \Phi^2 \,+ \, \lambda k
\, \delta(y) \, \Phi^2 \right) \, ,
\ee
where $m_\text{5d}^2 <0$ is the tachyonic mass squared in the bulk and
$\lambda$ measures the mass on the UV brane in units of the AdS scale
$k$. We use the same parametrization of AdS$_5$ as in Eqs.~\eqref{ansatz}
and~\eqref{solution}. In the following, we perform the KK
decomposition of such a scalar and determine the couplings of the
corresponding KK modes to a theory on the UV brane. For more details,
see~\cite{TachyonPaper}.

The equation of motion, which follows from the action Eq.~\eqref{5dAction}, is
\be
\frac{1}{\sqrt{g}} \partial_M \left( \sqrt{g} g^{MN} \partial_N \Phi \right)
\, - \, m_{\text{5d}}^2 \, \Phi \, - \, 2 \lambda k
\, \delta(y) \, \Phi= 0 \,.
\ee
The $\delta$-function can be absorbed into the boundary condition at the UV
brane. The KK expansion of the scalar is
\be
\label{TachyonKKexpansion}
\Phi(x,y) \, = \, \sum_n  \chi_n(x) \phi_n(y) \, ,
\ee
where the $\chi_n(x)$ are eigenmodes of the 4d d'Alembertian with eigenvalues
$m_n^2$. The equation of motion for the $\smash{\phi_n}$ then reads
\be
\frac{d}{d y} \left( e^{-4 k |y|} \frac{d \phi_n}{d y} \right) \, - \, 
e^{-4 k|y|} \, m_\text{5d}^2 \,\phi_n\, + \, m_n^2 
\, e^{-2 k |y|} \, \phi_n \, = \, 0 \,.\label{EOM2}
\ee
We consider a field which is even under the orbifold $\mathbb{Z}_2$-action,
i.e.~$\phi_n(y) = \phi_n(-y)$. The boundary conditions for the $\phi_n$ then
read
\be
\frac{d}{d y} \phi_n(0) \, = \, \lambda \, k \, \phi_n(0) \quad 
\text{and} \quad \frac{d}{d y} 
\phi_n(\ell) \, = \, 0 \,.\label{BC2}
\ee

Let us consider a scalar which just satisfies the
Breitenlohner-Freedman bound, i.e.~$m_\text{5d}^2 = - 4 k^2$. As we have
discussed in Section~\ref{tachyon}, for a vanishing mass on the UV
brane, i.e.~for $\lambda=0$, the 4d spectrum contains a tachyonic KK mode.
If the mass on the UV brane is switched on, this KK mode becomes `less
tachyonic' for growing $\lambda$ and for a certain value of $\lambda$, the
mass of the KK mode vanishes. Let us determine this value of
$\lambda$. The solution to Eq.~\eqref{EOM2} for $m_n=0$ is
\be
\phi_0 \, = \, A_0 \, e^{2 k |y|} \, + B_0 \,e^{2 k |y|} \, k \, |y| \, .
\ee
The overall prefactor of this solution is fixed by the normalization of
the wave function. The remaining constant is determined by the boundary
conditions on the UV brane and the IR brane in Eq.~\eqref{BC2}. This fixes
$\lambda$. We then find that a massless KK mode exists only if
\be
\lambda \, = \, \frac{4 \, \ell k}{1+2 \, \ell k} \, \simeq \, 2 \, .
\ee
The last step is valid since $k \ell \gg 1$. For this value of $\lambda$, the
tachyonic mode is lifted to a massless mode. However, in a
phenomenologically viable setup, no massless scalars should appear. We
therefore choose $\lambda$ somewhat larger than this value, say $\lambda
\approx 3$, which lifts the tachyonic mode to a very massive mode.

Solutions to Eqs.~\eqref{EOM} and \eqref{BC} for $m_n \neq 0$ are
\be
\label{BesselFunctions2}
\phi_n \, = \, \frac{e^{2 k |y|}}{N_n} \left[ J_0
\left(\frac{m_n}{k} e^{k |y|}\right) + B_n Y_0
\left(\frac{m_n}{k} e^{k |y|}\right)\right] \,,
\ee
where $J_0$ and $Y_0$ are Bessel functions. Using the boundary condition
at the UV brane in Eq.~\eqref{BC2}, we can determine the constant $B_n$:
\be
B_n \, = \,- \frac{( 2 - \lambda )\, J_0\left( \frac{m_n}{k} \right)-
\frac{m_n}{k} \, J_1\left( \frac{m_n}{k} \right)}{( 2 - \lambda
)\, Y_0\left( \frac{m_n}{k} \right)-
\frac{m_n}{k} \, Y_1\left( \frac{m_n}{k} \right)} \, .
\ee
We consider only light modes with masses $m_n \ll k$. Using the
asymptotic forms of the Bessel functions for small arguments, we find that
$\smash{B_n \sim \log(m_n/k)^{-1} \ll 1}$ for $m_n\ll k$ and $\lambda \approx
3$. We have used this fact in Section~\ref{tachyon}.
Alternatively, we can determine the constant $B_n$ from the boundary
condition at the IR brane in Eq.~\eqref{BC2}. Both results for $B_n$ have to
agree and we therefore get a condition on the masses $m_n$:
\be
\label{MassCondition}
\frac{( 2 - \lambda )\, J_0\left( \frac{m_n}{k} \right)-
\frac{m_n}{k} \, J_1\left( \frac{m_n}{k} \right)}{( 2 - \lambda
)\, Y_0\left( \frac{m_n}{k} \right)-
\frac{m_n}{k} \, Y_1\left( \frac{m_n}{k} \right)} \, = \, \frac{2\,
J_0\left( \frac{m_n}{k} e^{k \ell}\right)-
\frac{m_n}{k} \, e^{k \ell}  J_1\left( \frac{m_n}{k} e^{k \ell}\right)}{2\,
Y_0\left( \frac{m_n}{k}e^{k \ell} \right)-
\frac{m_n}{k} \, e^{k \ell} \, Y_1\left( \frac{m_n}{k} e^{k \ell}\right)} \, .
\ee
For $m_n \ll k$ and $\lambda \approx 3$, this condition simplifies
to
\begin{multline}
\label{CM}
\mathcal{O}(1) \left[ 2\,
Y_0\left( \frac{m_n}{k}e^{k \ell} \right)-
\frac{m_n}{k} \, e^{k \ell} \, Y_1\left( \frac{m_n}{k} e^{k
\ell}\right)\right]  \\ \sim \, \log \Bigl(\frac{m_n}{k}\Bigr) \left[ 2\,
J_0\left( \frac{m_n}{k} e^{k \ell}\right)- \frac{m_n}{k} \, e^{k \ell}
J_1\left( \frac{m_n}{k} e^{k \ell}\right)\right]\, .
\end{multline}
If we assume that the masses fulfill $\smash{m_n k^{-1} e^{k \ell} \gg 1}$,
we can use the asymptotic forms of the Bessel functions for
large arguments in Eq.~\eqref{CM}. Since $\log(m_n/k) \gg 1$, Eq.~\eqref{CM}
is then approximately solved for
\be
\label{TachyonicMasses}
J_1\left( \frac{m_n}{k} e^{k \ell} \right)\, \ll \, 1 \quad \Rightarrow \quad
m_n \, \approx \, \Bigl( n +\smash{\frac{1}{4}} \Bigr) \pi \,  e^{-k \ell} \,
k
\quad \text{for} \quad n \in \mathbb{N} \,.
\ee
For $n$ somewhat larger than 1, the assumption that $\smash{m_n k^{-1} e^{k
\ell} \gg 1}$ is indeed fulfilled. For smaller $n$, the masses differ from
the result in Eq.~\eqref{TachyonicMasses} but are still
quantized in units of the warped AdS scale, i.e.~$\smash{k e^{-k \ell}}$, as
we have discussed in Section~\ref{tachyon}.

In order to obtain a canonically normalized kinetic term for the KK modes, the
wave functions $\phi_n$ have to be normalized according to 
\be
\label{nn2}
\int_{-\ell}^\ell  dy \, e^{-2 k |y|} \left(\phi_n(y)\right)^2 \,
\overset{!}{=} \, 1 \, .
\ee 
This fixes the constant $N_n$ in Eq.~\eqref{BesselFunctions2} and we find
\be
N_n \, = \, \frac{1}{\sqrt{k}} \, \left[Z_0^2\left( \frac{m_n}{k} e^{k \ell}
\right)
\, \left(e^{2 k \ell}+  \frac{4\, k^2}{m_n^2} \right) \, - \, Z_0^2\left(
\frac{m_n}{k}\right) \, \left( 1 + \frac{k^2}{m_n^2} (\lambda -2)^2 \right)
\right]^{1/2} \,,
\ee
where $Z_0 \equiv J_0 + B_n Y_0$.

The coupling of the tachyonic scalar to a gauge theory on the UV brane is
given in Eq.~\eqref{TachyonicCoupling} and reads
\be
\label{TachyonicCoupling3}
S_\UV \, \supset \, \sim \frac{1}{M_{10}^4 \, R^{5/2}} \int d^4x \int_{-
\ell}^\ell dy \, \sqrt{g} \, F_{\mu \nu} F^{\mu \nu} \, \Phi \, \delta(y) \,,
\ee
where $F_{\mu \nu}$ is the gauge field strength. As we
have discussed in Section~\ref{tachyon}, the prefactor is motivated by the
DBI action and a dimensional reduction from 10d to 5d. Inserting the
KK expansion Eq.~\eqref{TachyonKKexpansion} in
Eq.~\eqref{TachyonicCoupling3}, the coupling of the n-th KK mode to the gauge
theory follows as
\be
\label{AppendixCoupling1}
S_\text{4d} \, \supset \, \sim \frac{g_n}{M_{10}^4 \, R^3} \int d^4x
 \, F_{\mu \nu} F^{\mu \nu} \, \chi_n \, .
\ee
The coupling strengths $g_n$ are determined by the value $\phi_n(0)$ of the
wave functions at the UV brane and read
\be
\label{AppendixCoupling2}
g_n \, = \, \left[\left(\frac{ (2-\lambda)
J_0\left(\frac{m_n}{k}\right)-\frac{m_n}{k}J_1\left(\frac{m_n}{k}\right)}{
2
J_0\left(\frac{m_n}{k}e^{k \ell}\right)-\frac{m_n}{k}e^{k \ell}
J_1\left(\frac{m_n}{k}e^{ k \ell}\right)}
\right)^2 
\left[e^{2 k \ell} + \frac{4 k^2}{m_n^2} \right]- \left[ 1 +
\frac{k^2}{m_n^2} (\lambda -2 )^2 \right]
 \right]^{-1/2} \, .
\ee
This rather complicated expression can be evaluated by using the asymptotic
forms of the Bessel functions and the mass quantization in
Eq.~\eqref{TachyonicMasses}. For $\lambda \approx 3$, we find
\be
g_n \, \sim \, \frac{1}{k \, \ell} \, \sqrt{\frac{m_n e^{-k \ell}}{k}} \, .
\ee

\chapter{Evaluation of a propagator}
\label{integral}
In Eq.~\eqref{propagator}, we had to evaluate the following propagator
in a mixed, energy-configuration-space representation:
\begin{equation}
\label{appendixgl}
\int \frac{d^6 \rho}{(2 \pi)^6 } \, \frac{ e^{ i \vec{A} \,
\vec{\rho}}}{m^2 -\vec{\rho}^2 +i \epsilon} \,.
\end{equation}
We perform the integral for imaginary values $m \rightarrow e^{i
\pi/2} m$ and use analytic continuation. The integral changes into
\begin{equation}
- \int \frac{d^6 \rho}{(2 \pi)^6 } \, \frac{ e^{ i \vec{A} \,
  \vec{\rho}}}{m^2 +\vec{\rho}^2 } \,.
\end{equation}
We can then employ the identity $\smash{c^{-1}=\int_0^\infty d \tau
e^{-c \tau}}$ for $\Re{c} > 0$ and get
\begin{equation}
\begin{split}
& \frac{-1}{(2\pi)^6} \int_0^\infty \hspace{-1.5mm} d \tau
\hspace{-.5mm} \int \hspace{-1mm} d^6 \rho \;\; e^{ i \vec{A} \,
\vec{\rho}} \, e^{ -(m^2 +\vec{\rho}^2) \tau } \\ = & \frac{-1}{(2
\pi)^6} \int_0^\infty \hspace{-1.5mm} d \tau \, \left( \left[ \int
\hspace{-1mm} d \rho_1 \; e^{ i A_1 \, \rho_1} \, e^{ -\rho_1^2 \tau }
\right] \cdots \left[ \int \hspace{-1mm} d \rho_6 \; e^{ i A_6 \,
\rho_6} \, e^{ -\rho_6^2 \tau } \right] e^{ -m^2 \tau} \right) \\ = &
\frac{-1}{(4 \pi)^3} \int_0^\infty \hspace{-1.5mm} d \tau \; \frac{1}{
\tau^3} \, e^{-A^2 / 4 \tau } \, e^{-m^2 \tau} \,.
\end{split}
\end{equation} 
We have used that $A^2 = A_1^2 + \cdots + A_6^2$. According to
Eq.~3.471.9 in \cite{GR}, this integral can be evaluated in terms of
the modified Bessel function $K_{-2} \equiv K_2$, which yields 
\be
\frac{-1}{(2\pi)^3} \frac{m^2}{A^2} \, K_2( m A) \,.  
\ee 
Following from
Eq.~9.6.4 in \cite{AS}, $K_2$ is related to the Hankel function
$H_2^+=J_2+i Y_2$. The above expression can be written as 
\be
\frac{i}{(4\pi)^2} \frac{m^2}{A^2} \, H^+_2( e^{i \pi/2} m A) \,.
\ee
The Hankel function has a branch cut along the negative real
axis. Therefore, one can analytically continue back to real values $m
\rightarrow e^{-i \pi/2} m$, which gives
\begin{equation}
\label{resultappendix}
\frac{-i}{(4\pi)^2} \frac{m^2}{A^2} \, H^+_2( m A) \,.
\end{equation}

\chapter{Additional processes in a thermalized situation}
\label{additional}
In Section~\ref{processes}, we have discussed several decay processes of
glueballs to lighter glueballs in the gauge theory sectors which are dual to
throats. There is another process in these sectors which can take place if the
gauge theory was initially in the deconfined phase. The glueballs
which are formed at the confinement phase transition then
interact with each other for a certain period of time. This leads to a
significant reduction of the abundances of all the states heavier than the
lightest glueball, including its superpartner if the mass splitting from
supersymmetry breaking is not too small. In this Appendix, we analyse
this process in some detail:

For simplicity, we focus on only two glueball species. Generically, the 
glueball effective action includes couplings of the type
\be
\label{quarticcouplings}
\mathcal{H} \mathcal{H} \, \mathcal{G} \mathcal{G} \,,
\ee
where $\mathcal{H}$ and $\mathcal{G}$ are the heavy and light glueball
respectively, and all Lorentz- and/or spinor-indices are appropriately 
contracted. By assumption, the masses of the two glueball species satisfy 
$\smash{m_\mathcal{G} < m_\mathcal{H}}$. As long as the two glueball
species 
are in equilibrium, the density $n_\mathcal{H}$ of the heavy glueballs is 
suppressed relative to the light glueball density $n_\mathcal{G}$ by an 
exponential factor 
\be
e^{-(m_\mathcal{H}-m_\mathcal{G})/\tilde{T}}
\ee
after the temperature $\tilde{T}$ of the glueball gas falls below $m_\IR$.
This
exponential decrease of the number density of $\mathcal{H}$ glueballs
continues until they are so dilute that they decouple. This happens, when
\be
\label{freeze-out}
n_\mathcal{H} \cdot \langle \sigma v \rangle \, \sim \, H \,.
\ee
Here $\langle \sigma v \rangle$ is the thermally averaged product of cross 
section and relative velocity for the $2 \cdot \mathcal{H} \leftrightarrow
2 
\cdot \mathcal{G}$ process, which evaluates to~\cite{swo}\footnote{If the 
mass difference $m_\mathcal{H}  - m_\mathcal{G}$ is very small, this 
cross section is kinematically suppressed. This may happen, for example,
for the superpartner of the lightest glueball. These glueballs are then
diluted to a lesser extent.}
\be
\langle \sigma v \rangle \, \sim \, m_\IR^{-2} \,.
\ee

Since $n_\mathcal{H}$ drops exponentially after the temperature
$\tilde{T}$ falls below $m_\IR$, the heavy glueballs decouple when the
temperature of the glueball gas is still of the order of $m_\IR$. We can
therefore derive the freezeout density of the heavy glueballs from
Eq.~(\ref{freeze-out}) by using the phase-transition Hubble rate
$H(T_\pt)$. Furthermore, we can approximate the light glueball density by
$m_\IR^3$. The ratio of heavy and light glueball densities directly after
freezeout, i.e. the dilution factor, is then given by 
\be
\label{dilutionfactor}
\frac{n_\mathcal{H}}{n_\mathcal{G}} \, \sim \, \frac{H(T_\pt)}{m_\IR} \, 
\sim \, \frac{g^{1/2} m_\IR M_4^{1/2}}{ N_\UV T_\RH^{3/2}}\,.
\ee
Here we have calculated the Hubble rate according to Eqs.~(\ref{rts}) and 
(\ref{Tpt}) and disregarded a small power of $N_\IR$. Our formula is valid 
if the right-hand side is smaller than 1. If, however, the right-hand side 
is formally larger than 1, the $\mathcal{H}$ glueballs are decoupled from
the 
beginning and not diluted at all.

\chapter*{Acknowledgements}
\addcontentsline{toc}{chapter}{Acknowledgements}
First and foremost, I would like to express my gratitude to Arthur Hebecker
for his supervision of this thesis. He helped me in many ways and always had
an open ear when I had questions. For
instance, I remember discussing with him on a sunday evening,
the day before Christmas Eve! There are certainly not many supervisors who
care so much for their students. Even more, I have also considerably benefited
from his knowledge and intuition in physics and from his pedagogical talent.
For all this, many thanks!

I am also indebted to Michael G. Schmidt for kindly agreeing to co-referee
this thesis, for teaching me a lot and for his support during the last years.

Part of the research for this thesis was done together with Sebastian
Halter and Tatsuya Noguchi (in addition to Arthur Hebecker). I would like to
thank them for an enjoyable collaboration. Furthermore, this thesis and
myself have benefited from interesting and helpful discussions with Andreas
Braun, Jens Braun, Wilfried Buchm\"uller, Xingang Chen, Jim Cline, Thomas
Dent, Dennis Dietrich, Hassan Firouzjahi, Andrew Frey, Christoph L\"udeling,
Stefan Groot Nibbelink, Kris Sigurdson, Hagen Triendl, Roberto Valandro and
Michele Trapletti. Special thanks go to Andreas Braun and Christoph L\"udeling
for their comments on a draft of this thesis. In addition, I would like to
thank my other colleagues in Heidelberg for creating a very pleasant
and stimulating atmosphere, especially Rainer Ebert, Sebastian Gerigk,
Christian Gross, Tae-Won Ha, Johannes
Held, Patrick Pl\"otz, Christian Viereck and Robert Ziegler.

My thesis was financially backed by a position at the Institute for
Theoretical Physics in Heidelberg, for which I am grateful. In this respect, I
would also like to thank the ``Ritterschaft des F\"urstentums L\"uneburg''
for additional financial support.

I am grateful to my parents for so much help and support. Et enfin, mais
s\^urement pas en dernier lieu, j'aimerais dire merci mille fois
\`a Emilie Omn\`es pour son soutien and sa compr\'ehension pendant les
derniers ann\'ees.

\end{document}